\documentclass[11pt]{article}
\pdfoutput=1
\usepackage{jheppub}
\usepackage[utf8]{inputenc}
\usepackage{graphicx}
\usepackage{amsfonts}
\usepackage[dvipsnames]{xcolor}
\usepackage{enumerate}
\usepackage{amsmath}
\usepackage{amssymb}
\usepackage[mathscr]{eucal}
\usepackage{slashed}
\usepackage{array}
\usepackage{amsthm}
\usepackage{epsfig}
\usepackage{wasysym} 
\usepackage{tikz-cd} 
\usepgfmodule{shapes}
\usetikzlibrary{backgrounds, arrows,calc,shapes,decorations.pathreplacing, decorations.markings, automata,positioning,calligraphy}
\usepackage{ytableau} 
\usepackage{dirtytalk} 
\usepackage{cancel} 
\usepackage{changepage} 
\usepackage{float} 
\usepackage{comment} 
\usepackage{stackengine}
\usepackage{dsfont} 

\tikzset{
  arrow/.pic={\path[tips,every arrow/.try,->,>=#1] (0,0) -- +(.1pt,0);},
  pics/arrow/.default={latex,very thick}
}

\stackMath
\newcommand\tsup[2][2]{%
 \def\useanchorwidth{T}%
  \ifnum#1>1%
    \stackon[-.5pt]{\tsup[\numexpr#1-1\relax]{#2}}{\scriptscriptstyle\sim}%
  \else%
    \stackon[.5pt]{#2}{\scriptscriptstyle\sim}%
  \fi%
}

\newcommand{\be}{\begin{equation}}
\newcommand{\ee}{\end{equation}}
\newcommand{\bea}{\begin{eqnarray}}
\newcommand{\eea}{\end{eqnarray}}
\newcommand{\ba}{\begin{array}}
\newcommand{\ea}{\end{array}}
\newcommand{\bpic}{\begin{tikzpicture}}
\newcommand{\epic}{\end{tikzpicture}}

\newcommand{\nn}{\nonumber}

\def\Tr{{\rm Tr}}
\newcommand{\tr}{\text{tr}\,}

\newcommand{\Flip}[1]{\text{Flip#1}}
\newcommand{\Flipper}[1]{\text{Flipper#1}}

\newcommand{\Llra}{\Longleftrightarrow}

\newcommand\pt{\tilde p}
\newcommand\qt{\tilde q}

\newcommand\Bt{\tilde B}

\newcommand\Qt{\tilde Q}
\newcommand\Pt{\tilde P}

\newcommand{\cA}{\mathcal{A}}

\newcommand{\cN}{\mathcal{N}}
\newcommand{\cO}{\mathcal{O}}

\newcommand{\cS}{\mathcal{S}}

\newcommand{\cU}{\mathcal{U}}

\newcommand{\cW}{\mathcal{W}}

\def\a{\alpha} 
\def\b{\beta}

\def\l{\lambda}

\def\s{\sigma}

\def\P{\Phi}

\def\Bt{\tilde{B}}

\newcommand{\udl}[1]{\mathrm{d} #1 \,}

\newcommand{\Gpq}[1]{\Gamma_e\left( #1\right)}

\newcommand{\PE}{\mathop{\rm PE}}

\title{S-confining gauge theories and supersymmetry enhancements}

\author[1]{Stephane Bajeot}
\author[2]{Sergio Benvenuti}
\author[3]{Matteo Sacchi}

\affiliation[1]{SISSA, Via Bonomea 265, 34136 Trieste, Italy}
\affiliation[2]{INFN, Sezione di Trieste, SISSA, via Bonomea 265, 34136 Trieste, Italy}
\affiliation[3]{Mathematical Institute, University of Oxford, Andrew-Wiles Building, Woodstock Road, Oxford, OX2 6GG, UK}
\emailAdd{sbajeot@sissa.it, benve79@gmail.com, matteo.sacchi@maths.ox.ac.uk}

\abstract{We propose new classes of $4d$ $\mathcal{N}\!=\!1$ S-confining gauge theories,  with a simple gauge group, rank-two matter and cubic superpotentials. The gauge group can be symplectic, orthogonal or special unitary. In some cases we derive the dualities via the deconfinement technique that uses iteratively known, more fundamental, dualities. In the symplectic case we discuss the $3d$ reduction to a confining unitary gauge theory with monopole superpotential. This $3d$ S-confinement provides an understanding of a recently proposed $4d$ $\mathcal{N}\!=\!1$ theory that flows to the conformal manifold of $\mathcal{N}\!=\!4$ SYM with $SU(2n+1)$ gauge group. The $3d$ perspective allows us to generalize this construction to another similar flow with supersymmetry enhancement: a $4d$ $\mathcal{N}\!=\!1$ theory that flows to the conformal manifold of a $4d$ $\mathcal{N}\!=\!2$ necklace theory with $SU(2n+1)^3$ gauge group.}

\begin{document}

\maketitle

\section{Introduction and summary}
\label{sec:intro}

During recent years, research in the area of four dimensional theories with minimal supersymmetry, $4d$ $\cN=1$, provided us with various instances of the phenomenon of \emph{supersymmetry enhancement}: $\cN=1$ theories which flow at strong coupling to superconformal fields theories (SCFTs) with $\cN=2,3,4$ supersymmetry \cite{Gadde:2015xta, Maruyoshi:2016tqk, Maruyoshi:2016aim, Agarwal:2016pjo, Benvenuti:2017lle, Benvenuti:2017kud, Agarwal:2017roi, Benvenuti:2017bpg, Giacomelli:2017ckh, Agarwal:2018ejn, Agarwal:2018oxb, Zafrir:2019hps, Garozzo:2020pmz, Zafrir:2020epd, Etxebarria:2021lmq, Kang:2023pot}. More precisely, $\mathcal{N}\geq 2$ theories might possess a conformal manifold of $\mathcal{N}=1$ exactly marginal deformations and one might be able to reach a point of this conformal manifold via Renormalization Group (RG) flow of some other $4d$ $\mathcal{N}=1$ ultraviolet (UV) theory.

Another recent line of research is concerned with the derivation of infra-red (IR) dualities involving rank-two matter applying iteratively known more basic dualities. An IR duality is a statement about two different microscopic theories flowing to the same fixed point at low energies. Not all IR dualities are independent of each other, in particular some of those with matter fields in a rank-two representation of the gauge group can be derived from more elementary ones with only fields in the fundamental representation by using a strategy that goes under the name of \emph{sequential deconfinement} \cite{Berkooz:1995km,Pouliot:1995me,Luty:1996cg,Benvenuti:2017kud, Giacomelli:2017vgk, Pasquetti:2019uop, Pasquetti:2019tix, Sacchi:2020pet, Benvenuti:2020gvy, Bajeot:2022kwt, Bottini:2022vpy, Bajeot:2022lah, Benvenuti:2021nwt, Bajeot:2022wmu}. The idea is to use a confining duality with fundamental matter fields to trade the rank-two field for a new gauge node. This gives a quiver gauge theory that can be further dualized using basic dualities so to reach the desired dual frame. A similar logic \cite{Hwang:2020wpd,Bottini:2021vms,Hwang:2021ulb,Comi:2022aqo} allowed to derive $4d$ and $3d$ mirror symmetry with eight and four supercharges from the Intriligator--Pouliot \cite{Intriligator:1995ne} and the Aharony duality \cite{Aharony:1997gp}, respectively. Interestingly, the deconfinement procedure also has an avatar in the math literature once implemented at the level of some supersymmetric partition function such as the $4d$ $\mathcal{N}=1$ supersymmetric index, see e.g.~\cite{2003math......9252R,spiridonov2004theta,Spiridonov:2009za,Spiridonov:2011hf}.

The present paper positions itself at the crossroad of these two lines of research.

\vspace{0.3cm}

In the first part of the paper, we propose new $4d$ $\cN=1$ S-confining theories, with non-zero, cubic, superpotential and a simple gauge group, that can be symplectic, orthogonal or special unitary. An S-confining theory is a theory that flows in the infrared to a Wess--Zumino (WZ) model, that is a theory with trivial gauge group \cite{Csaki:1996zb}. This means that all the gauge invariant degrees of freedom can be described by gauge singlet chiral fields that are free at low energies and thus the moduli space of the gauge theory is smooth, which is the reason for the name ``S-confinement".

The matter content of our gauge theories is given by a rank-two matter field $\phi$ (which for symplectic, orthogonal or special unitary gauge group sits in the antisymmetric, symmetric and adjoint representation, respectively) and fundamental matter $p$. The number of fundamental fields is tuned in such a way that the gauge theory is confining. Our examples generalize in particular those of \cite{Csaki:1996zb} by having a non-trivial superpotential for these fields. One class of such S-confining dualities  has a superpotential of the form $\phi pp$, and all the fundamentals enter the superpotential. Another class   has a superpotential of the form $\phi^3 + \phi pp$, and only a subset of the fundamentals enter the superpotential. Among this second class, the special unitary case is actually not S-confining, but more precisely it has a quantum deformed moduli space, similarly to what happens e.g.~for the $SU(2)$ SQCD with four fundamental chirals \cite{Seiberg:1994bz}. We discuss the details of the theories in Section \ref{4dSconfining}, while in Section \ref{sec:derivation} we derive some of the previously stated dualities, using deconfinement techniques as in \cite{Bajeot:2022kwt,Bottini:2022vpy} and/or Kutasov--Schwimmer-like dualities   \cite{Kutasov:1995ve,Kutasov:1995np,Intriligator:1995ff}. \emph{Pole pinching} in the superconformal index is also helpful \cite{Gaiotto:2012xa}. We provide additional checks of our proposals in Appendix \ref{app:anomalies}, where we match 't Hooft anomalies.

Reducing to $3d$ the S-confining duality for $USp(2n)$ with antisymmetric $a$ and $\cW \sim app$, using the methods of \cite{Benini:2017dud}, we obtain the following $3d$ $\cN=2$ S-confining duality:\footnote{Throughout this paper we color code quiver nodes as follows: $U(n)$ nodes are black, $SU(n)$ nodes are red, $USp(2n)$ nodes are blue, $SO(n)$ nodes are green and $O(n)$ nodes are purple.}
\be \label{3dSconfining} \scalebox{0.85}{\bpic[node distance=2cm,gUnode/.style={circle,black,draw,minimum size=8mm},gSUnode/.style={circle,red,draw,minimum size=8mm},gUSpnode/.style={circle,blue,draw,minimum size=8mm},gSOnode/.style={circle,ForestGreen,draw,minimum size=8mm},fnode/.style={rectangle,draw,minimum size=8mm}]
\begin{scope}[shift={(-3,0)}]    
\node[gUnode] (G1) at (0,0) {$n$};
\node[fnode,red] (F1) at (2.5,0) {$2n+1$};
\draw (0.4,0.2) to[out=30,in=150] pic[pos=0.5,sloped,very thick]{arrow=latex reversed} (1.8,0.2);
\draw (0.4,-0.2) to[out=-30,in=-150] pic[pos=0.5,sloped]{arrow} (1.8,-0.2);
\draw (-0.3,0.3) to[out=180,in=90]  (-0.8,0) to[out=-90,in=180] (-0.3,-0.3);
\node[right] at (-0.8,-1) {$ \cW= \phi q \qt +\mathfrak{M}^++\mathfrak{M}^-$};
\node at (-0.9,0.4) {$\phi$};
\node at (0.9,0.7) {$q$};
\node at (0.9,-0.1) {$\qt$};
\node at (5,0) {$\Llra$};
\end{scope}
\begin{scope}[shift={(4.5,0)}]
\node[fnode,red] at (0,0) {$2n+1$};
\draw (0.3,0.4) to[out=90,in=0]  (0,0.9) to[out=180,in=90] (-0.4,0.4);
\node[right] at (-0.8,-1) {$ \cW = \Phi^3$};
\node at (0.6,0.9) {$\Phi$};
\end{scope}
\epic} \ee 
On the left hand side of \eqref{3dSconfining} the superpotential includes monopole terms, while on the right side it is important that a cubic $SU(2n+1)$-invariant superpotential is present.

\vspace{0.3cm}

In the second part of the paper, we use the  $3d$ duality \eqref{3dSconfining} to explain and generalize the $4d$ susy enhancement recently proposed in \cite{Kang:2023pot}. Indeed, \cite{Kang:2023pot} provided compelling evidence that an $\cN=1$ $SU(2n+1)$ diagonal gauging of three copies of the $\cN=2$ non-Lagrangian theory $D_2(SU(2n+1))$ \cite{Xie:2012hs,Cecotti:2012jx,Cecotti:2013lda} flows to the conformal manifold of $4d$ $\cN=4$ SYM with gauge group $SU(2n+1)$. The work \cite{Kang:2023pot} was a result of a detailed analysis of the landscape of $4d$ $\mathcal{N}=1,2$ theories that have $a=c$ conformal central charges at finite $N$ \cite{Kang:2021lic,Kang:2021ccs,Kang:2022zsl,Kang:2022vab}. From their analysis, in particular the matching of anomalies, superconformal indices and certain operators in the spectrum, the intuition that one gets is that each copy of $D_2(SU(2n+1))$ is morally replacing an adjoint chiral in the SYM theory.

We reduce this proposed duality on a circle and use the fact that the $D_2(SU(2n+1))$ theory becomes Lagrangian in $3d$, namely it is the $3d$ $\cN=4$ $U(n)$ SQCD with $2n+1$ flavors \cite{Xie:2012hs,Buican:2015hsa,Giacomelli:2020ryy,Closset:2020afy}. Modulo the monopole superpotential, which as we will argue is dynamically generated once we compactify the $4d$ theory on a point of its conformal manifold that breaks all the abelian symmetries, this is the left hand side of \eqref{3dSconfining}. The $3d$ interpretation is summarized by the following diagram:
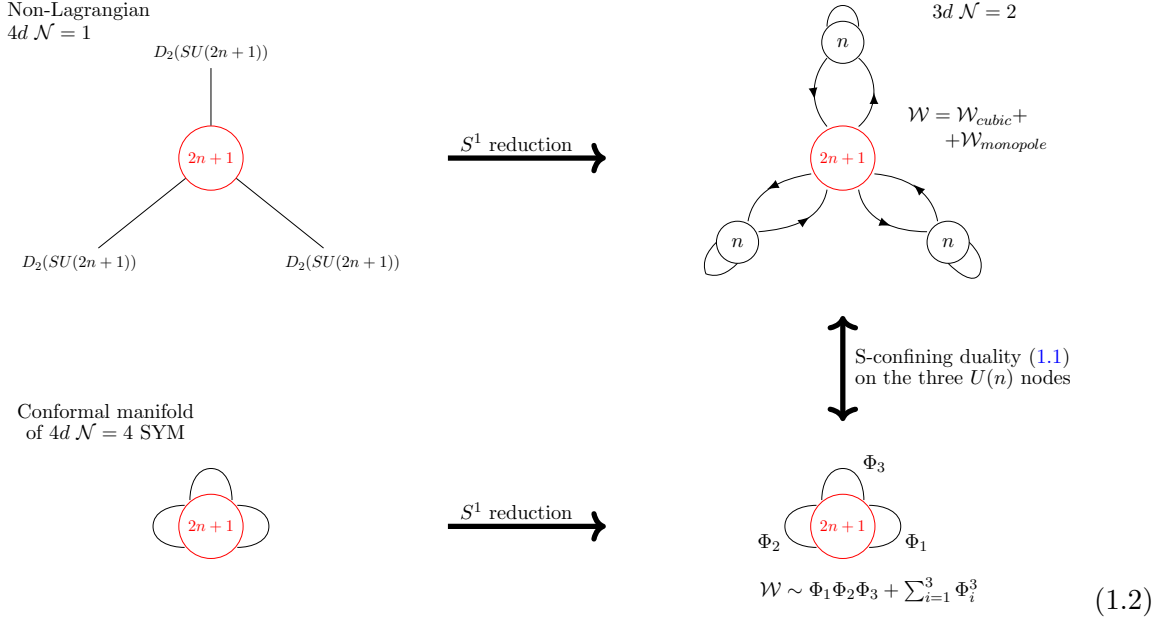
\be \label{expN=4} \scalebox{0.7}{\bpic[node distance=2cm,gUnode/.style={circle,black,draw,minimum size=8mm},gSUnode/.style={circle,red,draw,minimum size=8mm},gUSpnode/.style={circle,blue,draw,minimum size=8mm},gSOnode/.style={circle,ForestGreen,draw,minimum size=8mm},fnode/.style={rectangle,draw,minimum size=8mm}]
\begin{scope}[shift={(0,7)}]    
\node[right] at (-4,2.8) {Non-Lagrangian};
\node[right] at (-4,2.4) {$4d$ $\cN=1$};
\node[gSUnode] (G1) at (0,0) {\scalebox{0.8}{$2n+1$}};
\node (F1) at (2.5,-2) {\scalebox{0.8}{$D_2(SU(2n+1))$}};
\node (F2) at (-2.5,-2) {\scalebox{0.8}{$D_2(SU(2n+1))$}};
\node (F3) at (0,2) {\scalebox{0.8}{$D_2(SU(2n+1))$}};
\draw (G1) -- (F1);
\draw (G1) -- (F2);
\draw (G1) -- (F3);
\node at (5.8,0.3) {$S^1$ reduction};
\draw[->, line width=1mm] (4.5,0) -- (7.5,0);
\end{scope};
\begin{scope}[shift={(12,7)}]    
\node at (2.5,2.8) {$3d$ $\cN=2$};
\node[gSUnode] (G1) at (0,0) {\scalebox{0.8}{$2n+1$}};
\node[gUnode] (G2) at (2.0,-1.6) {$n$};
\node[gUnode] (G3) at (-2.0,-1.6) {$n$};
\node[gUnode] (G4) at (0,2.2) {$n$};
\node[right] at (1.1,0.8) {$\mathcal{W} = \mathcal{W}_{cubic} + $};
\node[right] at (1.8,0.3) {$ + \mathcal{W}_{monopole} $};
\draw (2.4,-1.6) to[out=-45,in=70] (2.6,-2.2) to[out=-120,in=-45] (2,-2);
\draw (-2.4,-1.6) to[out=-135,in=110]  (-2.6,-2.2) to[out=-20,in=-135] (-2,-2);
\draw (0.3,2.5) to[out=90,in=0]  (0,2.9) to[out=180,in=90] (-0.3,2.5);
\draw (0.6,-0.3) to[out=-05,in=100] pic[pos=0.6,sloped,very thick]{arrow=latex reversed} (1.8,-1.2);
\draw (0.3,-0.6) to[out=-80,in=180] pic[pos=0.6,sloped]{arrow} (1.6,-1.4);
\draw (-0.6,-0.3) to[out=-170,in=80] pic[pos=0.4,sloped,very thick]{arrow=latex reversed} (-1.8,-1.2);
\draw (-0.3,-0.6) to[out=-100,in=05] pic[pos=0.4,sloped]{arrow} (-1.6,-1.4);
\draw (0.3,0.6) to[out=40,in=-40] pic[pos=0.5,sloped]{arrow} (0.3,1.9);
\draw (-0.3,0.6) to[out=130,in=-130] pic[pos=0.4,sloped]{arrow} (-0.3,1.9);
\node[right] at (0.1,-3.8) {S-confining duality \eqref{3dSconfining}};
\node[right] at (0.1,-4.2) {on the three $U(n)$ nodes};
\draw[<->, line width=1mm] (0,-3) -- (0,-5);
\end{scope};
\begin{scope}[shift={(12,0)}]    
\node[gSUnode] (G1) at (0,0) {\scalebox{0.8}{$2n+1$}};
\draw (0.5,-0.4) to[out=0,in=-90] (1.1,0) to[out=90,in=0] (0.5,0.4);
\draw (-0.5,-0.4) to[out=180,in=-90] (-1.1,0) to[out=90,in=180] (-0.5,0.4);
\draw (0.4,0.5) to[out=90,in=0] (0,1.1) to[out=180,in=90] (-0.4,0.5);
\node at (1.4,-0.3) {$\P_1$};
\node at (-1.4,-0.3) {$\P_2$};
\node at (0.6,1.2) {$\P_3$};
\node at (0.5,-1.2) {$\mathcal{W} \sim \P_1 \P_2 \P_3 + \sum_{i=1}^3\P_i^3$};
\end{scope};
\begin{scope}[shift={(0,0)}]    
\node[gSUnode] (G1) at (0,0) {\scalebox{0.8}{$2n+1$}};
\draw (0.5,-0.4) to[out=0,in=-90] (1.1,0) to[out=90,in=0] (0.5,0.4);
\draw (-0.5,-0.4) to[out=180,in=-90] (-1.1,0) to[out=90,in=180] (-0.5,0.4);
\draw (0.4,0.5) to[out=90,in=0] (0,1.1) to[out=180,in=90] (-0.4,0.5);
\node at (-2,2.2) {Conformal manifold};
\node at (-2,1.8) {of $4d$ $\cN=4$ SYM};
\node at (5.8,0.3) {$S^1$ reduction};
\draw[->, line width=1mm] (4.5,0) -- (7.5,0);
\end{scope};
\epic} \ee
Since the $3d$ reduction of the non-Lagrangian $4d$ $\cN=1$ theory is, thanks to the S-confinement \eqref{3dSconfining}, dual to the $3d$ reduction of $\cN=4$ SYM, it is natural to expect that also the $4d$ theories on the left side of \eqref{expN=4} are dual to each other. Notice in particular that the $3d$ S-confining duality confirms the $4d$ intuition that each copy of $D_2(SU(2n+1))$ plays the role of one of the adjoint chirals of $\mathcal{N}=4$ SYM. We point out that leveraging the $S_3$ permutation symmetry of the $4d$ non-Lagrangian theory at $\cW=0$, it is possible to determine the R-charges without using a-maximization. Such symmetry also implies that the $4d$ non-Lagrangian theory at $\cW=0$ is dual to a point of the $\beta$-deformation \cite{Leigh:1995ep} line with $\beta=\pi$ in the conformal manifold of $\cN=4$ SYM.  We discuss this more in details in Subsection \ref{ENHN=4}.

Armed with this $3d$ understanding, we provide a new example of supersymmetry enhancement, which follows a very similar logic. We consider the $4d$ $\cN=2$ non-Lagrangian theory $D_2(SU(6n+3))$ and perform an $\cN=1$ gauging of an $SU(2n+1)^3$ subgroup of the $SU(6n+3)$ global symmetry. We provide strong evidence that this should flow to a point of the conformal manifold of the $4d$ $\mathcal{N}=2$ necklace quiver theory with three $SU(2n+1)$ gauge nodes. In this case, the $D_2(SU(6n+3))$ theory plays the role of the three adjoint and the six bifundamental chirals. These can be understood as coming from the decomposition of the moment map for the $SU(6n+3)$ global symmetry of $D_2(SU(6n+3))$ under the gauged subgroup, as it again becomes evident from the $3d$ perspective. 

Upon reduction to $3d$, indeed, we get a four node quiver, which after dualizing the middle unitary gauge group with the S-confining duality becomes an $SU(2n+1)^3$ gauge theory which is precisely the $3d$ reduction of the $4d$ $\cN=2$ necklace quiver with three nodes. The diagram summarizing the duality and the reductions if very similar to \eqref{expN=4}
\be \label{expN=2} \scalebox{0.7}{\bpic[node distance=2cm,gUnode/.style={circle,black,draw,minimum size=8mm},gSUnode/.style={circle,red,draw,minimum size=8mm},gUSpnode/.style={circle,blue,draw,minimum size=8mm},gSOnode/.style={circle,ForestGreen,draw,minimum size=8mm},fnode/.style={rectangle,draw,minimum size=8mm}]
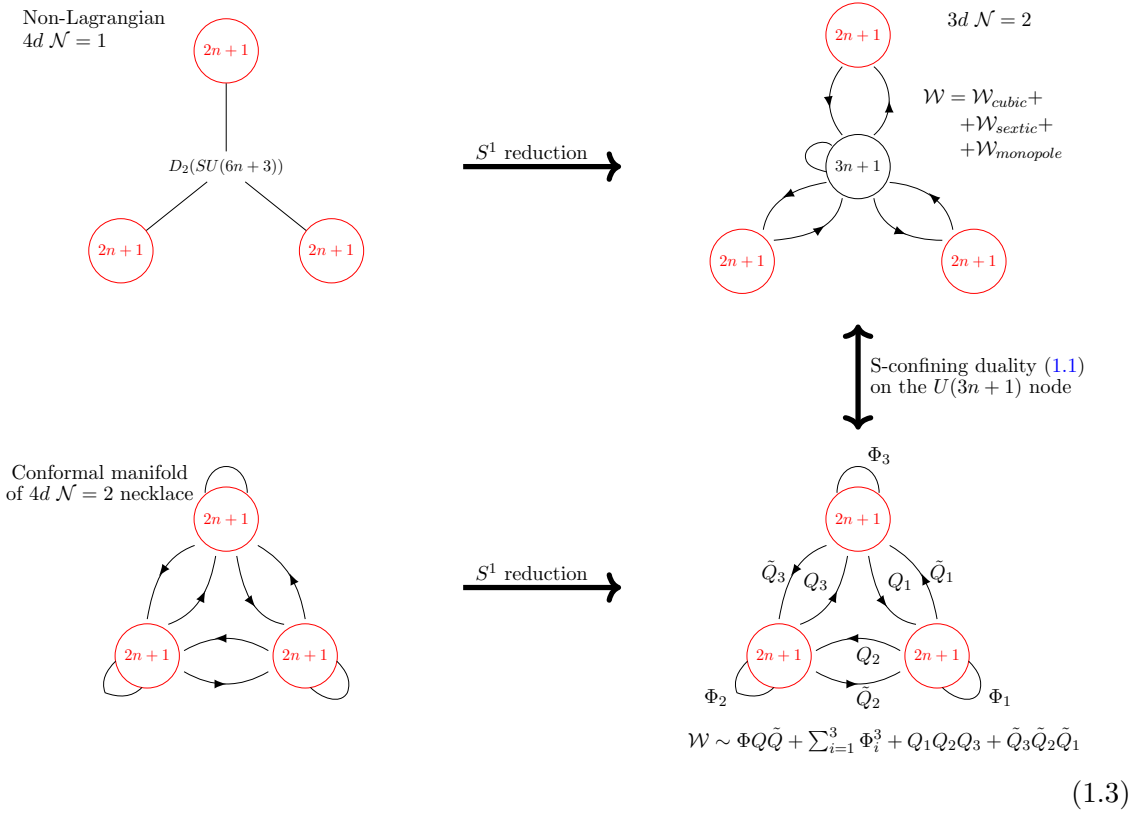
\begin{scope}[shift={(0,8)}]    
\node[right] at (-4,2.8) {Non-Lagrangian};
\node[right] at (-4,2.4) {$4d$ $\cN=1$};
\node (G1) at (0,0) {\scalebox{0.8}{$D_2(SU(6n+3))$}};
\node[gSUnode] (G2) at (2.0,-1.6) {\scalebox{0.8}{$2n+1$}};
\node[gSUnode] (G3) at (-2.0,-1.6) {\scalebox{0.8}{$2n+1$}};
\node[gSUnode] (G4) at (0,2.2) {\scalebox{0.8}{$2n+1$}};
\draw (G1) -- (G2);
\draw (G1) -- (G3);
\draw (G1) -- (G4);
\node at (5.8,0.3) {$S^1$ reduction};
\draw[->, line width=1mm] (4.5,0) -- (7.5,0);
\end{scope};
\begin{scope}[shift={(12,8)}]    
\node at (2.5,2.8) {$3d$ $\cN=2$};
\node[gUnode] (G1) at (0,0) {\scalebox{0.8}{$3n+1$}};
\node[gSUnode] (G2) at (2.2,-1.8) {\scalebox{0.8}{$2n+1$}};
\node[gSUnode] (G3) at (-2.2,-1.8) {\scalebox{0.8}{$2n+1$}};
\node[gSUnode] (G4) at (0,2.5) {\scalebox{0.8}{$2n+1$}};
\node[right] at (1.1,1.3) {$\mathcal{W} = \mathcal{W}_{cubic} + $};
\node[right] at (1.8,0.8) {$+ \mathcal{W}_{sextic} + $};
\node[right] at (1.8,0.3) {$ + \mathcal{W}_{monopole} $};
\draw (-0.5,0.4) to[out=135,in=90] (-1,0.2) to[out=-90,in=165] (-0.6,-0.1);
\draw (0.6,-0.3) to[out=-05,in=100] pic[pos=0.6,sloped,very thick]{arrow=latex reversed} (1.8,-1.2);
\draw (0.3,-0.6) to[out=-80,in=180] pic[pos=0.6,sloped]{arrow} (1.6,-1.4);
\draw (-0.6,-0.3) to[out=-170,in=80] pic[pos=0.4,sloped,very thick]{arrow=latex reversed} (-1.8,-1.2);
\draw (-0.3,-0.6) to[out=-100,in=05] pic[pos=0.4,sloped]{arrow} (-1.6,-1.4);
\draw (0.3,0.6) to[out=40,in=-40] pic[pos=0.5,sloped]{arrow} (0.3,1.9);
\draw (-0.3,0.6) to[out=130,in=-130] pic[pos=0.4,sloped]{arrow} (-0.3,1.9);
\node[right] at (0.1,-3.8) {S-confining duality \eqref{3dSconfining}};
\node[right] at (0.1,-4.2) {on the $U(3n+1)$ node};
\draw[<->, line width=1mm] (0,-3) -- (0,-5);
\end{scope};
\begin{scope}[shift={(12,0)}]    
\node[gSUnode] (G1) at (1.5,-1.3) {\scalebox{0.8}{$2n+1$}};
\node[gSUnode] (G2) at (-1.5,-1.3) {\scalebox{0.8}{$2n+1$}};
\node[gSUnode] (G3) at (0,1.3) {\scalebox{0.8}{$2n+1$}};
\draw (2.1,-1.4) to[out=-45,in=70] (2.3,-2) to[out=-120,in=-45] (1.6,-1.9);
\draw (-2.1,-1.4) to[out=-135,in=110] (-2.3,-2) to[out=-20,in=-135] (-1.6,-1.9);
\draw (0.4,1.8) to[out=90,in=0]  (0,2.3) to[out=180,in=90] (-0.4,1.8);
\draw (0.6,0.8) to[out=-30,in=100] pic[pos=0.6,sloped,very thick]{arrow=latex reversed} (1.5,-0.6);
\draw (0.2,0.6) to[out=-80,in=170] pic[pos=0.6,sloped]{arrow} (1.1,-0.7);
\draw (-0.8,-1.2) to[out=30,in=150] pic[pos=0.5,sloped,very thick]{arrow=latex reversed} (0.8,-1.2);
\draw (-0.8,-1.6) to[out=-30,in=-150] pic[pos=0.5,sloped]{arrow} (0.8,-1.6);
\draw (-0.6,0.8) to[out=-160,in=80] pic[pos=0.4,sloped,very thick]{arrow=latex reversed} (-1.5,-0.6);
\draw (-0.2,0.6) to[out=-100,in=20] pic[pos=0.4,sloped]{arrow} (-1.1,-0.7);
\node at (1.6,0.3) {$\Qt_1$};
\node at (0.8,0.1) {$Q_1$};
\node at (0.2,-2.1) {$\Qt_2$};
\node at (0.2,-1.3) {$Q_2$};
\node at (-1.6,0.3) {$\Qt_3$};
\node at (-0.8,0.1) {$Q_3$};
\node at (2.7,-2.1) {$\Phi_1$};
\node at (-2.7,-2.1) {$\Phi_2$};
\node at (0.4,2.5) {$\Phi_3$};
\node at (0.5,-2.9) {$\mathcal{W} \sim \P Q \Qt+ \sum_{i=1}^3\P_i^3+Q_1 Q_2 Q_3 + \Qt_3 \Qt_2 \Qt_1$};
\end{scope};
\begin{scope}[shift={(0,0)}]    
\node[gSUnode] (G1) at (1.5,-1.3) {\scalebox{0.8}{$2n+1$}};
\node[gSUnode] (G2) at (-1.5,-1.3) {\scalebox{0.8}{$2n+1$}};
\node[gSUnode] (G3) at (0,1.3) {\scalebox{0.8}{$2n+1$}};
\draw (2.1,-1.4) to[out=-45,in=70] (2.3,-2) to[out=-120,in=-45] (1.6,-1.9);
\draw (-2.1,-1.4) to[out=-135,in=110] (-2.3,-2) to[out=-20,in=-135] (-1.6,-1.9);
\draw (0.4,1.8) to[out=90,in=0]  (0,2.3) to[out=180,in=90] (-0.4,1.8);
\draw (0.6,0.8) to[out=-30,in=100] pic[pos=0.6,sloped,very thick]{arrow=latex reversed} (1.5,-0.6);
\draw (0.2,0.6) to[out=-80,in=170] pic[pos=0.6,sloped]{arrow} (1.1,-0.7);
\draw (-0.8,-1.2) to[out=30,in=150] pic[pos=0.5,sloped,very thick]{arrow=latex reversed} (0.8,-1.2);
\draw (-0.8,-1.6) to[out=-30,in=-150] pic[pos=0.5,sloped]{arrow} (0.8,-1.6);
\draw (-0.6,0.8) to[out=-160,in=80] pic[pos=0.4,sloped,very thick]{arrow=latex reversed} (-1.5,-0.6);
\draw (-0.2,0.6) to[out=-100,in=20] pic[pos=0.4,sloped]{arrow} (-1.1,-0.7);
\node at (-2.4,2.2) {Conformal manifold};
\node at (-2.4,1.8) {of $4d$ $\cN=2$ necklace};
\node at (5.8,0.3) {$S^1$ reduction};
\draw[->, line width=1mm] (4.5,0) -- (7.5,0);
\end{scope};
\epic} \ee
This logic strongly suggests that the $4d$ $\cN=1$ theory in the top left of \eqref{expN=2} flows to the same conformal manifold of the $4d$ $\cN=2$ theory on the bottom left  of \eqref{expN=2}. On top of the $3d$ analysis, we also check this statement by matching anomalies and superconformal indices. See Subsection \ref{ENHN=2} for more details. The $3d$ perspective also suggests that many examples of $4d$ susy enhancements can be produced in a similar way, essentially by replacing the chiral fields of a $4d$ $\mathcal{N}=2$ theory by one or multiple copies of $D_2(SU(\text{odd}))$ theories of which (part of) the global symmetry is gauged in an $\mathcal{N}=1$ fashion.

\vspace{0.3cm}The paper is organized as follows.
In Section \ref{4dSconfining} we propose the new S-confining theories, with simple gauge group and cubic superpotential. In the case of $USp(2n)$ gauge group and $\cW \sim a p p$, we discuss the reduction on a circle, which upon turning on appropriate real masses, leads to a $3d$ $U(n)$ gauge theory with $2n+1$ flavors and a monopole superpotential which is dual to an adjoint field $\Phi$ with cubic superpotential.
In Section \ref{sec:derivation} we derive some of the previously stated dualities, using deconfinement techniques and/or Kutasov--Schwimmer-like dualities. \emph{Pole pinching} in the superconformal index is also helpful. In Section \ref{sec:ENH} we turn to the study of various $4d$ $\mathcal{N}=1$ theories with susy enhancement and their understanding using the $3d$ S-confining duality upon circle compactification.
Specifically, in Subsection \ref{ENHN=4} we provide a $3d$ explanation of the susy enhancement of the theory proposed in \cite{Kang:2023pot}, that is an $\cN=1$ $SU(2n+1)$ gauging of three copies of the $\cN=2$ theory $D_2(SU(2n+1))$ which flows to a point of the conformal manifold of $4d$ $\cN=4$ SYM with gauge group $SU(2n+1)$.
Based on the $3d$ understanding of this case, in Subsection \ref{ENHN=2} we are then able to generalize it and give as a new example the $\cN=1$ $SU(2n+1)^3$ gauging of a single copy of the $\cN=2$ $D_2(SU(6n+3))$ theory which flows on the conformal manifold of the $4d$ $\cN=2$ necklace quiver with three $SU(2n+1)^3$ gauge group. Several appendices supplement the main text, with our conventions and additional checks of various statements. In particular, Appendix \ref{CMnecklace} discusses the $\mathcal{N}=1$ conformal manifold of the $4d$ $\mathcal{N}=2$ necklace quiver theory for generic number $k$ of nodes and generic rank of the $SU(N)$ gauge group. 

\section{$4d$ and $3d$ S-confining theories with cubic superpotential}
\label{4dSconfining}

The S-confining theories involving a simple gauge group and zero superpotential have been classified in \cite{Csaki:1996zb} and later derived by iterative use of elementary dualities in \cite{Bajeot:2022kwt,Bottini:2022vpy}. In this section we discuss new theories of this kind by lifting the assumption of zero superpotential. Here we will only state the results and some of the associated supersymmetric index identities, while in Section \ref{sec:derivation} we will provide a derivation of some of those and in Appendix \ref{app:anomalies} we will show additional tests based on anomaly matching for all of them.

\subsection{Theories with $\cW \sim \phi pp$}

We start in this subsection by considering the new S-confining dualities where the superpotential is of the form $\cW \sim \phi pp$, where $\phi$ is a generic chiral field in some rank-2 representation of the gauge group.

\subsubsection{$USp(2n)$ gauge group: $\cU_1[n]$ theories} 

The first example involves a $USp(2n)$ gauge group\footnote{For most of the paper we will only specify the Lie algebras and ignore issues related to the global structure of groups.} with an antisymmetric traceless chiral field $a$ and $4n+2$ chiral fields $p$ in the fundamental representation. We turn on the superpotential\footnote{Here $J^{(2n)}= \mathds{1}_{n} \otimes i\sigma_2 $ is the totally antisymmetric invariant tensor of $USp(2n)$. In the following we will often omit the contraction of indices.}
\be
\cW = app=a^{il}p^j_{\a}p^k_{\b}J^{(2n)}_{ij}J^{(2n)}_{kl}J_{(4n+2)}^{\a \b}\,.
\ee
This changes the non-abelian flavor symmetry into $USp(4n+2)$ and breaks one abelian global symmetry, so that the total non-anomalous global symmetry of the theory is
\be
USp(4n+2)\times U(1)_R\,.
\ee
The theory without superpotential and arbitrary number of fundamentals has been studied in \cite{Bajeot:2022lah}.

The Wess--Zumino (WZ) model that describes the IR physics of the gauge theory consists of a field in the antisymmetric traceless representation of the $USp(4n+2)$ flavor symmetry and the trace part\footnote{The trace part is the $USp(4n+2)$ singlet defined as $\tr A \equiv A^{i j} \, J_{ij}^{(4n+2)}$.}. The proposed duality can be summarized by the following quiver:
\be \label{U1} \scalebox{0.85}{\bpic[node distance=2cm,gSUnode/.style={circle,red,draw,minimum size=8mm},gUSpnode/.style={circle,blue,draw,minimum size=8mm},gSOnode/.style={circle,ForestGreen,draw,minimum size=8mm},fnode/.style={rectangle,draw,minimum size=8mm}]
\begin{scope}[shift={(-3,0)}]    
\node at (-2,1) {$\cU_1[n]:$};
\node[gUSpnode] (G1) at (0,0) {$2n$};
\node[fnode, blue] (F1) at (2.5,0) {$4n+2$};
\draw (G1) -- (F1) node[midway,above] {$p$};
\draw (0.3,0.3) to[out=90,in=0]  (0,0.7) to[out=180,in=90] (-0.3,0.3);
\node[right] at (0,-1.3) {$ \cW= a p p$};
\node at (0.6,0.9) {$a$, \textcolor{Green}{$\frac{4}{3}$}};
\node at (0.9,-0.4) {\textcolor{Green}{$\frac{1}{3}$}};
\node at (5,0) {$\Llra$};
\end{scope}
\begin{scope}[shift={(4.5,0)}]
\node[fnode, blue] (F2) at (0,0) {$4n+2$};
\draw (0.3,0.4) to[out=90,in=0]  (0,0.9) to[out=180,in=90] (-0.4,0.4);
\node at (0,-1.3) {$ \cW = A^3 + \tr(A) \, A^2$};
\node at (0.7,0.9) {$A$, \textcolor{Green}{$\frac{2}{3}$}};
\end{scope}
\epic} \ee 
Our convention for quivers is that boxes are flavor nodes and circles are gauge nodes, with lines connecting them denoting fields transorming in bifundamental representations of the corresponding groups and arcs fields in some rank-2 representation (in this case antisymmetric). Moreover, a blue node denotes a symplectic group.

The combination of the constraint from the vanishing of the $U(1)_R$ ABJ anomaly and the constraint from the superpotential fixes the R-charges of the fields to be 
\be\label{eq:cU1Rcharges}
R[a] = \frac{4}{3}\,,\qquad R[p] = \frac{1}{3}\,.
\ee
These are indicated in green in the quiver \eqref{U1}.
Another consequence of the superpotential is the truncation of the chiral ring of the theory. 
The mapping of the chiral ring generators is given by\footnote{We use the following notation: $\tr[ p \, p] \equiv p_{a_1}^{i_1} p_{a_2}^{i_2} J_{(2n)}^{a_1 a_2}  J_{i_1 i_2}^{(4n+2)}$.}
\be \label{mapDeconfinementUSp} 
\ba{c}
p \, p \\
\tr[p \, p] 
\ea
\quad \longleftrightarrow \quad
\ba{c}
A \\
\tr A 
\ea
\ee

At the level of the supersymmetric, this duality translates into the following non-trivial integral identity (see Appendix \ref{app:index} for our conventions):
\begin{align}\label{eq:indU1}
&\frac{(p;p)_\infty^n(q;q)_\infty^n}{2^n\, n!}\oint\prod_{a=1}^n\frac{\udl{z}_a}{2\pi i\,z_a}\frac{\prod_{a=1}^n\prod_{i=1}^{2n+1}\Gpq{(pq)^{\frac{1}{6}}z_a^{\pm1}f_i^{\pm1}}}{\prod_{a=1}^n\Gpq{z_a^{\pm2}}\prod_{a<b}^n\Gpq{z_a^{\pm1}z_b^{\pm1}}}\nn\\
&\times\Gpq{(pq)^{\frac{2}{3}}}^{n-1}\prod_{a<b}^n\Gpq{(pq)^{\frac{2}{3}}z_a^{\pm1}z_b^{\pm1}}
=\Gpq{(pq)^{\frac{1}{3}}}^{2n+1}\prod_{i<j}^{2n+1}\Gpq{(pq)^{\frac{1}{3}}f_i^{\pm1}f_j^{\pm1}}\,,
\end{align}
where $f_i$ for $i=1,\cdots,2n+1$ are the $USp(4n+2)$ flavor fugacities. 

We show the matching of anomalies for this duality in Appendix \ref{app:anomU1}, while in Subsection \ref{subsec:U1deriv} we will derive it by iterative applications of more elementary dualities. This latter derivation can be translated at the level of the index to prove the identity \eqref{eq:indU1}.

It would be interesting to understand whether it is possible to derive this S-confining duality from compactification of some 6d SCFT on a Riemann surface with flavor fluxes. For example, in \cite{Hwang:2021xyw} the duality between the $USp(2n)$ gauge theory with one antisymmetric, 6 fundamental chirals and no superpotential and the WZ model with $16n$ chirals and cubic superpotential of \cite{Csaki:1996zb} has been derived from compactification of the $6d$ $\mathcal{N}=(1,0)$ E-string SCFT on a sphere with flux.

\subsubsection{Reduction to $3d$ $U(n)$ S-confining gauge theory}
\label{sec3dSconfining}

It is interesting to consider the reduction to three dimensions of the duality $\cU_1[n]$. Similar dimensional reduction limits have been studied in \cite{Benini:2017dud,Pasquetti:2019hxf,Bottini:2021vms,Bajeot:2022lah,Comi:2022aqo}.

The limit we want to consider consists of two main steps. First, we consider the $S^1$ compactification of $\cU_1[n]$. This produces a similar $3d$ $\mathcal{N}=2$ duality, but with a dynamically generated monopole superpotential on the gauge theory side \cite{Aharony:2013dha,Aharony:2013kma}. Requiring that the fundamental monopole, i.e.~with minimal magnetic flux, of the $USp(2n)$ gauge group is exactly marginal in $3d$ indeed enforces the same constraint on the global symmetries as the cancellation of the ABJ anomaly in $4d$. In particular, it implies that, as in $4d$, also the $3d$ theory has no abelian global symmetry and the R-charges of the fields are as in \eqref{eq:cU1Rcharges}. We hence get the following $3d$ $\mathcal{N}=2$ duality:
\be \label{U13d} \scalebox{0.85}{\bpic[node distance=2cm,gSUnode/.style={circle,red,draw,minimum size=8mm},gUSpnode/.style={circle,blue,draw,minimum size=8mm},gSOnode/.style={circle,ForestGreen,draw,minimum size=8mm},fnode/.style={rectangle,draw,minimum size=8mm}]
\begin{scope}[shift={(-3,0)}]    
\node[gUSpnode] (G1) at (0,0) {$2n$};
\node[fnode, blue] (F1) at (2.5,0) {$4n+2$};
\draw (G1) -- (F1) node[midway,above] {$p$};
\draw (0.3,0.3) to[out=90,in=0]  (0,0.7) to[out=180,in=90] (-0.3,0.3);
\node[right] at (0,-1.3) {$ \cW= a p p+\mathfrak{M}$};
\node at (0.6,0.9) {$a$, \textcolor{Green}{$\frac{4}{3}$}};
\node at (0.9,-0.4) {\textcolor{Green}{$\frac{1}{3}$}};
\node at (5,0) {$\Llra$};
\end{scope}
\begin{scope}[shift={(4.5,0)}]
\node[fnode, blue] (F2) at (0,0) {$4n+2$};
\draw (0.3,0.4) to[out=90,in=0]  (0,0.9) to[out=180,in=90] (-0.4,0.4);
\node at (0,-1.3) {$ \cW = A^3 + \tr(A) \, A^2$};
\node at (0.7,0.9) {$A$, \textcolor{Green}{$\frac{2}{3}$}};
\end{scope}
\epic} \ee 

The second step consists of a combination of a real mass deformation and a Coulomb branch  vacuum expectation value (vev) that has the effect of breaking all the symplectic groups, both flavor and gauge, down to a unitary subgroup. Specifically, for each symplectic symmetry $USp(2N)$ we consider the subgroup
\be
USp(2N)\supset U(1)\times SU(N)\,,
\ee
where the embedding is
\be
{\bf 2N}\to {\bf N}^1\oplus\overline{\bf N}^{-1}\,,
\ee
and we perform a real mass deformation or a Coulomb branch vev for the $U(1)$ part, depending on whether it is a flavor or a gauge symmetry. This means that we turn on the scalar component for the background vector multiplet of such $U(1)$ subgroup if the symmetry is flavor and a vev for the scalar component of the dynamical vector multiplet of such a $U(1)$ subgroup if the symmetry is gauged, such that their values are tuned to be equal. Flowing to low energy, this combined deformation has the effect of integrating out some of the chirals and partially Higgsing the gauge group, so that we are left with a similar duality but where all the original symplectic symmetries are now unitary. In the process, a second monopole is generated in the superpotential, so that both of the fundamental monopoles of the $U(n)$ gauge group, i.e.~with magnetic flux equal to $\pm1$, are turned on. Eventually, we obtain the following $3d$ $\mathcal{N}=2$ duality:
\be \label{3dSconfiningQuiver} \scalebox{0.85}{\bpic[node distance=2cm,gUnode/.style={circle,black,draw,minimum size=8mm},gSUnode/.style={circle,red,draw,minimum size=8mm},gUSpnode/.style={circle,blue,draw,minimum size=8mm},gSOnode/.style={circle,ForestGreen,draw,minimum size=8mm},fnode/.style={rectangle,draw,minimum size=8mm}]
\begin{scope}[shift={(-3,0)}]    
\node at (-2,1) {$\cU_1^{(3d)}[n]:$};
\node[gUnode] (G1) at (0,0) {$n$};
\node[fnode,red] (F1) at (2.5,0) {$2n+1$};
\draw (0.4,0.2) to[out=30,in=150] pic[pos=0.5,sloped,very thick]{arrow=latex reversed} (1.8,0.2);
\draw (0.4,-0.2) to[out=-30,in=-150] pic[pos=0.5,sloped]{arrow} (1.8,-0.2);
\draw (-0.3,0.3) to[out=180,in=90]  (-0.8,0) to[out=-90,in=180] (-0.3,-0.3);
\node[right] at (-0.8,-1.4) {$ \cW= \phi q \qt +\mathfrak{M}^++\mathfrak{M}^-$};
\node at (-0.9,0.4) {$\phi$};
\node at (0.9,0.7) {$q$};
\node at (0.9,-0.8) {$\qt$};
\node at (5,0) {$\Llra$};
\end{scope}
\begin{scope}[shift={(4.5,0)}]
\node[fnode,red] at (0,0) {$2n+1$};
\draw (0.3,0.4) to[out=90,in=0]  (0,0.9) to[out=180,in=90] (-0.4,0.4);
\node[right] at (-0.8,-1.4) {$ \cW = \Phi^3$};
\node at (0.6,0.9) {$\Phi$};
\end{scope}
\epic} \ee 
where now the black circle node denotes a $U(n)$ gauge symmetry, the red box denotes an $SU(2n+1)$ flavor symmetry and the arcs denote adjoint chirals. In particular, on the l.h.s.~$\phi$ is taken to be in the adjoint representation of the $U(n)$ gauge symmetry including the trace part, while on the r.h.s.~$\Phi$ is in the adjoint representation of the $SU(2n+1)$ flavor symmetry and so it does not include the trace part.\footnote{In order to achieve such a configuration, we have to move one singlet corresponding to the trace part from one side of the duality to the other compared to what we had in $4d$.}

The global symmetry of both of the dual theories is just
\be
SU(2n+1)\,.
\ee
On the gauge theory side, the topological symmetry and the axial symmetry, i.e.~the symmetry assigning charge $+1$ to both $q$ and $\qt$ and $-2$ to $\phi$, are broken by the monopole superpotential, while as usual the baryonic symmetry acting with charges $+1$ and $-1$ on $q$ and $\qt$ respectively is part of the gauge symmetry. Since there is no abelian symmetry that can mix with the R-symmetry in the IR, the R-charges of the fields can just be determined by imposing that they are compatible with the superpotential. We get the same R-charges as in $4d$
\be\label{eq:cU13dRcharges}
R[\phi] = \frac{4}{3}\,,\qquad R[q] = R[\qt] = \frac{1}{3}\,.
\ee

The operator map works similarly to the original $4d$ duality, since the monopole superpotential completely lifts the Coulomb branch
\be \label{mapDeconfinementU3d} 
q\,\qt \quad \longleftrightarrow \quad \Phi\,,
\ee
where the trace part of the meson $q\,\qt$ is set to zero by the F-term equation of $\phi$.

Similarly to the comment we made at the end of the previous subsection for the $\cU_1[n]$ duality, it would be interesting to understand whether this $3d$ S-confining duality or some other related to it by RG flow can be derived from compactifications of 5d SCFTs on Riemann surfaces with flux, which have been recently investigated in \cite{Sacchi:2021afk,Sacchi:2021wvg,Sacchi:2023rtp}.

\subsubsection{$O(n)$ gauge group: $\cO_1[n]$ theories} 

The second theory is a $O(n)$ gauge theory
with a chiral $s$ in the symmetric traceless representation and $2n-2$ chirals $p$ in the vector representation. We turn on the superpotential\footnote{The indices are contracted with the totally symmetric invariant tensor $\delta^{(n)}_{ij}$ of $SO(n)$}
\be
\cW = s p p\,.
\ee
This changes the non-abelian flavor symmetry into $SO(2n-2)$ and breaks one abelian global symmetry, so that the total non-anomalous global symmetry of the theory is
\be
SO(2n-2)\times U(1)_R\,.
\ee

The WZ model that describes the IR physics of the gauge theory consists of a field in the symmetric traceless representation of the $SO(2n-2)$ flavor symmetry and the trace part\footnote{The trace part is the singlet defined as $\tr S \equiv S^{i j} \, \delta_{ij}^{(2n-2)}$.}. The proposed duality can be summarized by the following quiver:
\be \label{S1} \scalebox{0.85}{\bpic[node distance=2cm,gSUnode/.style={circle,red,draw,minimum size=8mm},gUSpnode/.style={circle,blue,draw,minimum size=8mm},gSOnode/.style={circle,ForestGreen,draw,minimum size=8mm},gOnode/.style={circle,purple,draw,minimum size=8mm},fnode/.style={rectangle,draw,minimum size=8mm}]
\begin{scope}[shift={(-3,0)}]    
\node at (-2,1) {$\cO_1[n]:$};
\node[gOnode] (G1) at (0,0) {$n$};
\node[fnode, ForestGreen] (F1) at (2.5,0) {$2n-2$};
\draw (G1) -- (F1) node[midway,above] {$p$};
\draw (0.3,0.3) to[out=90,in=0]  (0,0.7) to[out=180,in=90] (-0.3,0.3);
\node[right] at (0,-1.3) {$ \cW= s p p$};
\node at (0.6,0.9) {$s$, \textcolor{Green}{$\frac{4}{3}$}};
\node at (0.9,-0.4) {\textcolor{Green}{$\frac{1}{3}$}};
\node at (5,0) {$\Llra$};
\end{scope}
\begin{scope}[shift={(4.5,0)}]
\node[fnode, ForestGreen] (F2) at (0,0) {$2n-2$};
\draw (0.3,0.4) to[out=90,in=0]  (0,0.9) to[out=180,in=90] (-0.4,0.4);
\node at (0,-1.3) {$ \cW = S^3 + \tr(S) \, S^2$};
\node at (0.7,0.9) {$S$, \textcolor{Green}{$\frac{2}{3}$}};
\end{scope}
\epic} \ee 
where now a purple node denotes an orthogonal group and a green one a special orthogonal group, while an arc is a chiral in the symmetric representation.

The combination of the constraint from the vanishing of the $U(1)_R$ ABJ anomaly and the constraint from the superpotential fixes the R-charges of the fields to be 
\be
R[s] = \frac{4}{3}\,,\qquad R[p] = \frac{1}{3}\,. 
\ee
Another consequence of the superpotential is the truncation of the chiral ring of the theory. 
The mapping of the chiral ring generators is given by\footnote{We use the following notation: $\tr[ p \, p] \equiv p_{a_1}^{i_1} p_{a_2}^{i_2} \delta_{(n)}^{a_1 a_2}  \delta_{i_1 i_2}^{(2n-2)}$.}
\be \label{mapDeconfinementSO} 
\ba{c}
p \, p \\
\tr[p \, p] 
\ea
\quad \longleftrightarrow \quad
\ba{c}
S \\
\tr S 
\ea
\ee

From the operator map we can see that it is crucial to take the gauge group to be $O(n)$ rather than $SO(n)$. Indeed, in the $SO(n)$ gauge theory we would have an additional chiral operator corresponding to the baryon\footnote{The presence of this baryonic operator in the spectrum of the $SO(n)$ gauge theory can be explicitly checked with index computations for low values of $n$. We have checked that to low orders the index of the $SO(n)$ theory matches that of the WZ theory up to the contribution of the baryon.}
\be
\epsilon_{a_1\cdots a_n}p^{a_1}\cdots p^{a_n}
\ee
which would not be mapped under the duality. This operator is charged under the 0-form  $\mathbb{Z}^{\mathcal{C}}_2$ charge conjugation symmetry of the $SO(n)$ gauge theory. The $O(n)$ gauge theory is obtained by gauging $\mathbb{Z}_2^{\mathcal{C}}$, which has thus the effect of projecting out the baryonic operator and make the duality consistent.

Let us briefly comment on the structure of higher-form symmetries \cite{Gaiotto:2014kfa} in this duality. The $O(n)$ gauge theory has a $\mathbb{Z}_2$ 1-form symmetry coming from the center of the $SO(n)$ part that acts trivially on the matter fields. It also has a $\mathbb{Z}_2$ 2-form symmetry that arises from gauging the 0-form  $\mathbb{Z}^{\mathcal{C}}_2$ charge conjugation symmetry of the $SO(n)$ gauge theory. None of these symmetries is present on the WZ side. The two possibilities are that either these symmetries end up acting trivially at low energies or in the dual there should also be a topological sector that carries these symmetries. The latter is well-known to occur for example in dualities for $3d$ $\mathcal{N}=2$ theories, such as the duality appetizer of \cite{Jafferis:2011ns}. The simplest possibility for such a topological theory is a BF theory with action of the form
\be\label{eq:topth}
\pi\int_{X_4}\delta A_1\cup B_2\,,
\ee
where $A_1\in C^1(X_4,\mathbb{Z}_2)$ is a dynamical gauge field for a $\mathbb{Z}_2$ 0-form symmetry and $B_2\in C^2(X_4,\mathbb{Z}_2)$ is a dynamical gauge field for a $\mathbb{Z}_2$ 1-form symmetry. Such a BF theory indeed has both a $\mathbb{Z}_2$ 1-form symmetry whose charged line operators are $\exp\left(i\pi\int_L A_1\right)$ and a $\mathbb{Z}_2$ 2-form symmetry whose charged surface operators are $\exp\left(i\pi\int_S B_2\right)$. Each of these operators charged under one of the two higher-form symmetries is also the topological operator that generates the other symmetry, where the fact that they are topological is a consequence of the equations of motion, e.g.~the e.o.m.~of $B_2$ sets $\delta A_1=0$ and viceversa upon integration by parts. Nevertheless, such a topological symmetry cannot be detected with the tools at our disposal, e.g.~the anomalies for continuous symmetries and the index, and so we cannot make any conclusive statement about it. It would be interesting to clarify this issue by matching supersymmetric partition functions that are sensitive to such a topological sector.\footnote{There is also a question of whether the gauge group should be $O(n)$ or $Pin(n)$, both of which have the $\mathbb{Z}^{\mathcal{C}}_2$ charge conjugation symmetry gauged. The difference is that in the former the $\mathbb{Z}_2$ 1-form symmetry is electric, i.e.~it comes from its center, while in the latter it is magnetic, i.e.~it comes from the center of the dual group. One can move from one variant to the other by gauging these symmetries. If our guess about the presence of the topological sector in the duality is true, then as a consequence of the duality we can deduce that the two variants are actually equivalent, since under the gauging of the 1-form symmetry the topological theory \eqref{eq:topth} gives an identical theory, but expressed in terms of the dual field.}

%

\subsection{Theories with $\cW \sim \phi^3 + \phi qq$}

In this subsection, we present other examples of S-confining theories involving matter in a rank-2 representation under the gauge group, but with a different superpotential with respect to the previous examples. Specifically, we will also have a cubic superpotential term $\phi^3$ for the rank-2 field $\phi$.

\subsubsection{$USp(2n)$ gauge group: $\cU_2[n,h]$ theories} 
The first example is a two parameters family. It is the $USp(2n)$ gauge theory with a traceful antisymmetric chiral field $a$ and $2n+8-2h$ fundamental chirals that we split in two groups: $2n+8-4h$, $q$  and $2h$, $p$. The superpotential is
\be
\mathcal{W}=a^3+aqq+\Flip[pp; a p p; q p]\,.
\ee
Notice that we also added some gauge singlet chiral fields that flip\footnote{More precisely, we say that a singlet $F$ flips a gauge invariant operator $O$ if we have the superpotential term $\delta\mathcal{W}=\Flip[O]=F\,O$. The name is due to the fact that the e.o.m.~of $F$ sets $O=0$ in the chiral ring. We will also call the singlet $F$ \emph{flipper} field.} some gauge invariant operators involving $p$. The non-anomalous global symmetry of the theory that is preserved by the superpotential is
\be
SU(2h)\times USp(2n+8-4h)\times U(1)_R\,.
\ee

We claim that in the IR we obtain the trivial theory without any d.o.f.. We summarize this claim by the following quiver\footnote{To avoid cluttering the quiver, we don't draw the \emph{flipper} fields. Their presence can be inferred by looking at the superpotential.}
\be \label{U2} \scalebox{0.85}{\bpic[node distance=2cm,gSUnode/.style={circle,red,draw,minimum size=8mm},gUSpnode/.style={circle,blue,draw,minimum size=8mm},gSOnode/.style={circle,ForestGreen,draw,minimum size=8mm},fnode/.style={rectangle,draw,minimum size=8mm}]
\begin{scope}[shift={(-3,0)}]    
\node at (-2,1) {$\cU_2[n,h]:$};
\node[gUSpnode] (G1) at (0,0) {$2n$};
\node[fnode,blue] (F1) at (2.5,0) {$2n+8-4h$};
\node[fnode] (F2) at (0,-1.8) {$2h$};
\draw (G1) -- (F1) node[midway,above] {$q$};
\draw (G1) -- (F2) node[midway,left] {$p$};
\draw (0.3,0.3) to[out=90,in=0]  (0,0.7) to[out=180,in=90] (-0.3,0.3);
\node[right] at (-2,-2.9) {$ \cW= a^3 + a q q + \Flip[pp; a p p; q p]$};
\node at (0.6,0.9) {$a$, \textcolor{Green}{$\frac{2}{3}$}};
\node at (0.9,-0.4) {\textcolor{Green}{$\frac{2}{3}$}};
\node at (0.5,-1) {\textcolor{Green}{$\frac{h-n}{3h}$}};
\node at (5,0) {$\Llra$};
\end{scope}
\begin{scope}[shift={(4.5,0)}]
\node at (0,0) {Trivial theory};
\end{scope}
\epic} \ee 
The range of validity for the parameter $h$ is $0 \le h < \frac{n}{2} +2$. $h$ cannot be taken to be $\frac{n}{2}+2$ because otherwise we cannot turn on the deformation $a q q$ which is crucial. 

The associated index identity is
\begin{align}\label{eq:indU2}
&\prod_{k<l}^{2h}\Gpq{(pq)^{\frac{2h+n}{3h}}v_k^{-1}v_l^{-1}}\Gpq{(pq)^{\frac{h+n}{3h}}v_k^{-1}v_l^{-1}}\prod_{i=1}^{n+4-2h}\prod_{l=1}^{2h}\Gpq{(pq)^{\frac{3h+n}{6h}}f_i^{\pm1}v_l^{-1}}\nn\\
&\times\frac{(p;p)_\infty^n(q;q)_\infty^n}{2^n\, n!}\oint\prod_{a=1}^n\frac{\udl{z}_a}{2\pi i\,z_a}\frac{\prod_{a=1}^n\prod_{i=1}^{n+4-2h}\Gpq{(pq)^{\frac{1}{6}}z_a^{\pm1}f_i^{\pm1}}\prod_{l=1}^{2h}\Gpq{(pq)^{\frac{h-n}{6h}}z_a^{\pm1}v_l}}{\prod_{a=1}^n\Gpq{z_a^{\pm2}}\prod_{a<b}^n\Gpq{z_a^{\pm1}z_b^{\pm1}}}\nn\\
&\times\Gpq{(pq)^{\frac{1}{3}}}^{n}\prod_{a<b}^n\Gpq{(pq)^{\frac{1}{3}}z_a^{\pm1}z_b^{\pm1}}=1\,,
\end{align}
where $f_i$ for $i=1,\cdots,n+4-2h$ are the $USp(2n+8-4h)$ fugacities while $v_l$ for $l=1,\cdots,2h$ with $\prod_lv_l=1$ are the $SU(2h)$ fugacities.

We show the matching of anomalies for this duality in Appendix \ref{app:anomU2}, while in Subsection \ref{subsec:U2deriv} we will derive it by iterative applications of more elementary dualities.

We can formulate the duality in a little bit more general form where on the r.h.s.~we don't have a trivial theory. In order to do this, we have to flip on the l.h.s.~mesons involving the field $q$. For generic $n$ and $h$, the only meson that we can flip is $qq$. The reason is because the R-charge of the field $p$ becomes quickly very negative. Therefore, the meson built from this field $p$ will map to a fundamental field in the WZ with large negative R-charge and it becomes impossible to write down a superpotential. When we flip only $qq$ we get on the dual frame a field $A$ in the traceful antisymmetric representation of $USp(2n+8-4h)$:
\be \label{U2more} \scalebox{0.85}{\bpic[node distance=2cm,gSUnode/.style={circle,red,draw,minimum size=8mm},gUSpnode/.style={circle,blue,draw,minimum size=8mm},gSOnode/.style={circle,ForestGreen,draw,minimum size=8mm},fnode/.style={rectangle,draw,minimum size=8mm}]
\begin{scope}[shift={(-3,0)}]    
\node[gUSpnode] (G1) at (0,0) {$2n$};
\node[fnode,blue] (F1) at (2.5,0) {$2n+8-4h$};
\node[fnode] (F2) at (0,-1.8) {$2h$};
\draw (G1) -- (F1) node[midway,above] {$q$};
\draw (G1) -- (F2) node[midway,left] {$p$};
\draw (0.3,0.3) to[out=90,in=0]  (0,0.7) to[out=180,in=90] (-0.3,0.3);
\node[right] at (-2.2,-2.9) {$ \cW= a^3 + a q q + \Flip[pp; a p p; q p; q q]$};
\node at (0.6,0.9) {$a$, \textcolor{Green}{$\frac{2}{3}$}};
\node at (0.9,-0.4) {\textcolor{Green}{$\frac{2}{3}$}};
\node at (0.5,-1) {\textcolor{Green}{$\frac{h-n}{3h}$}};
\node at (5,0) {$\Llra$};
\end{scope}
\begin{scope}[shift={(4.5,0)}]
\node[fnode,blue] (F3) at (0,0) {$2n+8-4h$};
\draw (0.5,0.4) to[out=90,in=0]  (0,0.9) to[out=180,in=90] (-0.5,0.4);
\node at (0.8,1.1) {$A$, \textcolor{Green}{$\frac{2}{3}$}};
\node at (0,-1) {$ \cW= A^3 + \tr(A) \, A^2$};
\end{scope}
\epic} \ee 
The mapping is the following:
\be \label{mapU2} 
\ba{c}
\Flipper[q \, q]
\ea
\quad \longleftrightarrow \quad
\ba{c}
A
\ea
\ee

\paragraph{Special case of $h=1$. }
In this case, we can write the duality in yet another equivalent form with more fields on the r.h.s.. We obtain it by splitting the fundamental of $SU(2h) = SU(2)$, $p$ into two independent fields $p_1$ and $p_2$. The duality is the following:
\be \label{U2h1} \scalebox{0.85}{\bpic[node distance=2cm,gSUnode/.style={circle,red,draw,minimum size=8mm},gUSpnode/.style={circle,blue,draw,minimum size=8mm},gSOnode/.style={circle,ForestGreen,draw,minimum size=8mm},fnode/.style={rectangle,draw,minimum size=8mm}]
\begin{scope}[shift={(-3,0)}]    
\node[gUSpnode] (G1) at (0,0) {$2n$};
\node[fnode,blue] (F1) at (2.5,0) {$2n+4$};
\node[fnode,red] (F2) at (-1,-1.8) {$1$};
\node[fnode] (F3) at (1,-1.8) {$1$};
\draw (G1) -- (F1);
\draw (G1) -- (F2);
\draw (G1) -- (F3);
\draw (F2) -- (F3);
\draw (0.3,0.3) to[out=90,in=0]  (0,0.7) to[out=180,in=90] (-0.3,0.3);
\node at (0,-3) {$ \cW= a^3 + a q q + \Flip[p_1 p_2; a p_1 p_2; q p_2; a p_1 q; q q]$};
\node at (0,-3.8) {$+ \b \, \Flipper[a p_1 p_2] \, p_1 p_2 + \b \, \Flipper[a p_1 q] \, p_1 q$};
\node at (0.6,0.9) {$a$, \textcolor{Green}{$\frac{2}{3}$}};
\node at (1.1,0.4) {$q$, \textcolor{Green}{$\frac{2}{3}$}};
\node at (-1,-0.8) {\textcolor{Green}{$0$}, $p_1$};
\node at (1.4,-0.8) {$p_2$, \textcolor{Green}{$\frac{6-4n}{3}$}};
\node at (0,-2.2) {$\b$, \textcolor{Green}{$\frac{2}{3}$}};
\node at (5,0) {$\Llra$};
\end{scope}
\begin{scope}[shift={(4.5,0)}]
\node[fnode,blue] (F4) at (0,0) {$2n+4$};
\node[fnode,red] (F5) at (-1,-1.8) {$1$};
\node[fnode] (F6) at (1,-1.8) {$1$};
\draw (F4) -- (F5) -- (F6) -- (F4);
\draw (0.4,0.4) to[out=90,in=0]  (0,0.8) to[out=180,in=90] (-0.4,0.4);
\node at (0.8,1) {$A$, \textcolor{Green}{$\frac{2}{3}$}};
\node at (0,-3) {$ \cW= A^3 + A x y + x y z$};
\node at (-1,-0.8) {\textcolor{Green}{$\frac{2}{3}$}, $x$};
\node at (1,-0.8) {$y$, \textcolor{Green}{$\frac{2}{3}$}};
\node at (0,-2.2) {$z$, \textcolor{Green}{$\frac{2}{3}$}};
\end{scope}
\epic} \ee 
The mapping is the following:
\be \label{mapU2h1} 
\ba{c}
\Flipper[q \, q] \\
p_1 \, q \\
\Flipper[a p_1 q] \\
\b
\ea
\quad \longleftrightarrow \quad
\ba{c}
A \\
x \\
y \\
z
\ea
\ee 
We specified this duality because we will use this version in Subsection \ref{subsec:U1deriv} in the derivation of the dual of $\cU_1[n]$.

\subsubsection{$SO(n)$ gauge group: $\cS_2[n,h]$ theories} 

The second example consists again of a two parameters family, but this time we have the $SO(n)$ gauge theory with a traceful symmetric chiral field $s$ and $n-8-h$ chirals in the vector representation splitted in $n-8-2h$, $q$ and $h$, $p$. The superpotential is the same as in the previous subsection, but in this case the non-anomalous global symmetry is
\be
SU(h)\times SO(n-8-2h)\times U(1)_R\,.
\ee

We claim that in the IR this theory flows to a trivial theory without any d.o.f.
\be \label{S2} \scalebox{0.85}{\bpic[node distance=2cm,gSUnode/.style={circle,red,draw,minimum size=8mm},gUSpnode/.style={circle,blue,draw,minimum size=8mm},gSOnode/.style={circle,ForestGreen,draw,minimum size=8mm},gOnode/.style={circle,Purple,draw,minimum size=8mm},fnode/.style={rectangle,draw,minimum size=8mm}]
\begin{scope}[shift={(-3,0)}]    
\node at (-2,1) {$\cS_2[n,h]:$};
\node[gSOnode] (G1) at (0,0) {$n$};
\node[fnode,ForestGreen] (F1) at (2.5,0) {$n-8-2h$};
\node[fnode] (F2) at (0,-1.8) {$h$};
\draw (G1) -- (F1) node[midway,above] {$q$};
\draw (G1) -- (F2) node[midway,left] {$p$};
\draw (0.3,0.3) to[out=90,in=0]  (0,0.7) to[out=180,in=90] (-0.3,0.3);
\node[right] at (-2,-2.9) {$ \cW= s^3 + s q q + \Flip[pp; s p p; q p]$};
\node at (0.6,0.9) {$s$, \textcolor{Green}{$\frac{2}{3}$}};
\node at (0.9,-0.4) {\textcolor{Green}{$\frac{2}{3}$}};
\node at (0.5,-1) {\textcolor{Green}{$\frac{h-n}{3h}$}};
\node at (5,0) {$\Llra$};
\end{scope}
\begin{scope}[shift={(4.5,0)}]
\node at (0,0) {Trivial theory};
\end{scope}
\epic} \ee 
The range of validity for the parameter $h$ is $0 \le h < \frac{n}{2} - 4$. $h$ cannot be taken to be $\frac{n}{2}-4$ because otherwise we cannot turn on the deformation $s q q$ which is crucial. 

We show the matching of anomalies for this duality in Appendix \ref{app:anomS2}.

\subsubsection{$SU(n)$ gauge group: $\cA_2[n,h]$ theories} 
The last example is the $SU(n)$ gauge theory with a field $\phi$ in the adjoint plus the singlet trace and two sets of flavors, one of $n-2h$ flavors $q$, $\qt$ and a second of $h$ flavors $p$, $\pt$. The superpotential is
\be
\mathcal{W}=\phi^3+\phi q\qt +\Flip[p\pt; \phi p \pt; q \pt; \pt q]\,.
\ee
The non-anomalous global symmetry of the theory that is preserved by the superpotential is 
\be
SU(h) \times SU(h) \times SU(n-2h) \times U(1)_b \times U(1)_s \times U(1)_R\,,
\ee
where $U(1)_b$ acts with charges $\pm1$ on $q$, $\qt$ and $U(1)_s$ acts with charges $\pm1$ on $p$, $\pt$.

We claim that the IR of this theory is not trivial but is given by 3 singlets, two of R-charge equal to 0 and one of R-charge 2. The quiver summarizing this claim is the following:
\be \label{A2} \scalebox{0.85}{\bpic[node distance=2cm,gSUnode/.style={circle,red,draw,minimum size=8mm},gUSpnode/.style={circle,blue,draw,minimum size=8mm},gSOnode/.style={circle,ForestGreen,draw,minimum size=8mm},fnode/.style={rectangle,draw,minimum size=8mm}]
\begin{scope}[shift={(-3,0)}]    
\node at (-2,1) {$\cA_2[n,h]:$};
\node[gSUnode] (G1) at (0,0) {$n$};
\node[fnode] (F1) at (2.5,0) {$n-2h$};
\node[fnode] (F2) at (-0.8,-1.8) {$h$};
\node[fnode] (F3) at (0.8,-1.8) {$h$};
\draw (G1) -- pic[pos=0.4,sloped]{arrow} (F2);
\draw (G1) -- pic[pos=0.6,sloped]{arrow} (F3);
\draw (0.4,0.1) -- pic[pos=0.6,sloped]{arrow} (1.8,0.1);
\draw (0.4,-0.1) -- pic[pos=0.6,sloped, very thick]{arrow=latex reversed} (1.8,-0.1);
\draw (0.3,0.3) to[out=90,in=0]  (0,0.7) to[out=180,in=90] (-0.3,0.3);
\node at (0,-2.9) {$ \cW= \phi^3 + \phi q \qt + \Flip[p \pt; \phi p \pt; p \qt; \pt q]$};
\node at (0.5,1) {$\phi$, \textcolor{Green}{$\frac{2}{3}$}};
\node at (1.2,-0.4) {$q$, \textcolor{Green}{$\frac{2}{3}$}};
\node at (1.2,0.4) {$\qt$};
\node at (-1.2,-0.8) {\textcolor{Green}{$\frac{h-n}{3h}$}, $p$};
\node at (0.7,-0.9) {$\pt$};
\node at (5,0) {$\Llra$};
\end{scope}
\begin{scope}[shift={(6,0)}]
\node at (0,0) {	3 singlets with superpotential:};
\node at (0,-1) {$ \cW= N ( B \, \Bt + 1)$};
\end{scope}
\epic} \ee 
The range of validity for the parameter $h$ is $0 \le h < \frac{n}{2}$. $h$ cannot be taken to be $\frac{n}{2}$ because otherwise we cannot turn on the deformation $\phi q \qt$ which is crucial. 

In order to understand better the physical implications of this result, it is useful to look at the operator map
\be 
\ba{c}
\tr[p^h\phi^{n-h}] \\
\tr[\tilde{p}^h\phi^{n-h}]
\ea
\quad \longleftrightarrow \quad
\ba{c}
B \\
\tilde{B}
\ea
\ee 
where the flavor indices are contracted with the $h$-dimensional $\epsilon$ tensor so that the operator is a flavor singlet. The singlet $N$ instead is not in the chiral ring since it is turned on in the superpotential, but it is associated with the superpotential term $\phi q \qt$ on the l.h.s.~as it can be understood from the derivation of this result that we give in Subsection \ref{subsec:A2deriv}.

If we focus now on the r.h.s., we can understand the singlet $N$ of R-charge 2 as the analogue of the Lagrange multiplier that appears in the $SU(n)$ SQCD with $n$ flavors \cite{Seiberg:1994bz}. Indeed, its e.o.m.~forces $B$ and $\tilde{B}$ to have a non-vanishing vev and the theory has thus a quantum deformed moduli space of vacua. Remembering the operator map above, our result is then telling us that the theory on the l.h.s.~has a quantum deformed moduli space of vacua with a spontaneous breaking of the $U(1)_s$ symmetry, due to the fact that the operators $\tr[p^h\phi^{n-h}]$ and $\tr[\tilde{p}^h\phi^{n-h}]$ acquire a vev quantum mechanicaly triggered by the deformation $\phi q \qt$, while all the other operators trivialize at low energies.

At the level of the supersymmetric index, this duality translates into the following non-trivial integral identity:
\begin{align}
&\prod_{k,l=1}^h\Gpq{(pq)^{\frac{2h+n}{3h}}v_kw_l^{-1}}\Gpq{(pq)^{\frac{h+n}{3h}}v_kw_l^{-1}}\nn\\
&\times\prod_{i=1}^{n-2h}\prod_{l=1}^h\Gpq{(pq)^{\frac{3h+n}{6h}}b\,s^{-1}f_i^{-1}v_l}\Gpq{(pq)^{\frac{3h+n}{6h}}b^{-1}s\,f_iw_l^{-1}}\nn\\
&\times\frac{(p;p)_\infty^n(q;q)_\infty^n}{n!}\oint\prod_{a=1}^n\frac{\udl{z}_a}{2\pi i\,z_a}\frac{\prod_{a, b=1}^n\Gpq{(pq)^{\frac{2}{3}}z_az_b^{-1}}}{\prod_{a\neq b}^n\Gpq{z_az_b^{-1}}}\nn\\
&\times\prod_{a=1}^n\prod_{i=1}^{n-2h}\Gpq{(pq)^{\frac{1}{3}}\left(b\,z_af_i^{-1}\right)^{\pm1}}\prod_{l=1}^h\Gpq{(pq)^{\frac{h-n}{6h}}s\,z_av_l^{-1}}\Gpq{(pq)^{\frac{h-n}{6h}}s^{-1}z_a^{-1}w_l}\nn\\
&=\frac{2\pi i}{(p;p)_\infty(q;q)_\infty}\delta\left(s^h-1\right)\,,
\end{align}
where $f_i$ for $i=1,\cdots,n-2h$ with $\prod_if_i=1$ are the $SU(n-2h)$ flavor fugacities, while $v_l$, $w_l$ for $l=1,\cdots,h$ with $\prod_lv_l=\prod_lw_l=1$ are the fugacities for the two $SU(h)$ symmetries. The singular behaviour of the index is typical of theories with a quantum deformed moduli space of vacua, as discussed in \cite{Spiridonov:2014cxa,Bottini:2021vms}. It can be understood as a singular limit of the index of the three chirals $N$, $B$, $\tilde{B}$ interacting with a cubic superpotential when the linear term in $N$ is turned on
\be
\lim_{\epsilon\to1}\underbrace{\Gpq{pq\,\epsilon^2}}_{N}\underbrace{\Gpq{\epsilon^{-1}s^h}}_{B}\underbrace{\Gpq{\epsilon\,s^{-h}}}_{\tilde{B}}=\frac{2\pi i}{(p;p)_\infty(q;q)_\infty}\delta\left(s^h-1\right)\,.
\ee

\section{Derivation of the S-confining dualities}
\label{sec:derivation}

In this section we provide a derivation of the dualities for the theories $\cU_1[n]$ and $\cU_2[n,h]$ with symplectic gauge group. This is done by combining the study of various Higgs mechanisms and the deconfinement technique \cite{Berkooz:1995km,Bajeot:2022kwt,Bottini:2022vpy,Bajeot:2022lah}, which is based on an iterative application of some more fundamental dualities. The proof that we will give for $\cU_2[n,h]$\footnote{We will use an abuse of language by calling the duality statement in the same way as the original theory.} is also working both for $\cS_2[n,h]$ and $\cA_2[n,h]$. However the proof for $\cU_1[n]$ does not generalize to $\cO_1[n]$.

\subsection{Derivation of $\cU_2[n,h]$}
\label{subsec:U2deriv}

We will start from the duality $\cU_2[n,h]$ \eqref{U2}, since we will then use it in the derivation of $\cU_1[n]$. We will first show how this can be understood as a consequence of a duality proposed by Intriligator in \cite{Intriligator:1995ff}, which generalizes the Kutasov--Schwimmer duality \cite{Kutasov:1995ve,Kutasov:1995np} to the case of symplectic gauge group and so in the following we will refer to it as ``symplectic KS duality". Then we will show how for the case $h=1$ the same result can be obtained by studying the Higgsing due to a vev for some operator, which we do at the level of the supersymmetric index.

\paragraph{Higgsing via symplectic KS duality.}

We first show how to the derive the $\cU_2[n,h]$ duality \eqref{U2} for generic $n$ and $h$ from the symplectic KS duality of \cite{Intriligator:1995ff}, whose statement we recall in \eqref{KutasovUSpAntisym}. Applying it to our case, we obtain the following:
\be \label{ProofU2a} \scalebox{0.85}{\bpic[node distance=2cm,gSUnode/.style={circle,red,draw,minimum size=8mm},gUSpnode/.style={circle,blue,draw,minimum size=8mm},gSOnode/.style={circle,ForestGreen,draw,minimum size=8mm},fnode/.style={rectangle,draw,minimum size=8mm}]
\begin{scope}[shift={(-3,0)}]    
\node[gUSpnode] (G1) at (0,0) {$2n$};
\node[fnode] (F1) at (2.5,0) {$2n+8-4h$};
\node[fnode] (F2) at (0,-1.8) {$2h$};
\draw (G1) -- (F1) node[midway,above] {$q$};
\draw (G1) -- (F2) node[midway,left] {$p$};
\draw (0.3,0.3) to[out=90,in=0]  (0,0.7) to[out=180,in=90] (-0.3,0.3);
\node at (0,-2.9) {$ \cW= a^3$};
\node at (0.6,0.9) {$a$, \textcolor{Green}{$\frac{2}{3}$}};
\node at (1.2,-0.9) {\textcolor{Green}{$1-\frac{2(n+2)}{3(n+4-h}$}};
\node at (5,0) {$\Llra$};
\node at (5,-0.4) {KS};
\end{scope}
\begin{scope}[shift={(4.5,0)}]
\node[gUSpnode] (G2) at (0,0) {\scalebox{0.85}{$2n+8-4h$}};
\node[fnode] (F3) at (3,0) {$2n+8-4h$};
\node[fnode] (F4) at (0,-2.3) {$2h$};
\draw (G2) -- (F3) node[midway,above] {$Q$};
\draw (G2) -- (F4) node[midway,left] {$P$};
\draw (0.4,0.9) to[out=90,in=0]  (0,1.3) to[out=180,in=90] (-0.4,0.9);
\node[right] at (-3,-3.4) {$\cW= A^3 + \Flip[PP; QQ; PQ; PAP; QAQ; PAQ]$};
\node at (0.7,1.5) {$A$, \textcolor{Green}{$\frac{2}{3}$}};
\node at (1.5,-1.3) {\textcolor{Green}{$1-\frac{2(n+6-2h)}{3(n+4-h}$}};
\end{scope}
\epic} \ee 
Here we consider the antisymmetric fields on both sides to be traceful, with the trace parts being mapped to each other under the duality.

Now we deform this duality by the superpotential term $a q q$ and some flippers on the l.h.s.. Following the mapping of the KS duality shown in \eqref{mapKutasovUSpAntisym}, the term $a q q$ is mapped to the term $\Flipper[Q Q]$. We also use the mapping to understand the flipping terms. We obtain:
\be \label{ProofU2b} \scalebox{0.85}{\bpic[node distance=2cm,gSUnode/.style={circle,red,draw,minimum size=8mm},gUSpnode/.style={circle,blue,draw,minimum size=8mm},gSOnode/.style={circle,ForestGreen,draw,minimum size=8mm},fnode/.style={rectangle,draw,minimum size=8mm}]
\begin{scope}[shift={(-3,0)}]    
\node[gUSpnode] (G1) at (0,0) {$2n$};
\node[fnode, blue] (F1) at (2.5,0) {$2n+8-4h$};
\node[fnode] (F2) at (0,-1.8) {$2h$};
\draw (G1) -- (F1) node[midway,above] {$q$, \textcolor{Green}{$\frac{2}{3}$}};
\draw (G1) -- (F2) node[midway,left] {$p$};
\draw (0.3,0.3) to[out=90,in=0]  (0,0.7) to[out=180,in=90] (-0.3,0.3);
\node at (0,-2.9) {$ \cW= a^3 + aqq + \Flip[pp; app; pq]$};
\node at (0.6,0.9) {$a$, \textcolor{Green}{$\frac{2}{3}$}};
\node at (0.5,-1) {\textcolor{Green}{$\frac{h-n}{3h}$}};
\node at (5,0) {$\Llra$};
\end{scope}
\begin{scope}[shift={(4.5,0)}]
\node[gUSpnode] (G2) at (0,0) {\scalebox{0.85}{$2n+8-4h$}};
\node[fnode, blue] (F3) at (3.2,0) {$2n+8-4h$};
\node[fnode] (F4) at (0,-2.3) {$2h$};
\draw (G2) -- (F3) node[midway,above] {$Q$, \textcolor{Green}{$0$}};
\draw (G2) -- (F4) node[midway,left] {$P$};
\draw (0.4,0.9) to[out=90,in=0]  (0,1.3) to[out=180,in=90] (-0.4,0.9);
\node[right] at (-3,-3.4) {$\cW= A^3 + \Flip[PP; QQ; PQ; PAP; QAQ; PAQ]$};
\node[right] at (-4.5,-4.2) {$+ \Flipper[Q Q] + \Flip[\Flipper[PAP]; \Flipper[PP]; \Flipper[PAQ]]$};
\node at (0.7,1.5) {$A$, \textcolor{Green}{$\frac{2}{3}$}};
\node at (0.7,-1.5) {\textcolor{Green}{$\frac{h+n}{3h}$}};
\end{scope}
\epic} \ee 
We see that a lot of flippers get a mass. After integrating out the massive ones and naming the others, we get
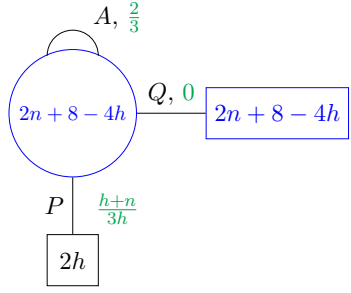
\be \label{ProofU2c} \scalebox{0.85}{\bpic[node distance=2cm,gSUnode/.style={circle,red,draw,minimum size=8mm},gUSpnode/.style={circle,blue,draw,minimum size=8mm},gSOnode/.style={circle,ForestGreen,draw,minimum size=8mm},fnode/.style={rectangle,draw,minimum size=8mm}]
\begin{scope}[shift={(0,0)}]
\node[gUSpnode] (G2) at (0,0) {\scalebox{0.85}{$2n+8-4h$}};
\node[fnode, blue] (F3) at (3.2,0) {$2n+8-4h$};
\node[fnode] (F4) at (0,-2.3) {$2h$};
\draw (G2) -- (F3) node[midway,above] {$Q$, \textcolor{Green}{$0$}};
\draw (G2) -- (F4) node[midway,left] {$P$};
\draw (0.4,0.9) to[out=90,in=0]  (0,1.3) to[out=180,in=90] (-0.4,0.9);
\node at (0,-3.4) {$\cW= A^3 + \eta_{QQ} + \eta_{QQ} QQ + \eta_{PQ} PQ + \eta_{QAQ} QAQ$};
\node at (0.7,1.5) {$A$, \textcolor{Green}{$\frac{2}{3}$}};
\node at (0.7,-1.5) {\textcolor{Green}{$\frac{h+n}{3h}$}};
\end{scope}
\epic} \ee

At this point we notice that the e.o.m.~of the flipper $\eta_{QQ}$ implies that the operator $QQ$ takes a vev which initiates an Higgsing of the gauge group. In the color-flavor space the field $Q$ can be taken to be diagonal
\be \label{Qdiag}
Q_\a^I = \begin{pmatrix}
\l_1 &  &  &\\
&  \ddots &  &\\
& & & \l_{n+4-2h}
\end{pmatrix} \otimes i\s_2
\ee 
Due to the e.o.m.~of the flipper $\eta_{QQ}$, the operator $QQ$ satisfies
\be \label{eom}
Q Q \equiv  Q_{\a}^{I} Q_{\b}^{J} \, J^{\a \b}_{Gauge} = - J_{Flav} \, \left(\equiv - \mathds{1}_{n+4-2h} \otimes i\s_2 \right)
\ee 
Using the presentation \eqref{Qdiag} for $Q$, the equation \eqref{eom} becomes
\be 
\begin{pmatrix}
\l_1^2 &  &  &\\
&  \ddots &  &\\
& & & \l_{n+4-2h}^2
\end{pmatrix} \otimes i\s_2 = \mathds{1}_{n+4-2h} \otimes i\s_2
\ee 
We obtain that the field $Q$ takes the following vev:
\be 
Q = \mathds{1}_{n+4-2h} \otimes i\s_2 \label{VEVQ}
\ee 
The conclusion of this vev is that the $USp(2n+8-4h)$ gauge group and the $USp(2n+8-4h)$ flavor group are broken to the diagonal $USp(2n+8-4h)$ flavor group. It is the mechanism of color-flavor locking. If we forget for a moment the two flippers $\eta_{PQ}$ and $\eta_{QAQ}$, we would say that the result of the Higgsing is the following:
\be \label{ProofU2Higgsing} \scalebox{0.85}{\bpic[node distance=2cm,gSUnode/.style={circle,red,draw,minimum size=8mm},gUSpnode/.style={circle,blue,draw,minimum size=8mm},gSOnode/.style={circle,ForestGreen,draw,minimum size=8mm},fnode/.style={rectangle,draw,minimum size=8mm}]
\begin{scope}[shift={(-3,0)}]    
\node[gUSpnode] (G2) at (0,0) {\scalebox{0.85}{$2n+8-4h$}};
\node[fnode, blue] (F3) at (3.2,0) {$2n+8-4h$};
\node[fnode] (F4) at (0,-2.3) {$2h$};
\draw (G2) -- (F3) node[midway,above] {$Q$, \textcolor{Green}{$0$}};
\draw (G2) -- (F4) node[midway,right] {$P$, \textcolor{Green}{$\frac{h+n}{3h}$}};
\draw (0.4,0.9) to[out=90,in=0]  (0,1.3) to[out=180,in=90] (-0.4,0.9);
\node at (0,-3.4) {$\cW= A^3 + \eta_{QQ} + \eta_{QQ} QQ$};
\node at (0.7,1.5) {$A$, \textcolor{Green}{$\frac{2}{3}$}};
\node at (5.5,0) {$\longrightarrow$};
\node at (5.5,-0.5) {Higgsing};
\end{scope}
\begin{scope}[shift={(5,0)}]
\node[fnode, blue] (F5) at (0,0) {$2n+8-4h$};
\node[fnode] (F6) at (0,-2.3) {$2h$};
\draw (F5) -- (F6);
\draw (0.4,0.4) to[out=90,in=0]  (0,0.8) to[out=180,in=90] (-0.4,0.4);
\node at (0.7,1) {$A$, \textcolor{Green}{$\frac{2}{3}$}};
\draw (F5) -- (F6) node[midway,right] {$P$, \textcolor{Green}{$\frac{h+n}{3h}$}};
\node at (0,-3.4) {$\cW= A^3$};
\end{scope}
\epic} \ee 
The fields on the r.h.s.~are obtained as follows. The field $Q$ gets a vev so it disapears as well as the flipper $\eta_{QQ}$. The fields $A$ and $P$ are instead unaffected and transform under the remaining $USp(2n+8-4h)$ flavor symmetry.


Starting from the result \eqref{ProofU2Higgsing}, we can go back to \eqref{ProofU2c} and study the effect of adding the two flippers $\eta_{PQ}$ and $\eta_{QAQ}$. Once we plug the vev of $Q$ \eqref{VEVQ} in the term $\eta_{PQ} P Q$, it becomes a mass term and then both $P$ and $\eta_{PQ}$ are integrated out. Similarly, $\eta_{QAQ} Q A Q$ becomes a mass term and both $A$ and $\eta_{QAQ}$ disappear\footnote{Here the assumption of the antisymmetric to be traceful is important. Indeed if $A$ is traceless, the trace part of the flipper $\eta_{QAQ}$ would not receive a mass and would stay in the IR.}. On the dual side, since $A$ and $P$ disappeared, we are left with no d.o.f.~and we obtain a trivial theory as claimed in \eqref{U2}. 

This proof using the KS duality can be applied in the same way in the case of the $SO$ duality \eqref{S2}. The specific form of the KS duality is recalled in \eqref{KutasovSOSym}. The $SU$ duality \eqref{A2} can also be obtain in a very similar way but there is a little difference that we are going to explain in the next subsection. 


\paragraph{Higgsing via the index.} For the special case of $h=1$ we can also derive the $\cU_2[n,h=1]$ duality by directly studying a Higgsing that is triggered by a vev for a specific operator, without having to apply any other more fundamental duality. We point out that the duality $\cU_2[n,h=1]$ is the one we will need to derive the duality $\cU_1[n]$, as we will explain in the Subsection \ref{subsec:U1deriv}.

In order to understand which operator is taking a vev and study the associated Higgs mechanism, it is more convenient to use the perspective of the index, which for the theory $\cU_2[n,h=1]$ is given on the l.h.s.~of \eqref{eq:indU2} with $h=1$
\begin{align}\label{eq:indU2h1}
&\Gpq{(pq)^{\frac{n+2}{3}}}\Gpq{(pq)^{\frac{n+1}{3}}}\Gpq{(pq)^{\frac{n+3}{6}}}\frac{(p;p)_\infty^n(q;q)_\infty^n}{2^n\, n!}\nn\\
&\times\oint\prod_{a=1}^n\frac{\udl{z}_a}{2\pi i\,z_a}\frac{\prod_{a=1}^n\prod_{i=1}^{n+2}\Gpq{(pq)^{\frac{1}{6}}z_a^{\pm1}f_i^{\pm1}}\Gpq{(pq)^{\frac{1-n}{6}}z_a^{\pm1}v^{\pm1}}}{\prod_{a=1}^n\Gpq{z_a^{\pm2}}\prod_{a<b}^n\Gpq{z_a^{\pm1}z_b^{\pm1}}}\nn\\
&\times\Gpq{(pq)^{\frac{1}{3}}}^{n-1}\prod_{a<b}^n\Gpq{(pq)^{\frac{1}{3}}z_a^{\pm1}z_b^{\pm1}}\,.
\end{align}
Let us focus on the following combination of Gamma functions:
\be\label{eq:gammasvev}
\Gpq{(pq)^{\frac{1-n}{6}}z_1^{-1}v}\prod_{a=1}^{n-1}\Gpq{(pq)^{\frac{1}{3}}z_az_{a+1}^{-1}}\Gpq{(pq)^{\frac{1-n}{6}}z_nv^{-1}}\,.
\ee
These provide the following sets of poles:
\begin{align}
&z_1=v(pq)^{\frac{1-n}{6}}p^{k_1}q^{l_1}\,,\nn\\
&z_{a+1}=z_a(pq)^{\frac{1}{3}}p^{k_{a+1}}q^{l_{a+1}}\,,\quad a=1,\cdots,n-1\,,\nn\\
&z_n=v(pq)^{\frac{n-1}{6}}p^{-k_{n+1}}q^{-l_{n+1}}\,,
\end{align}
where $k_a,l_a=0,\cdots,\infty$ for $a=1,\cdots,n+1$. These collide so to pinch the $n$-dimensional integration contour at the points corresponding to $k_a=l_a=0$ (for more details about this type of pinching see \cite{BADpaper})
\begin{align}\label{eq:pinch}
&z_1=v(pq)^{\frac{1-n}{6}}\,,\nn\\
&z_2=z_1(pq)^{\frac{1}{3}}=v(pq)^{\frac{3-n}{6}}\,,\nn\\
&\cdots\nn\\
&z_{n-1}=z_{n-2}(pq)^{\frac{1}{3}}=v(pq)^{\frac{n-3}{6}}\,,\nn\\
&z_n=z_{n-1}(pq)^{\frac{1}{3}}=v(pq)^{\frac{n-1}{6}}\,.
\end{align}
Following \cite{Gaiotto:2012xa}, the interpretation of this pinching is that there is an operator that is taking a vev. Specifically, such operator is the one constructed from the chirals whose index contribution is \eqref{eq:gammasvev} (this type of vev has been studied also in \cite{Comi:2022aqo,BADpaper})
\be
\langle\tr[a^{n-1}p^2]\rangle\neq 0\,.
\ee
Notice indeed that this operator has zero R-charge and it is also uncharged under any global symmetry, which is consistent with it getting a vev. The prescription of \cite{Gaiotto:2012xa} is to then take the residue at the points \eqref{eq:pinch}, which implements at the level of the index the Higgs mechanism triggered by such a vev. Observe that in this way we completely get rid of the $n$-dimensional integral, which means that the gauge group is Higgsed completely by the vev. Evaluating the residue we can figure out what massless fields survive at the end of the Higgsing and we find in this case that there are none, which is compatible with our claim that the theory flows to a trivial theory.

\subsection{Derivation of $\cA_2[n,h]$}
\label{subsec:A2deriv}

\paragraph{Higgsing via KS duality.} The derivation is similar to the symplectic case so we will be brief and just highlight the key differences. We start by using the KS duality \eqref{KutasovSUadjoint} deformed by the cubic superpotential $\phi q \qt$. We get the following duality:
\be \label{ProofA2} \scalebox{0.85}{\bpic[node distance=2cm,gSUnode/.style={circle,red,draw,minimum size=8mm},gUSpnode/.style={circle,blue,draw,minimum size=8mm},gSOnode/.style={circle,ForestGreen,draw,minimum size=8mm},fnode/.style={rectangle,draw,minimum size=8mm}]
\begin{scope}[shift={(-3,0)}]    
\node at (-2,1) {$\cA_2[n,h]:$};
\node[gSUnode] (G1) at (0,0) {$n$};
\node[fnode] (F1) at (2.5,0) {$n-2h$};
\node[fnode] (F2) at (-0.8,-1.8) {$h$};
\node[fnode] (F3) at (0.8,-1.8) {$h$};
\draw (G1) -- pic[pos=0.4,sloped]{arrow} (F2);
\draw (G1) -- pic[pos=0.6,sloped]{arrow} (F3);
\draw (0.4,0.1) -- pic[pos=0.6,sloped]{arrow} (1.8,0.1);
\draw (0.4,-0.1) -- pic[pos=0.6,sloped, very thick]{arrow=latex reversed} (1.8,-0.1);
\draw (0.3,0.3) to[out=90,in=0]  (0,0.7) to[out=180,in=90] (-0.3,0.3);
\node at (0,-2.9) {$ \cW= \phi^3 + \phi q \qt + \Flip[p \pt; \phi p \pt; p \qt; \pt q]$};
\node at (0.5,1) {$\phi$, \textcolor{Green}{$\frac{2}{3}$}};
\node at (1.2,-0.4) {$q$, \textcolor{Green}{$\frac{2}{3}$}};
\node at (1.2,0.4) {$\qt$};
\node at (-1.2,-0.8) {\textcolor{Green}{$\frac{h-n}{3h}$}, $p$};
\node at (0.7,-0.9) {$\pt$};
\node at (5,0) {$\Llra$};
\end{scope}
\begin{scope}[shift={(4.5,0)}]
\node[gSUnode] (G2) at (0,0) {\scalebox{0.8}{$n-2h$}};
\node[fnode] (F4) at (2.5,0) {$n-2h$};
\node[fnode] (F5) at (-0.8,-1.8) {$h$};
\node[fnode] (F6) at (0.8,-1.8) {$h$};
\draw (G2) -- pic[pos=0.4,sloped]{arrow} (F5);
\draw (G2) -- pic[pos=0.6,sloped]{arrow} (F6);
\draw (0.6,0.1) -- pic[pos=0.6,sloped]{arrow} (1.8,0.1);
\draw (0.6,-0.1) -- pic[pos=0.6,sloped, very thick]{arrow=latex reversed} (1.8,-0.1);
\draw (0.4,0.5) to[out=90,in=0]  (0,1) to[out=180,in=90] (-0.4,0.5);
\node at (0.5,-2.9) {$ \cW= \Phi^3 + \eta_{Q \Qt} + \eta_{Q \Qt} Q \Qt + \Flip[\Phi Q \Qt; P \Qt; \Pt Q]$};
\node at (0.8,1.1) {$\Phi$, \textcolor{Green}{$\frac{2}{3}$}};
\node at (1.2,-0.4) {$Q$, \textcolor{Green}{$0$}};
\node at (1.2,0.4) {$\Qt$};
\node at (-1.2,-0.8) {\textcolor{Green}{$\frac{n+h}{3h}$}, $P$};
\node at (0.7,-0.9) {$\Pt$};
\end{scope}
\epic} \ee 
On the r.h.s., we have named the $\Flipper[Q \, \Qt]$ as $\eta_{Q \Qt}$, while we didn't give a name to the other flippers. 

Now, as in the $USp$ case, the e.o.m.~of $\eta_{Q \Qt}$ gives a vev to $Q$ and $\Qt$ equal to the identity.  This vev triggers the color-flavor locking mechanism. The $SU(n-2h)$ gauge group and the $SU(n-2h)$ flavor group are broken to the diagonal $SU(n-2h)$ flavor symmetry. In terms of this diagonal symmetry, the fields $Q$, $\Qt$ and $\eta_{Q \Qt}$ decompose as adjoint plus singlet (each field has $(n-2h)^2$ d.o.f.'s), so we have 3 adjoints and 3 singlets. We denote the singlets by $B, \Bt$ and $N$, respectively. In the Higgsing one adjoint recombines with the broken generators of the gauge group, while the other two become massive. On the other hand, the 3 singlets ($B, \Bt$ have R-charge $0$ and $N$ has R-charge $2$) survive the Higgsing\footnote{This is the main difference with respect to the $USp$ case, where only two singlets are produced, and they give a mass to each other, disappearing in the IR.}.

We are interested in $\cA_2[n,h]$, where there are additional flippers with respect to \eqref{ProofA2}. In $\cA_2[n,h]$, after the KS duality and the Higgsing, the fields $P, \Pt$ combine with the flippers of $P \Qt, \Pt Q$  and become massive, hence disappearing in the IR. The field $\Phi$ combines with the flipper of $\Phi Q \Qt$ and become massive (the adjoint field $\Phi$ is traceful, hence all of the components of the $\Flipper[\Phi Q \Qt]$ become massive).

To summarize, $\cA_2[n,h]$ is dual to a theory of 3 singlets, two of R-charge 0 ($B$ and $\Bt$, coming from $Q$ and $\Qt$) and one of R-charge 2 ($N$, coming from $\eta_{QQ}$). From the superpotential $\tr(\eta_{Q \Qt}) + \tr(\eta_{Q \Qt} Q \Qt)$ in the KS dual, after the Higgsing only the following superpotential survives:  
\be 
\cW = N  + N B \Bt 
\ee 
This is precisely the claim of \eqref{A2}.

\subsection{Derivation of $\cU_1[n]$}
\label{subsec:U1deriv}
In this subsection, we show how to obtain the $\cU_1[n]$ duality of \eqref{U1}\footnote{We assume that $n\ge3$, the $n=2$ case is a bit different and is done in the Appendix \ref{ProofU1n2}.}. We use a combination of the deconfinement technique, the iteration of more fundamental dualities and a recursive argument. Indeed from now on, we assume that the duality $\cU_1[n-1]$ is correct and we want to obtain the statement for $\cU_1[n]$. 

The first step is the splitting of the flavors into $4n+1+1$ fundamentals. Of course this step doesn't change anything but it is the correct form to apply the deconfinement of \cite{Bajeot:2022kwt,Bottini:2022vpy, Bajeot:2022lah}. We get
\be \label{U1na} \scalebox{0.85}{\bpic[node distance=2cm,gSUnode/.style={circle,red,draw,minimum size=8mm},gUSpnode/.style={circle,blue,draw,minimum size=8mm},gSOnode/.style={circle,ForestGreen,draw,minimum size=8mm},fnode/.style={rectangle,draw,minimum size=8mm}]
\begin{scope}[shift={(-3,0)}]    
\node at (-2,1) {$\cU_1[n]:$};
\node[gUSpnode] (G1) at (0,0) {$2n$};
\node[fnode, blue] (F1) at (2.5,0) {$4n+2$};
\draw (G1) -- (F1) node[midway,above] {$p$};
\draw (0.3,0.3) to[out=90,in=0]  (0,0.7) to[out=180,in=90] (-0.3,0.3);
\node[right] at (0,-1.3) {$ \cW= a p p$};
\node at (0.6,0.9) {$a$, \textcolor{Green}{$\frac{4}{3}$}};
\node at (0.9,-0.4) {\textcolor{Green}{$\frac{1}{3}$}};
\node at (5,0) {$\equiv$};
\end{scope}
\begin{scope}[shift={(4.5,0)}]
\node[gUSpnode] (G2) at (0,0) {$2n$};
\node[fnode, blue] (F2) at (2.5,0) {$4n$};
\node[fnode,Maroon] (F3) at (1,-1.8) {$1$};
\node[fnode,red] (F4) at (-1,-1.8) {$1$};
\draw (G2) -- (F2);
\draw (G2) -- (F3);
\draw (G2) -- (F4);
\draw (0.3,0.3) to[out=90,in=0]  (0,0.7) to[out=180,in=90] (-0.3,0.3);
\node[right] at (-0.8,-3) {$ \cW= a q q + a p_1 p_2$};
\node at (0.6,0.9) {$a$, \textcolor{Green}{$\frac{4}{3}$}};
\node at (1.3,0.4) {$q$, \textcolor{Green}{$\frac{1}{3}$}};
\node at (1.1,-0.7) {$p_2$, \textcolor{Green}{$\frac{1}{3}$}};
\node at (-1.1,-0.7) {\textcolor{Green}{$\frac{1}{3}$}, $p_1$};
\end{scope}
\epic} \ee 

Now we can deconfine the antisymmetric traceless field $a$. The basic idea \cite{Berkooz:1995km,Luty:1996cg} is to trade the antisymmetric field by a gauge node which confines. More details and the specific form of the deconfinement that we are using here can be found in \cite{Bajeot:2022kwt,Bajeot:2022lah}. We obtain
\be \label{U1nb} \scalebox{0.85}{\bpic[node distance=2cm,gSUnode/.style={circle,red,draw,minimum size=8mm},gUSpnode/.style={circle,blue,draw,minimum size=8mm},gSOnode/.style={circle,ForestGreen,draw,minimum size=8mm},fnode/.style={rectangle,draw,minimum size=8mm}]
\begin{scope}[shift={(0,0)}]
\node[gUSpnode] (G1) at (-2,0) {\scalebox{0.8}{$2n-2$}};
\node[gUSpnode] (G2) at (0.3,0) {$2n$};
\node[fnode, blue] (F1) at (2.5,0) {$4n$};
\node[fnode,red] (F2) at (-2,-1.9) {$1$};
\node[fnode] (F3) at (0.3,-1.9) {$1$};
\node[fnode,Maroon] (F4) at (1.9,-1.9) {$1$};
\draw (G1) -- (G2);
\draw (G2) -- (F1);
\draw (G1) -- (F2);
\draw (G1) -- (F3);
\draw (G2) -- (F3);
\draw (G2) -- (F4);
\node at (-0.8,0.4) {$b$, \textcolor{Green}{$\frac{2}{3}$}};
\node at (1.5,0.4) {$q$, \textcolor{Green}{$\frac{1}{3}$}};
\node at (1.6,-0.9) {$p_2$, \textcolor{Green}{$\frac{1}{3}$}};
\node at (-2.7,-1) {\textcolor{Green}{$-\frac{1}{3}$}, $v$};
\node at (-0.8,-0.8) {$d$};
\node at (-0.9,-1.4) {\textcolor{Green}{$\frac{7-4n}{3}$}};
\node at (0.1,-0.8) {\textcolor{Green}{$\frac{4n-3}{3}$}};
\node at (0.5,-1.2) {$e$};
\node at (0,-3) {$ \cW= bq qb + v b^3 p_2 + b d e + \Flip[v d; b b]$};
\end{scope}
\epic} \ee 
Notice indeed that we can apply the basic Seiberg-like duality for symplectic gauge group due to Intriligator and Pouliot (IP) \cite{Intriligator:1995ne} to the new $USp(2n-2)$ node to confine it and go back to the theory in \eqref{U1na}.

At this stage, the $USp(2n)$ gauge group is coupled only to fields in the fundamental representation. We can therefore apply the IP duality to it. After the dualization, the term $bq qb$ becomes a mass term and then no link is created between the $USp(2n-2)$ gauge node and the $USp(4n-4)$ flavor node. The term $b d e$ becomes a mass term for the field $d$. After integrating it out, we obtain
\be \label{U1nc} \scalebox{0.85}{\bpic[node distance=2cm,gSUnode/.style={circle,red,draw,minimum size=8mm},gUSpnode/.style={circle,blue,draw,minimum size=8mm},gSOnode/.style={circle,ForestGreen,draw,minimum size=8mm},fnode/.style={rectangle,draw,minimum size=8mm}]
\begin{scope}[shift={(0,0)}]
\node[gUSpnode] (G1) at (-2,0) {\scalebox{0.8}{$2n-2$}};
\node[gUSpnode] (G2) at (0,0) {\scalebox{0.8}{$4n-4$}};
\node[fnode, blue] (F1) at (2.5,0) {$4n$};
\node[fnode,red] (F2) at (-2,-2) {$1$};
\node[fnode,Maroon] (F3) at (0,-2) {$1$};
\node[fnode] (F4) at (1.6,-2) {$1$};
\draw (G1) -- (G2);
\draw (G2) -- (F1);
\draw (G1) -- (F2);
\draw (G1) -- (F3);
\draw (G2) -- (F3);
\draw (G2) -- (F4);
\draw (-1.6,0.5) to[out=90,in=0]  (-2,1) to[out=180,in=90] (-2.4,0.5);
\node at (-1,0.4) {$B$, \textcolor{Green}{$\frac{1}{3}$}};
\node at (1.3,0.4) {$Q$, \textcolor{Green}{$\frac{2}{3}$}};
\node at (1.6,-0.9) {$E$, \textcolor{Green}{$\frac{6-4n}{3}$}};
\node at (-2.7,-1) {\textcolor{Green}{$-\frac{1}{3}$}, $v$};
\node at (-1.1,-0.8) {\textcolor{Green}{$1$}, $R$};
\node at (0,-1) {\textcolor{Green}{$\frac{2}{3}$}, $P$};
\node at (-2.8,1) {\textcolor{Green}{$\frac{4}{3}$}, $A$};
\node at (0,-3.1) {$ \cW= BQ QB + A B B + v A R + B R P + \Flip[v B E; Q Q; P E; E Q; P Q]$};
\end{scope}
\epic} \ee 
Here the antisymmetric field $A$ is traceless because the trace part has been killed by the superpotential term $\Flip[b b]$. 

We can now use our recursive hypothesis. In the splitted form, the statement of the $\cU_1[n-1]$ duality is the following:
\be \label{U1n2Split} \scalebox{0.85}{\bpic[node distance=2cm,gSUnode/.style={circle,red,draw,minimum size=8mm},gUSpnode/.style={circle,blue,draw,minimum size=8mm},gSOnode/.style={circle,ForestGreen,draw,minimum size=8mm},fnode/.style={rectangle,draw,minimum size=8mm}]
\begin{scope}[shift={(-3,0)}]    
\node[gUSpnode] (G1) at (0,0) {\scalebox{0.8}{$2n-2$}};
\node[fnode, blue] (F1) at (2,0) {\scalebox{0.8}{$4n-4$}};
\node[fnode, red] (F2) at (-1,-1.8) {$1$};
\node[fnode, Maroon] (F3) at (1,-1.8) {$1$};
\draw (G1) -- (F1) node[midway,above] {$B$};
\draw (G1) -- (F2);
\draw (G1) -- (F3);
\draw (0.4,0.5) to[out=90,in=0]  (0,1) to[out=180,in=90] (-0.4,0.5);
\node at (0,-2.8) {$\cW= A B B + v A R$};
\node at (0.6,0.9) {$A$};
\node at (0.8,-0.7) {$R$};
\node at (-0.7,-0.7) {$v$};
\node at (4.5,0) {$\Llra$};
\end{scope}
\begin{scope}[shift={(4.5,0)}]
\node[fnode, blue] (F4) at (0,0) {\scalebox{0.8}{$4n-4$}};
\node[fnode, red] (F5) at (-1,-1.8) {$1$};
\node[fnode, Maroon] (F6) at (1,-1.8) {$1$};
\draw (0.3,0.4) to[out=90,in=0]  (0,0.9) to[out=180,in=90] (-0.4,0.4);
\draw (F4) -- (F5);
\draw (F4) -- (F6);
\draw (F5) -- (F6);
\node at (0,-2.8) {$ \cW = B^3 + BXY + XYZ $};
\node at (0,-3.8) {$+ \tr(B) \, B^2 + \tr(B) \, XY$};
\node at (0.7,0.9) {$B$};
\node at (-0.9,-0.9) {$X$};
\node at (0.9,-0.9) {$Y$};
\node at (0,-1.5) {$Z$};
\end{scope}
\epic} \ee 
Applying it to \eqref{U1nc} we get
\be \label{U1nIntermediate} \scalebox{0.85}{\bpic[node distance=2cm,gSUnode/.style={circle,red,draw,minimum size=8mm},gUSpnode/.style={circle,blue,draw,minimum size=8mm},gSOnode/.style={circle,ForestGreen,draw,minimum size=8mm},fnode/.style={rectangle,draw,minimum size=8mm}]
\begin{scope}[shift={(0,0)}]
\node[gUSpnode] (G1) at (0,0) {\scalebox{0.8}{$4n-4$}};
\node[fnode, blue] (F1) at (2.5,0) {$4n$};
\node[fnode,red] (F2) at (-1.6,-2) {$1$};
\node[fnode,Maroon] (F3) at (0,-2) {$1$};
\node[fnode] (F4) at (1.6,-2) {$1$};
\draw (G1) -- (F1);
\draw (G1) -- (F2);
\draw (G1) -- (F4);
\draw (0.4,0.5) to[out=90,in=0]  (0,1) to[out=180,in=90] (-0.4,0.5);
\draw (0.1,-0.6) -- (0.1,-1.6);
\draw (-0.1,-0.6) -- (-0.1,-1.6);
\node at (0,-3) {$ \cW= B q q +Y P + \Flip[X E; q q; E P; E q; P q; X Y]$};
\node at (0,-3.8) {$ + B^3 + B X Y + \tr(B) B^2 + \tr(B) X Y$};
\node at (0.5,1.1) {$B$};
\node at (1.3,0.3) {$q$};
\node at (1.1,-0.7) {$E$};
\node at (-1.1,-0.7) {$X$};
\node at (-0.4,-1) {$Y$};
\node at (0.4,-1) {$P$};
\end{scope}
\epic} \ee 
At this stage, the antisymmetric field $B$ is traceful. We also see that the fields $Y$ and $P$ are massive. After integrating them out, we obtain
\be \label{U1nd} \scalebox{0.85}{\bpic[node distance=2cm,gSUnode/.style={circle,red,draw,minimum size=8mm},gUSpnode/.style={circle,blue,draw,minimum size=8mm},gSOnode/.style={circle,ForestGreen,draw,minimum size=8mm},fnode/.style={rectangle,draw,minimum size=8mm}]
\begin{scope}[shift={(0,0)}]
\node[gUSpnode] (G1) at (0,0) {\scalebox{0.8}{$4n-4$}};
\node[fnode, blue] (F1) at (2.5,0) {$4n$};
\node[fnode,red] (F2) at (-1.2,-1.8) {$1$};
\node[fnode] (F3) at (1.2,-1.8) {$1$};
\draw (G1) -- (F1);
\draw (G1) -- (F2);
\draw (G1) -- (F3);
\draw (0.4,0.5) to[out=90,in=0]  (0,1) to[out=180,in=90] (-0.4,0.5);
\draw (-0.8,-1.8) -- (0.8,-1.8);
\node at (0,-3.4) {$ \cW= B^3 + \tr(B) B^2 + B q q + \Flip[X E; B X E; E q; B X q; qq]$};
\node at (0,-4.2) {$+ \b \, \Flipper[B X E] \, X E  + \b \, \Flipper[B X q] \, X q  + \Flipper[B X E] \, \tr(B) X E + \Flipper[B X q] \, \tr(B) X q$};
\node at (0.7,1.1) {$B$, \textcolor{Green}{$\frac{2}{3}$}};
\node at (1.3,0.3) {$q$, \textcolor{Green}{$\frac{2}{3}$}};
\node at (1.3,-0.7) {$E$, \textcolor{Green}{$\frac{6-4n}{3}$}};
\node at (-0.9,-0.7) {\textcolor{Green}{$0$}, $X$};
\node at (0,-2.3) {$\b$, \textcolor{Green}{$\frac{2}{3}$}};
\end{scope}
\epic} \ee 

The theory that we obtained is precisely of the form of the one appearing in the l.h.s of \eqref{U2h1}, the special form of the $\cU_2[2n-2,h=1]$ duality. We saw that the dual is given by
\be \label{U1ne} \scalebox{0.85}{\bpic[node distance=2cm,gSUnode/.style={circle,red,draw,minimum size=8mm},gUSpnode/.style={circle,blue,draw,minimum size=8mm},gSOnode/.style={circle,ForestGreen,draw,minimum size=8mm},fnode/.style={rectangle,draw,minimum size=8mm}]
\begin{scope}[shift={(-2,0)}]
\node[fnode,blue] (F4) at (0,0) {$4n$};
\node[fnode,red] (F5) at (-1,-1.8) {$1$};
\node[fnode] (F6) at (1,-1.8) {$1$};
\draw (F4) -- (F5) -- (F6) -- (F4);
\draw (0.3,0.4) to[out=90,in=0]  (0,0.8) to[out=180,in=90] (-0.3,0.4);
\node at (0.8,1) {$a$, \textcolor{Green}{$\frac{2}{3}$}};
\node at (0,-3) {$ \cW= a^3 + a x y + x y z$};
\node at (-1,-0.8) {\textcolor{Green}{$\frac{2}{3}$}, $x$};
\node at (1,-0.8) {$y$, \textcolor{Green}{$\frac{2}{3}$}};
\node at (0,-2.2) {$z$, \textcolor{Green}{$\frac{2}{3}$}};
\node at (3,0) {$\equiv$};
\end{scope}
\begin{scope}[shift={(4,0)}]
\node[fnode, blue] (F2) at (0,0) {$4n+2$};
\draw (0.3,0.4) to[out=90,in=0]  (0,0.9) to[out=180,in=90] (-0.4,0.4);
\node at (0,-1.3) {$ \cW = A^3 + \tr(A) \, A^2$};
\node at (0.7,0.9) {$A$, \textcolor{Green}{$\frac{2}{3}$}};
\end{scope}
\epic} \ee 
Where on the r.h.s.~we simply recombined the fields into a $USp(4n+2)$ traceful antisymmetric.

The conclusion of this chain of dualities is that the l.h.s.~of \eqref{U1na} is dual to the r.h.s.~of \eqref{U1ne}, which is precisely the statement of the $\cU_1[n]$ duality claimed in \eqref{U1}.

%

\section{Understanding $4d$ susy enhancements using $3d$ S-confining dualities}\label{sec:ENH}

Recently in \cite{Kang:2023pot} a $4d$ $\mathcal{N}=1$ theory that flows on a point of the $\mathcal{N}=1$ conformal manifold of the $\mathcal{N}=4$ SYM with gauge group $SU(2n+1)$ was constructed. In this section we propose an understanding of this result from the perspective of the $3d$ reduction of these theories, which uses one of the S-confining dualities we previously discussed. The $3d$ perspective naturally suggests how to generalize the result of \cite{Kang:2023pot} and as an appetizer we propose a $4d$ $\mathcal{N}=1$ theory that flows on a point of the same $\mathcal{N}=1$ conformal manifold of an $\mathcal{N}=2$ necklace quiver theory.

\subsection{$4d$ $\mathcal{N}=1$ theory flowing to $\mathcal{N}=4$ SYM}\label{ENHN=4}

The theory considered in \cite{Kang:2023pot} is obtained via an $\mathcal{N}=1$ gauging of the global symmetry of an $\mathcal{N}=2$ SCFT (the same theory and similar ones were previously studied in \cite{Kang:2021ccs}). The $\mathcal{N}=2$ SCFT in question is the non-Lagrangian $D_2(SU(2n+1))$ theory \cite{Xie:2012hs,Cecotti:2012jx,Cecotti:2013lda}, which can be engineered either as Type IIB on a 3-fold Calabi--Yau hypersurface singularity or as a class $\mathcal{S}$ theory of type $A_{2n}$ on a sphere with one irregular and one regular puncture. The latter realization makes it manifest that the theory has an $SU(2n+1)$ flavor symmetry. The construction of \cite{Kang:2023pot} consists of taking three copies of this $D_2(SU(2n+1))$ theory and performing an $\mathcal{N}=1$ gauging of their diagonal $SU(2n+1)$ symmetry. The theory can be summarized by the following quiver:
\be \label{N1N4} \scalebox{0.85}{\bpic[node distance=2cm,gSUnode/.style={circle,red,draw,minimum size=8mm},gUSpnode/.style={circle,blue,draw,minimum size=8mm},gSOnode/.style={circle,ForestGreen,draw,minimum size=8mm},fnode/.style={rectangle,draw,minimum size=8mm}]
\begin{scope}[shift={(0,0)}]    
\node[gSUnode] (G1) at (0,0) {\scalebox{0.8}{$2n+1$}};
\node (F1) at (2.5,-2) {\scalebox{0.8}{$D_2(SU(2n+1))$}};
\node (F2) at (-2.5,-2) {\scalebox{0.8}{$D_2(SU(2n+1))$}};
\node (F3) at (0,2) {\scalebox{0.8}{$D_2(SU(2n+1))$}};
\node[right] at (2,0) {$ \cW= 0$};
\draw (G1) -- (F1);
\draw (G1) -- (F2);
\draw (G1) -- (F3);
\end{scope}
\epic} \ee
The claim of \cite{Kang:2023pot} is that the resulting model flows to a point of the $\mathcal{N}=1$ conformal manifold of the $\mathcal{N}=4$ SYM with gauge group $SU(2n+1)$.

The $\mathcal{N}=1$ theory has naively three $U(1)$ symmetries. In fact, on top of the $SU(2n+1)$ flavor symmetry, each $D_2(SU(2n+1))$ theory has an $\mathcal{N}=2$ R-symmetry $SU(2)_R\times U(1)_r$, which when thought of as an $\mathcal{N}=1$ theory decomposes to $U(1)_{R_0}\times U(1)_F$ where $U(1)_{R_0}$ is the $\mathcal{N}=1$ R-symmetry while $U(1)_F$ is a flavor symmetry from the $\mathcal{N}=1$ perspective. After the $\mathcal{N}=1$ gauging of three copies of $D_2(SU(2n+1))$, the $U(1)_{R_0}$ symmetries of each of them get identified and give the reference R-symmetry of the resulting theory, while the three $U(1)_F$ symmetries remain as flavor symmetries. Nevertheless, one combination of them is gauge anomalous, so that the full continuos non-anomalous symmetry is only $U(1)^2$. This is then identified with the Cartan of the $SU(3)$ flavor symmetry of the $\mathcal{N}=4$ SYM when considered as an $\mathcal{N}=1$ theory. 

The conclusion of \cite{Kang:2023pot} is that the $\mathcal{N}=1$ theory obtained by gauging three copies of $D_2(SU(2n+1))$ flows to a point of the conformal manifold of the $\mathcal{N}=4$ SYM with gauge group $SU(2n+1)$ where only $\mathcal{N}=1$ supersymmetry and the $U(1)^2$ flavor symmetry is manifest. We next want to review the structure of the conformal manifold of the SYM theory and clarify which one is the specific point reached by the $\mathcal{N}=1$ theory of \cite{Kang:2023pot}.

The $\mathcal{N}=4$ SYM when considered as an $\mathcal{N}=1$ theory consists of an $SU(2n+1)$ gauge group and three adjoint chiral fields $\Phi_i$ with $i=1,2,3$. These have $U(1)_R$ R-charge $\frac{2}{3}$ and are rotated by the $SU(3)$ flavor symmetry. The $\mathcal{N}=1$ superpotential compatible preserving these symmetries is
\begin{equation}\label{eq:N4suppot}
\mathcal{W}^{\mathcal{N}=4}_{\text{SYM}}=\lambda\,\mathrm{Tr}\left(\Phi_1\left[\Phi_2,\Phi_3\right]\right)\,,
\end{equation}
where $\lambda$ is related to the gauge coupling $\tau$ and the trace is taken over the $SU(2n+1)$ color indices.

The $\mathcal{N}=1$ conformal manifold is three dimensional \cite{Leigh:1995ep}. One direction is parametrized by the gauge coupling $\tau$ and preserves the full $\mathcal{N}=4$ supersymmetry as well as the $SU(3)$ flavor symmetry. The other two directions, along which only $\mathcal{N}=1$ is preserved, are parametrized by some cubic invariants constructed from the adjoint chirals $\Phi_i$ that correspond to exactly marginal deformations. These can be found by looking at space of all the marginal deformations, which are in the ${\bf 10}\oplus{\bf 1}$ of $SU(3)$ where $\bf 1$ is the gauge coupling, and quotienting by the complexified flavor symmetry \cite{Kol:2002zt,Benvenuti:2005wi,Green:2010da,Kol:2010ub}. The generic point of the $\mathcal{N}=1$ conformal manifold is reached by turning on the superpotential
\begin{equation}\label{eq:N4N1suppot}
\mathcal{W}^{\mathcal{N}=1}_{\text{SYM}}=\lambda\,\mathrm{Tr}\left(\Phi_1\left(\mathrm{e}^{i\beta}\Phi_2\Phi_3-\mathrm{e}^{-i\beta}\Phi_3\Phi_2\right)\right)+\eta\,\mathrm{Tr}\left(\Phi_1^3+\Phi_2^3+\Phi_3^3\right)\,.
\end{equation}
When $\beta=\eta=0$ we go back to the $\mathcal{N}=4$ line. When $\eta=0$ but $\beta\neq 0$ we have the so-called $\beta$-deformation, which breaks supersymmetry to $\mathcal{N}=1$ and also the $SU(3)$ flavor symmetry to its Cartan $U(1)^2$. When $\eta\neq0$ the continuous flavor symmetry is broken completely. 

The direction of the conformal manifold to which the $\mathcal{N}=1$ theory of \cite{Kang:2023pot} flows is the one parametrized by the $\beta$-deformation. In particular, since the $\mathcal{N}=1$ theory \eqref{N1N4} with $\cW=0$ has an $S_3$ symmetry that permutes the three $D_2(SU(2n+1))$ building blocks, this implies that it flows to a point of the conformal manifold of $\mathcal{N}=4$ SYM with $\beta=\pi$. Indeed, $\beta=\pi$ is the only value of $\beta$ for which \eqref{eq:N4N1suppot} is $S_3$-symmetric in the $\Phi_i$ (at $\beta=0$ there is the Weyl group of $SU(3)$, which is an $S_3$ acting as the \emph{signed} permutations of the three $\Phi_i$'s, so it does not seem possible that \eqref{N1N4} with $\cW=0$ flows to a point on the line $\beta=0$, which is the $\cN=4$ supersymmetric locus). Notice that the $S_3$ permutation symmetry imposes that the moment maps $\mu_i$ from the $SU(2n+1)$ symmetry of the three $D_2(SU(2n+1))$ theories share the same R-charge, which can be determined by the vanishing of the ABJ anomaly. Hence a-maximization \cite{Intriligator:2003jj} is not really required to conclude that under the superconformal R-symmetry the operators $\mu_i$ have the canonical R-charge $\frac{2}{3}$, which is crucial for the claim that this theory is related to $\cN=4$ SYM. 

We would like to reduce both the $\mathcal{N}=1$ theory of \cite{Kang:2023pot} and the dual $\beta,\eta$-deformed $\mathcal{N}=4$ SYM on a circle to three dimensions. Before doing the dimensional reduction, we thus further deform \eqref{N1N4} with also the two marginal $\eta$ and $\beta$ deformations, so to go on the most generic point of the conformal manifold. On the side of the $\mathcal{N}=1$ gauging of three copies of $D_2(SU(2n+1))$ this means that we are turning on the superpotential deformation
\begin{equation}\label{eq:defsuperpot}
\delta\mathcal{W}=\hat{\lambda}\,\mathrm{Tr}\left(\mu_1\left(\mathrm{e}^{i\hat{\beta}}\mu_2\mu_3-\mathrm{e}^{-i\hat{\beta}}\mu_3\mu_2\right)\right)+\hat{\eta}\,\mathrm{Tr}\left(\mu_1^3+\mu_2^3+\mu_3^3\right)\,,
\end{equation}
where again $\mu_i$ is the moment map operator for the $SU(2n+1)$ flavor symmetry of the $i$-th $D_2(SU(2n+1))$ theory and it is mapped to $\Phi_i$ on the SYM side. The variables $\hat{\eta}, \hat{\beta}, \hat{\lambda}\}$ should map to the variables $\{\eta, \beta, \lambda\}$. We do not know the precise mapping, but for symmetry reasons it must be that $\hat{\eta}=0$ if and only if $\eta=0$, and that when $\eta=0$, $\hat{\beta}=0, \pi$ if and only if $\beta=0, \pi$, which is all that we need to know for our purposes.


\subsubsection*{Reduction to $3d$}

Our next goal is to show that the resulting $3d$ $\mathcal{N}=2$ theory is dual to the $3d$ $\mathcal{N}=8$ SYM with gauge group $SU(2n+1)$ on a generic point of the three-dimensional $4d$ $\cN=1$ conformal manifold, where supersymmetry is broken to $3d$ $\mathcal{N}=2$ and the flavor symmetry is completely broken.

In order to show this, we use the remarkable fact that the circle reduction of the $4d$ $\mathcal{N}=2$ $D_2(SU(2n+1))$ SCFT is Lagrangian and it is given by the $3d$ $\mathcal{N}=4$ SQCD with $U(n)$ gauge group and $2n+1$ fundamental hypermultiplets \cite{Xie:2012hs,Buican:2015hsa,Giacomelli:2020ryy,Closset:2020afy}. Hence, the circle reduction of the $4d$ $\mathcal{N}=1$ theory of \cite{Kang:2023pot} obtained by gauging three copies of $D_2(SU(2n+1))$ is a $3d$ $\mathcal{N}=2$ Lagrangian theory that can be summarized with the following quiver:

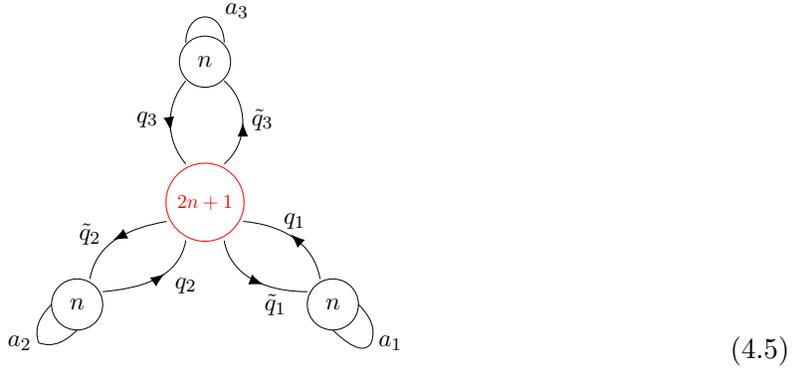
\be \label{3dN1N4} \scalebox{0.85}{\bpic[node distance=2cm,gUnode/.style={circle,black,draw,minimum size=8mm},gSUnode/.style={circle,red,draw,minimum size=8mm},gUSpnode/.style={circle,blue,draw,minimum size=8mm},gSOnode/.style={circle,ForestGreen,draw,minimum size=8mm},fnode/.style={rectangle,draw,minimum size=8mm}]
\begin{scope}[shift={(0,0)}]    
\node[gSUnode] (G1) at (0,0) {\scalebox{0.8}{$2n+1$}};
\node[gUnode] (G2) at (2.0,-1.6) {$n$};
\node[gUnode] (G3) at (-2.0,-1.6) {$n$};
\node[gUnode] (G4) at (0,2.2) {$n$};
\draw (2.4,-1.6) to[out=-45,in=70] (2.6,-2.2) to[out=-120,in=-45] (2,-2);
\draw (-2.4,-1.6) to[out=-135,in=110]  (-2.6,-2.2) to[out=-20,in=-135] (-2,-2);
\draw (0.3,2.5) to[out=90,in=0]  (0,2.9) to[out=180,in=90] (-0.3,2.5);
\draw (0.6,-0.3) to[out=-05,in=100] pic[pos=0.6,sloped,very thick]{arrow=latex reversed} (1.8,-1.2);
\draw (0.3,-0.6) to[out=-80,in=180] pic[pos=0.6,sloped]{arrow} (1.6,-1.4);
\draw (-0.6,-0.3) to[out=-170,in=80] pic[pos=0.4,sloped,very thick]{arrow=latex reversed} (-1.8,-1.2);
\draw (-0.3,-0.6) to[out=-100,in=05] pic[pos=0.4,sloped]{arrow} (-1.6,-1.4);
\draw (0.3,0.6) to[out=40,in=-40] pic[pos=0.5,sloped]{arrow} (0.3,1.9);
\draw (-0.3,0.6) to[out=130,in=-130] pic[pos=0.4,sloped]{arrow} (-0.3,1.9);
\node at (1.4,-0.3) {$q_1$};
\node at (1.1,-1.6) {$\qt_1$};
\node at (-0.3,-1.3) {$q_2$};
\node at (-1.8,-0.5) {$\qt_2$};
\node at (-0.9,1.3) {$q_3$};
\node at (0.9,1.3) {$\qt_3$};
\node at (2.9,-2.2) {$a_1$};
\node at (-2.9,-2.2) {$a_2$};
\node at (0.5,3) {$a_3$};
\end{scope}
\epic} \ee 

We claim that, after starting from the $4d$ theory with the deformation \eqref{eq:defsuperpot}, the superpotential of this $3d$ $\mathcal{N}=2$ theory is
\begin{align}\label{eq:3dsuppot}
\mathcal{W}_{\text{3d}}=\hat{\lambda} \left(\mu_1\left(\mathrm{e}^{i\hat{\beta}}\mu_2\mu_3-\mathrm{e}^{-i\hat{\beta}}\mu_3\mu_2\right)\right)+\hat{\eta}\left(\sum_{i=1}^3\mu_i^3\right)+\sum_{i=1}^3\left(a_iq_i\tilde{q}_i+\mathfrak{M}_i^++\mathfrak{M}_i^-\right)\,,
\end{align}
where $\mu_i=q_i\tilde{q}_i$ with the contraction of the $U(n)$ indices being understood and $\mathfrak{M}_i^\pm$ are the fundamental monopoles of the $i$-th $U(n)$ gauge group. The terms $a_iq_i\tilde{q}_i$ come from the superpotential of the $3d$ $\mathcal{N}=4$ theory when considered as a $3d$ $\mathcal{N}=2$ theory. Instead, we claim that the monopole superpotentials $\mathfrak{M}_i^++\mathfrak{M}_i^-$ are dynamically generated in the compactification by the same arguments of \cite{Aharony:2013dha,Aharony:2013kma}. The above superpotential sets the R-charges of the $3d$ fields to the values
\begin{equation}
R[a_i]=\frac{4}{3}\,,\qquad R[q_i]=R[\tilde{q}_i]=\frac{1}{3}\,,
\end{equation}
which implies that the monopoles $\mathfrak{M}_i^\pm$ have R-charge 2 and can thus be dynamically generated in the superpotential.

This monopole superpotential is crucial, since now we can apply the S-confining duality $\cU_1^{(3d)}[n]$ in \eqref{3dSconfiningQuiver} to each of the three $U(n)$ nodes. The net effect is to replace each of these gauge nodes with an adjoint chiral $\Phi_i$ for the middle $SU(2n+1)$ node, which are mapped to $\mu_i=q_i\tilde{q}_i$ before applying the duality. In the dualization we also produce the $\Phi_i^3$ terms in the superpotential, but these were already present due to the original $\eta$-deformation. To summarize, we obtain the following quiver\footnote{Here $\tilde{\eta}$ is different from $\hat{\eta}$ in \eqref{eq:3dsuppot} due to the application of the duality.} 
\be \label{3dN1N4} \scalebox{0.85}{\bpic[node distance=2cm,gSUnode/.style={circle,red,draw,minimum size=8mm},gUSpnode/.style={circle,blue,draw,minimum size=8mm},gSOnode/.style={circle,ForestGreen,draw,minimum size=8mm},fnode/.style={rectangle,draw,minimum size=8mm}]
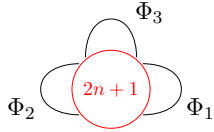
\begin{scope}[shift={(0,0)}]    
\node[gSUnode] (G1) at (0,0) {\scalebox{0.8}{$2n+1$}};
\draw (0.5,-0.4) to[out=0,in=-90] (1.1,0) to[out=90,in=0] (0.5,0.4);
\draw (-0.5,-0.4) to[out=180,in=-90] (-1.1,0) to[out=90,in=180] (-0.5,0.4);
\draw (0.4,0.5) to[out=90,in=0] (0,1.1) to[out=180,in=90] (-0.4,0.5);
\node at (1.4,-0.3) {$\P_1$};
\node at (-1.4,-0.3) {$\P_2$};
\node at (0.6,1.2) {$\P_3$};
\node at (0.5,-1.5) {$\mathcal{W}_{\text{3d}}=\hat{\lambda}\mathrm{Tr}\left(\P_1\left(\mathrm{e}^{i\hat{\beta}} \P_2\P_3- \mathrm{e}^{-i\hat{\beta}}\P_3\P_2\right)\right)+\tilde{\eta} \mathrm{Tr}\left(\sum_{i=1}^3\P_i^3\right)$};
\end{scope}
\epic} \ee 
It corresponds precisely to the 3d $\mathcal{N}=8$ SYM with gauge group $SU(2n+1)$ on a generic point of the $4d$ conformal manifold, in agreement with the claim of \cite{Kang:2023pot}.

\subsection{$4d$ $\mathcal{N}=1$ theory flowing to an $\mathcal{N}=2$ necklace quiver}\label{ENHN=2}

Using the $3d$ perspective we can quite naturally understand how to possibly generalize the findings of \cite{Kang:2023pot} to other $\mathcal{N}=1$ gaugings of $D_2(SU(2n+1))$ so to get an $\mathcal{N}=2$ Lagrangian theory. The idea is to replace adjoint chirals with copies of the $D_2(SU(2n+1))$ theory, and also bifundamental hypermultiplets, if we consider gauging only a subgroup of the flavor symmetry, which would come from the related decomposition of the moment map $\mu$. From our previous analysis it seems that a crucial necessary requirement in order for the construction to work is that the $\mathcal{N}=2$ theory admits an $\mathcal{N}=1$ preserving exactly marginal deformation which completely breaks the continuos flavor symmetry and that can be eventually related to the $\Phi^3$ superpotential in the $3d$ S-confining duality.

As an example, let us consider the $4d$ $\mathcal{N}=2$ necklace quiver theory with three $SU(2n+1)$ gauge nodes
\be \label{Necklace} \scalebox{0.85}{\bpic[node distance=2cm,gSUnode/.style={circle,red,draw,minimum size=8mm},gUSpnode/.style={circle,blue,draw,minimum size=8mm},gSOnode/.style={circle,ForestGreen,draw,minimum size=8mm},fnode/.style={rectangle,draw,minimum size=8mm}]
\begin{scope}[shift={(0,0)}]    
\node[gSUnode] (G1) at (1.5,-1.3) {\scalebox{0.8}{$2n+1$}};
\node[gSUnode] (G2) at (-1.5,-1.3) {\scalebox{0.8}{$2n+1$}};
\node[gSUnode] (G3) at (0,1.3) {\scalebox{0.8}{$2n+1$}};
\draw (2.1,-1.4) to[out=-45,in=70] (2.3,-2) to[out=-120,in=-45] (1.6,-1.9);
\draw (-2.1,-1.4) to[out=-135,in=110] (-2.3,-2) to[out=-20,in=-135] (-1.6,-1.9);
\draw (0.4,1.8) to[out=90,in=0]  (0,2.3) to[out=180,in=90] (-0.4,1.8);
\draw (0.6,0.8) to[out=-30,in=100] pic[pos=0.6,sloped,very thick]{arrow=latex reversed} (1.5,-0.6);
\draw (0.2,0.6) to[out=-80,in=170] pic[pos=0.6,sloped]{arrow} (1.1,-0.7);
\draw (-0.8,-1.2) to[out=30,in=150] pic[pos=0.5,sloped,very thick]{arrow=latex reversed} (0.8,-1.2);
\draw (-0.8,-1.6) to[out=-30,in=-150] pic[pos=0.5,sloped]{arrow} (0.8,-1.6);
\draw (-0.6,0.8) to[out=-160,in=80] pic[pos=0.4,sloped,very thick]{arrow=latex reversed} (-1.5,-0.6);
\draw (-0.2,0.6) to[out=-100,in=20] pic[pos=0.4,sloped]{arrow} (-1.1,-0.7);
\node at (1.6,0.3) {$\Qt_1$};
\node at (0.8,0.1) {$Q_1$};
\node at (0.2,-2.1) {$\Qt_2$};
\node at (0.2,-1.3) {$Q_2$};
\node at (-1.6,0.3) {$\Qt_3$};
\node at (-0.8,0.1) {$Q_3$};
\node at (2.7,-2.1) {$\Phi_1$};
\node at (-2.7,-2.1) {$\Phi_2$};
\node at (0.4,2.5) {$\Phi_3$};
\end{scope}
\epic} \ee 

\noindent which we drew in $\mathcal{N}=1$ notation. The superpotential is
\begin{equation}
\mathcal{W}^{\cN=2}_{\text{necklace}}=\sum_{i=1}^3 \lambda_i \mathrm{Tr}(\Phi_i(Q_i\tilde{Q}_i-Q_{i+1}\tilde{Q}_{i+1}))\,,
\end{equation}
where $Q_4=Q_1$ and $\tilde{Q}_4=\tilde{Q}_1$. 

The continuous global symmetries of the theory depend on the position on the conformal manifold. There is a $3$-complex dimensional conformal manifold with $\cN=2$ supersymmetry, parameterized by the $3$ gauge couplings. On this $\cN=2$ conformal manifold the superpotential couplings $\lambda_i$ are set by supersymmetry in terms of the gauge couplings. The global symmetry is $\prod_{i=1}^3 U(1)_{i, baryonic} \times SU(2)_R \times U(1)_r$, where the $i^{th}$ baryonic $U(1)$ acts on the fields $\{Q_i, \Qt_i\}$ with charges $\{+1, -1\}$ and with zero charge on the other fields.

In $\cN=1$ language the visible global symmetry is  
\be \label{globsymm1} \prod_{i=1}^3 U(1)_{i, baryonic} \times U(1)_F \times U(1)_{R_0}\,,\ee
where $U(1)_F$ acts with charges $+2,-1,-1$ on $\Phi_i, Q_i, \Qt_i$, respectively, while $U(1)_{R_0}$ is the canonical $\mathcal{N}=1$ R-symmetry under which all the fields $\Phi_i, Q_i, \Qt_i$ have R-charge $\frac{2}{3}$.

\subsubsection*{$\cN=1$ conformal manifold of the necklace quiver}

We are interested in the full $\cN=1$ conformal manifold of the $SU(N)^3$ necklace (in Appendix \ref{CMnecklace} we discuss the general $\mathbb{Z}_k$ orbifold of $\cN=4$ SYM, which as far as we are aware was never determined in the literature). We use the method of \cite{Kol:2002zt}, of quotienting the space of marginal deformations (that is the chiral ring operators with $U(1)_{R_0}$ R-charge $R_0=2$) by the broken global symmetries.

There are $6$ operators with $R_0=2$ in the chiral ring
\be\label{R=2ops} 
\mathrm{Tr}(\Phi Q \Qt)\,, \quad \mathrm{Tr}(\Phi_1^3)\,,\quad  \mathrm{Tr}(\Phi_2^3)\,,\quad \mathrm{Tr}(\Phi_3^3)\,,\quad  \mathrm{Tr}(Q_1 Q_2 Q_3)\,,\quad \mathrm{Tr}(\Qt_3 \Qt_2 \Qt_1)\,,
\ee
where $\mathrm{Tr}(\Phi Q \Qt)$ represents the only operator in the chiral ring of the form $\Phi Q \Qt$, since all of the operators of this form are equal in the chiral ring due to the F-terms relations, more precisely we can write this operator as
\be 
\mathrm{Tr}(\Phi Q \Qt)=\sum_{i=1}^3 \mathrm{Tr}(\Phi_i Q_i\tilde{Q}_i + \Phi_i Q_{i+1}\tilde{Q}_{i+1}) \,,
\ee
This is the operator which drives the so-called $\beta$-deformation, which exists in all SCFTs on D3-branes at toric Calabi--Yau singularities \cite{Benvenuti:2005wi}.

Turining on the operators in \eqref{R=2ops}, the global symmetry \eqref{globsymm1} is broken to $U(1)^2_{baryonic} \times U(1)_{R_0}$. The rank of the global symmetry decreases by $2$ units. This implies that there are $6-2=4$ directions in the full $\cN=1$ conformal manifold, which is thus $7$-dimensional.

One submanifold which is useful for us has a special point where the $S_3$ exchange symmetry of the $3$ nodes is preserved. Such submanifold has superpotential
\begin{align}\label{eq:suppotneckCM}
\mathcal{W}^{\cN=1}_{\text{necklace}}&= \lambda \sum_{i=1}^3 \mathrm{Tr}(\Phi_i(e^{i \beta}Q_i\tilde{Q}_i-e^{- i \beta}Q_{i+1}\tilde{Q}_{i+1})) \nn\\
&+ \eta_1 \sum_{i=1}^3 \mathrm{Tr}(\Phi_i^3) + \eta_2 (\mathrm{Tr}(Q_1 Q_2 Q_3) + \mathrm{Tr}(\Qt_3 \Qt_2 \Qt_1))\,.
\end{align}
This is a $3$-dimensional manifold parameterized by the gauge coupling, $\lambda$, $\beta$ (in terms of which the couplings $\eta_1$ and $\eta_2$ are set). In particular, we observe that the point $\beta=\pi$ is invariant under the $S_3$ exchange symmetry of the three gauge nodes. The superpotential \eqref{eq:suppotneckCM} will also be useful when compactifying to $3d$ because the last two terms force the R-charges of the elementary fields to be $\frac{2}{3}$ also in $3d$, which as we have seen in the previous subsection is crucial for our analysis.

\subsubsection*{Construction via $\cN=1$ gauging of  the $D_2(SU(6n+3))$ theory}
We claim that we can construct an $\mathcal{N}=1$ theory that flows to point on the direction of the conformal manifold of the $\mathcal{N}=2$ necklace quiver theory corresponding to the $\beta$-deformation by taking one copy of the $D_2(SU(6n+3))$ theory and gauging its $SU(2n+1)^3$ subgroup. The symbolic quiver representing this theory is as follows:
\be \label{N1N2} \scalebox{0.85}{\bpic[node distance=2cm,gSUnode/.style={circle,red,draw,minimum size=8mm},gUSpnode/.style={circle,blue,draw,minimum size=8mm},gSOnode/.style={circle,ForestGreen,draw,minimum size=8mm},fnode/.style={rectangle,draw,minimum size=8mm}]
\begin{scope}[shift={(0,0)}]    
\node (G1) at (0,0) {\scalebox{0.8}{$D_2(SU(6n+3))$}};
\node[gSUnode] (G2) at (2.0,-1.6) {\scalebox{0.8}{$2n+1$}};
\node[gSUnode] (G3) at (-2.0,-1.6) {\scalebox{0.8}{$2n+1$}};
\node[gSUnode] (G4) at (0,2.2) {\scalebox{0.8}{$2n+1$}};
\draw (G1) -- (G2);
\draw (G1) -- (G3);
\draw (G1) -- (G4);
\node[] at (3,0.3) {\scalebox{0.8}{$\mathcal{W}=\Flip[s_1;s_2]$}};
\end{scope}
\epic} \ee

The moment map $\mu$ in the adjoint representation of the $SU(6n+3)$ flavor symmetry decomposes under the $SU(2n+1)^3\times U(1)^2$ subgroup as
\begin{align}\label{eq:BRSU6n+3}
\textbf{adj}_{[\mu]}&\to 2\times({\bf 1},{\bf 1},{\bf 1})^{(0,0)}_{[s_1,s_2]}\oplus(\textbf{adj},{\bf 1},{\bf 1})^{(0,0)}_{[\P_1]}\oplus ({\bf 1},\textbf{adj},{\bf 1})^{(0,0)}_{[\P_2]}\oplus ({\bf 1},{\bf 1},\textbf{adj})^{(0,0)}_{[\P_3]}\nonumber\\
&\oplus ({\bf 2n+1},{\bf 1},\overline{\bf 2n+1})^{(0,-1)}_{[Q_1]}\oplus (\overline{\bf 2n+1},{\bf 1},{\bf 2n+1})^{(0,1)}_{[\Qt_1]}\oplus (\overline{\bf 2n+1},{\bf 2n+1},{\bf 1})^{(-1,0)}_{[Q_2]}\nonumber\\
&\oplus ({\bf 2n+1},\overline{\bf 2n+1},{\bf 1})^{(1,0)}_{[\Qt_2]}\oplus ({\bf 1},\overline{\bf 2n+1},{\bf 2n+1})^{(1,-1)}_{[Q_3]}\oplus ({\bf 1},{\bf 2n+1},\overline{\bf 2n+1})^{(-1,1)}_{[\Qt_3]}\,,
\end{align}
where the superscripts are the $U(1)^2$ charges while the subscripts indicate the field in the $\mathcal{N}=2$ necklace quiver to which each state corresponds.
Hence, the intuition from the $3d$ perspective is that after applying the S-confining duality $\mu$ will decompose into the three adjoint chirals and the three pairs of bifundamental chirals that form the necklace quiver. Notice that $\mu$ also supplements two extra singlets that we will need to flip in order to get the $\mathcal{N}=2$ theory and we will see momentarily that this is further justified by the fact that if we don't do so they would give a decoupled free sector. 

We claim that the theory \eqref{N1N2} with superpotential $\mathcal{W}=\Flip[s_1;s_2]$ flows to a point on the conformal manifold of the $\mathcal{N}=2$ necklace quiver that sits on the $\beta$-deformation line. If we want instead to go to a more generic point of the conformal manifold corresponding to \eqref{eq:suppotneckCM}, then the superpotential is (the name of the fields are the ones appearing in the decomposition of the moment map \eqref{eq:BRSU6n+3})
\begin{align} \label{WN1N2}
\mathcal{W} &= \hat{\lambda} \sum_{i=1}^3 \mathrm{Tr}(\Phi_i(e^{i \hat{\beta}}Q_i\tilde{Q}_i-e^{- i \hat{\beta}}Q_{i+1}\tilde{Q}_{i+1})) + \hat{\eta}_1 \sum_{i=1}^3 \mathrm{Tr}(\Phi_i^3) \nn\\
&+ \hat{\eta}_2 (\mathrm{Tr}(Q_1 Q_2 Q_3) + \mathrm{Tr}(\Qt_3 \Qt_2 \Qt_1)) + \Flip[s_1;s_2]\,.
\end{align}

Let us analyze this $\mathcal{N}=1$ model more in details. First of all, it naively has three abelian symmetries, where two of them come from the decomposition $SU(6n+3)\to SU(2n+1)^3\times U(1)^2$, while the third $U(1)_F$ is the commutant of the $\mathcal{N}=1$ $U(1)_{R_0}$ R-symmetry inside the $\mathcal{N}=2$ $SU(2)_R\times U(1)_r$ R-symmetry
\begin{align}\label{eq:CartanN2Rsym}
R_0=\frac{1}{3}r+\frac{4}{3}I_3\,,\qquad F=-r+2I_3\,,
\end{align}
where $R_0$, $F$ and $r$ are the generators of $U(1)_{R_0}$, $U(1)_F$ and $U(1)_r$ respectively, while $I_3$ is the generator of the Cartan of $SU(2)_R$. We use here the usual parametrization where for a hypermultiplet scalar $r=0$ and $I_3=\frac{1}{2}$ while for a gaugino $r=1$ and $I_3=0$. Nevertheless, the $U(1)_F$ symmetry is anomalous and so the actual symmetry of the model is only $U(1)^2$. 

We also observe that the $\mathcal{N}=1$ theory has an $S_3$ symmetry that comes from the Weyl group of the original $SU(6n+3)$ symmetry which is not part of the Weyl of the gauged $SU(3n+1)^3$ symmetry and which acts by permuting these three factors. These considerations about the continuous and discrete global symmetries of the model, remembering our observation below \eqref{eq:suppotneckCM}, suggest that the particular point of the conformal manifold of the $4d$ $\mathcal{N}=2$ necklace theory to which the $\mathcal{N}=1$ theory in \eqref{N1N2} with only the flipping superpotential flows is the one with $\beta=\pi$. At this point indeed the theory has $\mathcal{N}=1$ supersymmetry, $U(1)^2$ continuous symmetry and $S_3$ discrete symmetry.

In order to that the $U(1)_F$ symmetry is anomalous, we first define the trial $U(1)_{R_{\text{trial}}}$ R-symmetry obtained by mixing $U(1)_{R_0}$ with $U(1)_F$
\begin{equation}
R_{\text{trial}}=R_0+\epsilon\,F\,.
\end{equation}
Notice that we didn't consider a mixing with the other $U(1)^2$ symmetries. This is because they come from the non-abelian symmetry $SU(6n+3)$ of the $D_2(SU(6+3))$ symmetry whose cubic anomaly vanishes, and so all anomalies that do not involve these $U(1)^2$ symmetries quadratically vanish implying that any mixing coefficient with them would be set to zero by the $a$-maximization. Then we want to compute its ABJ anomaly 
\begin{equation}
\mathrm{Tr}\,R_{\text{trial}}SU(2n+1)^2=2n+1+\mathrm{Tr}\,R_{\text{trial}}SU(2n+1)^2\Big|_{D_2(SU(6n+3))}\,,
\end{equation}
where the first part is the contribution of the gauginos while the second part is the contribution of the $D_2(SU(6n+3))$ theory. The latter can be obtained as follows:
\begin{align}
\mathrm{Tr}\,R_{\text{trial}}SU(2n+1)^2\Big|_{D_2(SU(6n+3))}&=\left(\frac{1}{3}-\epsilon\right)\mathrm{Tr}\,rSU(6n+3)^2\Big|_{D_2(SU(6n+3))}\nonumber\\
&+\left(\frac{4}{3}+2\epsilon\right)\mathrm{Tr}\,I_3SU(6n+3)^2\Big|_{D_2(SU(6n+3))}\nonumber\\
&=-\left(\frac{1}{3}-\epsilon\right)\frac{k_{4d}^{SU(6n+3)}}{2}=-\left(\frac{1}{3}-\epsilon\right)\frac{6n+3}{2}\,,
\end{align}
where we used that the embedding index fo $SU(2n+1)$ inside $SU(6n+3)$ is trivial and that for a $4d$ $\mathcal{N}=2$ SCFT
\begin{equation}
\mathrm{Tr}\,rF^2=-\frac{k_{4d}^F}{2}\,,\qquad \mathrm{Tr}\,I_3F^2=0\,,
\end{equation}
with $k_{4d}^F$ is the flavor symmetry central charge, which for the $SU(N)$ symmetry of a $D_p(SU(N))$ theory is \cite{Xie:2016evu}
\begin{equation}
k_{4d}^{SU(N)}=2N-\frac{2N}{p}\,.
\end{equation}
Overall, we then find the ABJ anomaly
\begin{equation}
\mathrm{Tr}\,R_{\text{trial}}SU(2n+1)^2=2n+1-\left(\frac{1}{3}-\epsilon\right)\frac{6n+3}{2}
\end{equation}
and requiring that it vanishes fixes
\begin{equation}
\epsilon=-\frac{1}{3}\,.
\end{equation}

As we mentioned before the other $U(1)^2$ symmetries cannot mix with the R-symmetry, so the previous computation determines what will be the superconformal R-symmetry of the theory
\begin{equation}
R_{\text{s.c.}}=R_0-\frac{1}{3}F=\frac{2}{3}(r+I_3)\,.
\end{equation}
Notice in particular that this gives the R-charge $\frac{2}{3}$ to the moment map $\mu$, which is necessary to connect with the $\mathcal{N}=2$ necklace theory where as we commented above the components of $\mu$ are associated to the fields $\Phi_i$, $Q_i$ and $\tilde{Q}_i$. On the other hand, we see that the two singlets in the decomposition \eqref{eq:BRSU6n+3} are gauge invariants with the free R-charge and so they should be flipped.

We now want to compute the $a$ and $c$ central charges and check that they match those of the $\mathcal{N}=2$ necklace theory. For this, we first compute
\begin{align}
\mathrm{Tr}\,R_{\text{s.c.}}=3\times4n(n+1)+2\left(\frac{4}{3}-1\right)+\frac{2}{3}\mathrm{Tr}\,r\Big|_{D_2(SU(6n+3))}=-2
\end{align}
where the first term is the contribution of the gauginos, the second one is from the flipper fields and the last term is the contribution of $D_2(SU(6n+3))$ which was computed using \cite{Cecotti:2013lda,Xie:2015rpa,Giacomelli:2017ckh}
\begin{equation}
\mathrm{Tr}\,r\Big|_{D_2(SU(6n+3))}=48\left(a_{D_2(SU(6n+3))}-c_{D_2(SU(6n+3))}\right)=-2\left(9 n^2+9 n+2\right)\,.
\end{equation}
We then compute 
\begin{align}
\mathrm{Tr}\,R_{\text{s.c.}}^3&=3\times4n(n+1)+2\left(\frac{4}{3}-1\right)^3+\frac{8}{9}\mathrm{Tr}\,rI_3^2\Big|_{D_2(SU(6n+3))}+\frac{8}{27}\mathrm{Tr}\,r^3\Big|_{D_2(SU(6n+3))}\nonumber\\
&=\frac{2}{9} \left(48 n^2+48 n-1\right)\,,
\end{align}
where we used that
\begin{align}
\mathrm{Tr}\,rI_3^2\Big|_{D_2(SU(6n+3))}&=4a_{D_2(SU(6n+3))}-2c_{D_2(SU(6n+3))}=\frac{1}{2} \left(9 n^2+9 n+2\right)\,,\nonumber\\
\mathrm{Tr}\,r^3\Big|_{D_2(SU(6n+3))}&=48\left(a_{D_2(SU(6n+3))}-c_{D_2(SU(6n+3))}\right)=-2\left(9 n^2+9 n+2\right)\,.
\end{align}
We finally find the conformal $a$ and $c$ central charges \cite{Anselmi:1997am}
\begin{align}
a&=\frac{3}{32}\left(3\mathrm{Tr}\,R_{\text{s.c.}}^3-\mathrm{Tr}\,R_{\text{s.c.}}\right)=3 n^2+3 n+\frac{1}{8}\,,\nonumber\\
c&=\frac{1}{32}\left(9\mathrm{Tr}\,R_{\text{s.c.}}^3-5\mathrm{Tr}\,R_{\text{s.c.}}\right)=3 n^2+3 n+\frac{1}{4}\,,
\end{align}
which precisely coincide with those of the $\mathcal{N}=2$ necklace quiver theory
\begin{align}
a&=\frac{5n_v+n_h}{24}=3 n^2+3 n+\frac{1}{8}\,,\nonumber\\
c&=\frac{2n_v+n_h}{12}=3 n^2+3 n+\frac{1}{4}\,,
\end{align}
where $n_v=12n(n+1)$ is the number of $\mathcal{N}=2$ vectors and $n_h=3(2n+1)^2$ is the number of hypers.

\subsubsection*{Reduction to $3d$}
Now we want to reduce the theory \eqref{N1N2} with superpotential \eqref{WN1N2} to $3d$. Using once again the Lagrangian theory corresponding to the circle reduction of $D_2(SU(6n+3))$ we get the following quiver:
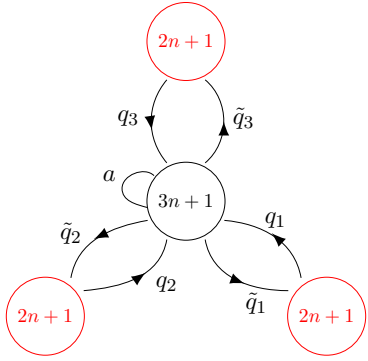
\be \label{3dN1N2} \scalebox{0.85}{\bpic[node distance=2cm,gUnode/.style={circle,black,draw,minimum size=8mm},gSUnode/.style={circle,red,draw,minimum size=8mm},gUSpnode/.style={circle,blue,draw,minimum size=8mm},gSOnode/.style={circle,ForestGreen,draw,minimum size=8mm},fnode/.style={rectangle,draw,minimum size=8mm}]
\begin{scope}[shift={(0,0)}]    
\node[gUnode] (G1) at (0,0) {\scalebox{0.8}{$3n+1$}};
\node[gSUnode] (G2) at (2.2,-1.8) {\scalebox{0.8}{$2n+1$}};
\node[gSUnode] (G3) at (-2.2,-1.8) {\scalebox{0.8}{$2n+1$}};
\node[gSUnode] (G4) at (0,2.5) {\scalebox{0.8}{$2n+1$}};
\draw (-0.5,0.4) to[out=135,in=90] (-1,0.2) to[out=-90,in=165] (-0.6,-0.1);
\draw (0.6,-0.3) to[out=-05,in=100] pic[pos=0.6,sloped,very thick]{arrow=latex reversed} (1.8,-1.2);
\draw (0.3,-0.6) to[out=-80,in=180] pic[pos=0.6,sloped]{arrow} (1.6,-1.4);
\draw (-0.6,-0.3) to[out=-170,in=80] pic[pos=0.4,sloped,very thick]{arrow=latex reversed} (-1.8,-1.2);
\draw (-0.3,-0.6) to[out=-100,in=05] pic[pos=0.4,sloped]{arrow} (-1.6,-1.4);
\draw (0.3,0.6) to[out=40,in=-40] pic[pos=0.5,sloped]{arrow} (0.3,1.9);
\draw (-0.3,0.6) to[out=130,in=-130] pic[pos=0.4,sloped]{arrow} (-0.3,1.9);
\node at (1.4,-0.3) {$q_1$};
\node at (1.1,-1.6) {$\qt_1$};
\node at (-0.3,-1.3) {$q_2$};
\node at (-1.8,-0.5) {$\qt_2$};
\node at (-0.9,1.3) {$q_3$};
\node at (0.9,1.3) {$\qt_3$};
\node at (-1.2,0.4) {$a$};
\node at (0,-3.2) {$\cW= \sum_{i=1}^3 \mathrm{Tr}(\left(q_i \qt_i \right)^3) +  (\mathrm{Tr}([q_3 \qt_1] [q_1 \qt_2] [q_2 \qt_3]) + \mathrm{Tr}([\qt_3 q_1] [\qt_1 q_2] [\qt_2 q_3])) $};
\node at (0,-4.2) {$+ \Flip[\mathrm{Tr}(q_1 \qt_1); \mathrm{Tr}(q_2 \qt_2)] +a\sum_{i=1}^3q_i\tilde{q}_i+\mathfrak{M}^++\mathfrak{M}^-$};
\end{scope}
\epic} \ee 
Now we can apply the S-confining duality $\cU_1^{(3d)}[n]$ of Section \ref{sec3dSconfining} to the central $U(3n+1)$ node to obtain
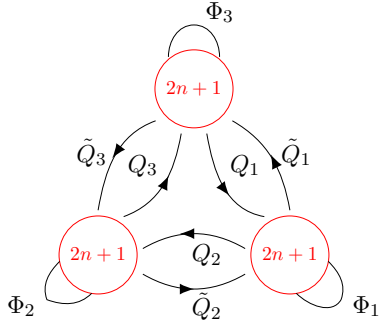
\be \label{3dNecklace} \scalebox{0.85}{\bpic[node distance=2cm,gSUnode/.style={circle,red,draw,minimum size=8mm},gUSpnode/.style={circle,blue,draw,minimum size=8mm},gSOnode/.style={circle,ForestGreen,draw,minimum size=8mm},fnode/.style={rectangle,draw,minimum size=8mm}]
\begin{scope}[shift={(0,0)}]    
\node[gSUnode] (G1) at (1.5,-1.3) {\scalebox{0.8}{$2n+1$}};
\node[gSUnode] (G2) at (-1.5,-1.3) {\scalebox{0.8}{$2n+1$}};
\node[gSUnode] (G3) at (0,1.3) {\scalebox{0.8}{$2n+1$}};
\draw (2.1,-1.4) to[out=-45,in=70] (2.3,-2) to[out=-120,in=-45] (1.6,-1.9);
\draw (-2.1,-1.4) to[out=-135,in=110] (-2.3,-2) to[out=-20,in=-135] (-1.6,-1.9);
\draw (0.4,1.8) to[out=90,in=0]  (0,2.3) to[out=180,in=90] (-0.4,1.8);
\draw (0.6,0.8) to[out=-30,in=100] pic[pos=0.6,sloped,very thick]{arrow=latex reversed} (1.5,-0.6);
\draw (0.2,0.6) to[out=-80,in=170] pic[pos=0.6,sloped]{arrow} (1.1,-0.7);
\draw (-0.8,-1.2) to[out=30,in=150] pic[pos=0.5,sloped,very thick]{arrow=latex reversed} (0.8,-1.2);
\draw (-0.8,-1.6) to[out=-30,in=-150] pic[pos=0.5,sloped]{arrow} (0.8,-1.6);
\draw (-0.6,0.8) to[out=-160,in=80] pic[pos=0.4,sloped,very thick]{arrow=latex reversed} (-1.5,-0.6);
\draw (-0.2,0.6) to[out=-100,in=20] pic[pos=0.4,sloped]{arrow} (-1.1,-0.7);
\node at (1.6,0.3) {$\Qt_1$};
\node at (0.8,0.1) {$Q_1$};
\node at (0.2,-2.1) {$\Qt_2$};
\node at (0.2,-1.3) {$Q_2$};
\node at (-1.6,0.3) {$\Qt_3$};
\node at (-0.8,0.1) {$Q_3$};
\node at (2.7,-2.1) {$\Phi_1$};
\node at (-2.7,-2.1) {$\Phi_2$};
\node at (0.4,2.5) {$\Phi_3$};
\node at (0,-3.2) {$\cW= \sum_{i=1}^3 \mathrm{Tr}(\Phi_i(\mathrm{e}^{i\hat{\beta}}Q_i\tilde{Q}_i-\mathrm{e}^{-i\hat{\beta}}Q_{i+1}\tilde{Q}_{i+1}))$};
\node at (0,-4.2) {$+ \sum_{i=1}^3 \mathrm{Tr}(\Phi_i^3) + (\mathrm{Tr}(Q_1 Q_2 Q_3) + \mathrm{Tr}(\Qt_3 \Qt_2 \Qt_1))$};
\end{scope}
\epic} \ee
It corresponds precisely to the $3d$ reduction of the $4d$ necklace quiver \eqref{Necklace} on a generic point of its conformal manifold which sustains our proposed $4d$ duality.

\subsubsection*{Superconfomal index}

We can perform a further check of our claim that the $\mathcal{N}=1$ gauging of the $SU(2n+1)^3$ subgroup of one copy of $D_2(SU(6n+3))$ flows to a direction of the conformal manifold of the $\mathcal{N}=2$ necklace quiver theory that preserves a $U(1)^2$ flavor symmetry. This was also done in \cite{Kang:2023pot} for their $\mathcal{N}=4$ example and it consists of matching a limit of the superconformal index.
We review more in details our conventions for the $4d$ superconformal index in Appendix \ref{app:index}, while here we only use some specific facts that are useful for our analysis.

The superconformal index of a $4d$ $\mathcal{N}=1$ theory is defined as
\be
\mathcal{I}_{\mathcal{N}=1}=\mathrm{Tr}_{\delta=0}(-1)^F\left(\frac{p}{q}\right)^{j_1}(p\,q)^{j_2+\frac{R}{2}}\prod_if_i^{T_i}\,,
\ee
where $\delta=2j_2+\frac{3}{2}R$, $R$ is the generator of the IR superconformal R-symmetry and $f_i$, $T_i$ are fugacities and generators for additional global symmetries of the theory. As we have seen previously, for our $\mathcal{N}=1$ model the superconformal R-symmetry is given by $R=R_0+\epsilon F$ with $\epsilon=-\frac{1}{3}$ and the global symmetry is $U(1)^2$, so we can write
\be\label{eq:indN1}
\mathcal{I}_{\mathcal{N}=1}=\mathrm{Tr}_{\delta=0}(-1)^F\left(\frac{p}{q}\right)^{j_1}(p\,q)^{j_2+\frac{R_0}{2}+\frac{\epsilon F}{2}}\prod_{i=1}^2f_i^{T_i}\,.
\ee

Instead, the superconformal index of a $4d$ $\mathcal{N}=2$ theory is defined as
\be
\mathcal{I}_{\mathcal{N}=2}=\mathrm{Tr}_{\delta=0}(-1)^F\left(\frac{p}{q}\right)^{j_1}(p\,q)^{j_2+\frac{r}{2}}t^{I_3-\frac{r}{2}}\prod_if_i^{T_i}\,,
\ee
where now $\delta=2j_2+2I_3+\frac{r}{2}$ and we have one additional fugacity $t$ due to the fact that the R-symmetry is larger. Remembering \eqref{eq:CartanN2Rsym}, we can rewrite this as
\be\label{eq:indN2}
\mathcal{I}_{\mathcal{N}=2}=\mathrm{Tr}_{\delta=0}(-1)^F\left(\frac{p}{q}\right)^{j_1}(p\,q)^{j_2+\frac{R_0}{2}-\frac{F}{3}}t^{\frac{F}{2}}\prod_if_i^{T_i}\,.
\ee
As we have seen above, when we perform the $\mathcal{N}=1$ gauging of the $\mathcal{N}=2$ $D_2(SU(6n+3))$ theory one combination of $U(1)_{R_0}$ and $U(1)_F$ is broken by the ABJ anomaly. At the level of the index this means that we should lose one fugacity and indeed we can go from \eqref{eq:indN2} to \eqref{eq:indN1} by setting $t=(p\,q)^{\frac{2}{3}+\epsilon}=(p\,q)^{\frac{1}{3}}$.

The above considerations lead us to write the index of the $\mathcal{N}=1$ gauging of the $SU(2n+1)^3$ subgroup of one copy of $D_2(SU(6n+3))$ as
\begin{align}
\mathcal{I}_{\mathcal{N}=1}(f_1,f_2,p,q)&=\left(\mathcal{I}_{\text{chir}}^{\mathcal{N}=1}\left((p\,q)^{\frac{2}{3}};p,q\right)\right)^2\frac{1}{(2n+1)!^3}\oint\left[\prod_{a=1}^3\left(\prod_{i=1}^{2n+1}\frac{\mathrm{d}z_i^{(a)}}{2\pi iz_i^{(a)}}\right)\mathcal{I}_{\text{vec}}^{\mathcal{N}=1}\left(\vec{z}^{(a)};p,q\right)\right]\nn\\
&\times\mathcal{I}_{D_2(SU(6n+3))}\left(\vec{z}^{(1)},\vec{z}^{(2)},\vec{z}^{(3)},f_1,f_2;p,q,t=(p\,q)^{\frac{1}{3}}\right)\,.
\end{align}
The first factor is the contribution of the flipper fields, which is written in terms of the index of a chiral multiplet with R-charge $\Delta$
\be\label{eq:indchir}
\mathcal{I}_{\text{chir}}^{\mathcal{N}=1}\left((p\,q)^{\frac{\Delta}{2}}x;p,q\right)=\mathrm{PE}\left[\frac{(p\,q)^{\frac{\Delta}{2}}x-(p\,q)^{\frac{2-\Delta}{2}}x^{-1}}{(1-p)(1-q)}\right]\,.
\ee
The integral is over the Cartan of the gauge group $SU(2n+1)^3$, where the fugacities satisfy $\prod_iz_i^{(a)}=1$. The contribution of an $\mathcal{N}=1$ vector multiplet is
\be
\mathcal{I}_{\text{vec}}^{\mathcal{N}=1}\left(\vec{z};p,q\right)=\mathrm{PE}\left[\frac{2p\,q-p-q}{(1-p)(1-q)}\chi_{\textbf{adj}}^{SU(2n+1)}(\vec{z})\right]\,,
\ee
where $\chi_{\bf\text{adj}}^{SU(2k+1)}(\vec{u})$ is the character of the adjoint representation of $SU(2k+1)$, which in the following we are going to parametrize as
\be
\chi_{\textbf{adj}}^{SU(2n+1)}(\vec{z})=\sum_{i,j=1}^{2n+1}\frac{z_i}{z_j}-1\,.
\ee
Finally, $\mathcal{I}_{D_2(SU(6n+3))}$ is the $\mathcal{N}=2$ index of the $D_2(SU(6n+3))$ theory, where the fugacities for its $SU(6n+3)$ haven been decomposed in terms of those of the $SU(2n+1)^3\times U(1)^2$ subgroup.

The $D_2(SU(6n+3))$ theory is non-Lagrangian and so its full index is in general not known. The only case for which it is known is for the $D_2(SU(3))=(A_1,D_4)$ theory, for which one can compute it using the $\mathcal{N}=1$ Lagrangians of \cite{Maruyoshi:2016tqk,Maruyoshi:2016aim,Agarwal:2016pjo,Garozzo:2020pmz}. This corresponds to the case $n=1$ which is degenarate since the $\mathcal{N}=2$ necklace quiver doesn't have any gauge symmetry and only consists of 3 free hypers or 6 free chirals. In particular, since there is no gauge symmetry there is no reason to impose the constraint on the fugacities $t=(p\,q)^{\frac{1}{3}}$ due to the ABJ anomaly. Nevertheless, we curiously observe that the index of the $D_2(SU(3))$ theory computed in eq.~(5.12) of \cite{Agarwal:2016pjo} still reduces to the index of eight chirals of R-charge $\frac{2}{3}$ after the specialization $t=(p\,q)^{\frac{1}{3}}$, where the two extra chirals are removed once we add the flipper fields.

For the general $n$ case, we can still match a particular limit of the full index. Indeed, there is a famous limit of the $\mathcal{N}=2$ index known as the \emph{Schur limit} \cite{Gadde:2011uv} which is known for a generic $D_2(SU(2k+1))$ theory \cite{Xie:2016evu,Song:2017oew}
\be
\mathcal{I}_{D_2(SU(2k+1))}(\vec{u};p,q,t=q)=\mathrm{PE}\left[\frac{q}{1-q^2}\chi_{\textbf{adj}}^{SU(2k+1)}(\vec{u})\right]\,,
\ee
with $\prod_iu_i=1$. Hence, in our case we have
\be\label{eq:D2ind}
\mathcal{I}_{D_2(SU(6n+3))}\left(\vec{z}^{(1)},\vec{z}^{(2)},\vec{z}^{(3)},f_1,f_2;p,q,t=q\right)=\mathrm{PE}\left[\frac{q}{1-q^2}\chi_{\textbf{adj}}^{SU(6n+3)}\left(\vec{z}^{(1)},\vec{z}^{(2)},\vec{z}^{(3)},f_1,f_2\right)\right]\,,
\ee
where we decompose the character of the adjoint representation of $SU(6n+3)$ according to the branching rule \eqref{eq:BRSU6n+3}
\begin{align}\label{eq:BRSU6n+3char}
&\chi_{\textbf{adj}}^{SU(6n+3)}\left(\vec{z}^{(1)},\vec{z}^{(2)},\vec{z}^{(3)},f_1,f_2\right)=2+\sum_{a=1}^3\chi_{\textbf{adj}}^{SU(2n+1)}\left(\vec{z}^{(a)}\right)\nn\\
&\qquad\qquad+f_2^{-1}\chi_{\textbf{fund}}^{SU(2n+1)}\left(z^{(1)}\right)\chi_{\overline{\textbf{fund}}}^{SU(2n+1)}\left(z^{(3)}\right)+f_2\chi_{\overline{\textbf{fund}}}^{SU(2n+1)}\left(z^{(1)}\right)\chi_{\textbf{fund}}^{SU(2n+1)}\left(z^{(3)}\right)\nn\\
&\qquad\qquad+f_1^{-1}\chi_{\overline{\textbf{fund}}}^{SU(2n+1)}\left(z^{(1)}\right)\chi_{\textbf{fund}}^{SU(2n+1)}\left(z^{(2)}\right)+f_1\chi_{\textbf{fund}}^{SU(2n+1)}\left(z^{(1)}\right)\chi_{\overline{\textbf{fund}}}^{SU(2n+1)}\left(z^{(2)}\right)\nn\\
&\qquad\qquad+\frac{f_1}{f_2}\chi_{\overline{\textbf{fund}}}^{SU(2n+1)}\left(z^{(2)}\right)\chi_{\textbf{fund}}^{SU(2n+1)}\left(z^{(3)}\right)+\frac{f_2}{f_1}\chi_{\textbf{fund}}^{SU(2n+1)}\left(z^{(2)}\right)\chi_{\overline{\textbf{fund}}}^{SU(2n+1)}\left(z^{(3)}\right)\,.
\end{align}

This means that we can study the limit of the index of the $\mathcal{N}=1$ theory in which $q=t=(p\,q)^{\frac{1}{3}}$, that is $p=q^2$. First of all, in this limit the contribution of the flipper fields becomes
\be
\left.\left(\mathcal{I}_{\text{chir}}^{\mathcal{N}=1}\left((p\,q)^{\frac{2}{3}};p,q\right)\right)^2\right|_{p=q^2}=\mathrm{PE}\left[-\frac{2q}{1-q^2}\right]
\ee
and so it simplifies against that of the first two singlets in \eqref{eq:BRSU6n+3char}, as expected since their purpose was to remove this decoupled free sector from the theory. Instead, the remaining contribution of the $D_2(SU(6n+3))$ theory in \eqref{eq:D2ind} becomes equivalent to the index of chiral multiplets in the representations appearing in \eqref{eq:BRSU6n+3char} except the singlets and with R-charge $\frac{2}{3}$. This is because the index of a chiral of R-charge $\frac{2}{3}$ in the limit $p=q^2$ is (remember \eqref{eq:indchir})
\be
\left.\mathcal{I}_{\text{chir}}^{\mathcal{N}=1}\left((p\,q)^{\frac{1}{3}};p,q\right)\right|_{p=q^2}=\mathrm{PE}\left[\frac{q}{1-q^2}\right]\,.
\ee
Hence, we end up precisely with the index of the $\mathcal{N}=2$ necklace quiver theory in the limit $p=q^2$
\begin{align}
\mathcal{I}_{\mathcal{N}=1}\left(f_1,f_2,p=q^2,q\right)&=\frac{1}{(2n+1)!^3}\oint\left[\prod_{a=1}^3\left(\prod_{i=1}^{2n+1}\frac{\mathrm{d}z_i^{(a)}}{2\pi iz_i^{(a)}}\right)\mathcal{I}_{\text{vec}}^{\mathcal{N}=1}\left(\vec{z}^{(a)};p=q^2,q\right)\right]\nn\\
&\times\frac{1}{\left(\mathcal{I}_{\text{chir}}^{\mathcal{N}=1}\left((p\,q)^{\frac{1}{3}};p=q^2,q\right)\right)^3}\prod_{a=1}^3\prod_{i,j=1}^{2n+1}\mathcal{I}_{\text{chir}}^{\mathcal{N}=1}\left((p\,q)^{\frac{1}{3}}\frac{z^{(a)}_i}{z^{(a)}_j};p=q^2,q\right)\nn\\
&\times\prod_{i,j=1}^{2n+1}\mathcal{I}_{\text{chir}}^{\mathcal{N}=1}\left((p\,q)^{\frac{1}{3}}f_2^{-1}\frac{z^{(1)}_i}{z^{(3)}_j};p=q^2,q\right)\mathcal{I}_{\text{chir}}^{\mathcal{N}=1}\left((p\,q)^{\frac{1}{3}}f_2\frac{z^{(3)}_i}{z^{(1)}_j};p=q^2,q\right)\nn\\
&\times\prod_{i,j=1}^{2n+1}\mathcal{I}_{\text{chir}}^{\mathcal{N}=1}\left((p\,q)^{\frac{1}{3}}f_1^{-1}\frac{z^{(2)}_i}{z^{(1)}_j};p=q^2,q\right)\mathcal{I}_{\text{chir}}^{\mathcal{N}=1}\left((p\,q)^{\frac{1}{3}}f_1\frac{z^{(1)}_i}{z^{(2)}_j};p=q^2,q\right)\nn\\
&\times\prod_{i,j=1}^{2n+1}\mathcal{I}_{\text{chir}}^{\mathcal{N}=1}\left((p\,q)^{\frac{1}{3}}\frac{f_1}{f_2}\frac{z^{(3)}_i}{z^{(2)}_j};p=q^2,q\right)\mathcal{I}_{\text{chir}}^{\mathcal{N}=1}\left((p\,q)^{\frac{1}{3}}\frac{f_2}{f_1}\frac{z^{(2)}_i}{z^{(3)}_j};p=q^2,q\right)\nn\\
&=\mathcal{I}_{\text{necklace}}(f_1,f_2,p=q^2,q)\,.
\end{align}

\section*{Acknowledgments}
We would like to thank Lea Bottini, Sara Pasquetti and Shlomo Razamat for useful discussions. MS is also grateful to the Scuola Internazionale Superiore di Studi Avanzati (SISSA) for ospitality when this project was initiated. MS is partially supported by the ERC Consolidator Grant \#864828 “Algebraic Foundations of Supersymmetric Quantum Field Theory (SCFTAlg)” and by the Simons Collaboration for the Nonperturbative Bootstrap under grant \#494786 from the Simons Foundation. Stephane Bajeot and Sergio Benvenuti are partially supported by the INFN Research Project GAST.

\appendix

\section{$4d$ $\mathcal{N}=1$ supersymmetric index conventions}
\label{app:index}

In this appendix we briefly summarize our conventions for the supersymmetric index of a $4d$ $\mathcal{N}=1$ theory, which coincides with the superconformal index \cite{Kinney:2005ej,Romelsberger:2005eg,Dolan:2008qi} when computed with the superconformal R-symmetry (see also \cite{Rastelli:2016tbz} for a review).  

The index of a $4d$ $\mathcal{N}=1$ SCFT is a refined Witten index of the theory quantized on $S^3\times {\mathbb R}$
\be
\mathcal{I}_{\mathcal{N}=1}=\mathrm{Tr}_{\delta=0}(-1)^F\left(\frac{p}{q}\right)^{j_1}(p\,q)^{j_2+\frac{R}{2}}\prod_if_i^{T_i}\,,
\ee
where 
\be
\delta=\frac{1}{2} \left\{\mathcal{Q},\mathcal{Q}^{\dagger}\right\}=2j_2+\frac{3}{2}R
\ee
with $\mathcal{Q}=\widetilde{\mathcal{Q}}_{\dot{-}}$ one of the Poincar\'e supercharges and $\mathcal{Q}^{\dagger}=\mathcal{S}$ the conjugate conformal supercharge, while $j_1$, $j_2$ are the Cartan generators of the isometry group $SO(4)=SU(2)_1 \times SU(2)_2$ of $S^3$, $R$ is the generator of the IR superconformal R-symmetry and $T_i$ are $\mathcal{Q}$-closed generators of additional global symmetries of the theory. The parameters $p$ and $q$ are fugacities associated with the supersymmetry preserving squashing of the $S^3$ \cite{Dolan:2008qi}, while $f_i$ are fugacities for the symmetries associated with the generators $T_i$. 

The index counts gauge invariant operators (``words") that can be constructed from modes of the fields (``letters"). The single letter indices for a vector multiplet and a chiral multiplet transforming in the representation ${\bf R}$ of the gauge and flavor group and with R-charge $R$ are
\begin{align}
i_{\text{vec}} \left(p,q,U\right) & =  \frac{2pq-p-q}{(1-p)(1-q)} \chi_{adj}\left(U\right), \nonumber\\
i_{\text{chir}}^{{\bf R}}\left(p,q,U,V,R\right) & =  
\frac{(pq)^{\frac{R}{2}} \chi_{{\bf R}} \left(U,V\right) - (pq)^{\frac{2-R}{2}} \chi_{\overline{{\bf R}}} \left(U,V\right)}{(1-p)(1-q)}\,,
\end{align}
where $\chi_{{\bf R}} \left(U,V\right)$ and $\chi_{\overline{{\bf R}}} \left(U,V\right)$ are the characters of the representation ${\bf R}$ and the conjugate representation $\overline{{\bf R}}$, with $U$ and $V$ gauge and flavor group matrices, respectively.

The index is obtained by symmetrizing of all of such letters into words and then projecting them to the gauge invariant ones by integrating over the Haar measure of the gauge group. This takes the general form
\be
\mathcal{I}_{\mathcal{N}=1} \left(p,q,V\right)=\int \left[\udl{U}\right] \prod_{k} \PE\left[i_k\left(p,q,U,V\right)\right]\,,
\ee
where $k$ labels the different multiplets in the theory and $\PE[i_k]$ is the plethystic exponential of the single letter index of the $k$-th multiplet
\be
\PE\left[i_k\left(p,q,U,V\right)\right] = \exp \left[  \sum_{m=1}^{\infty} \frac{1}{m} i_k\left(p^m,q^m,U^m,V^m\right) \right]\,,
\ee
which implements the symmetrization of the letters.

Let us consider the example of an $SU(n)$ gauge group. The contribution of a chiral superfield in the fundamental representation $\bf n$ or anti-fundamental representation $\overline{\bf n}$ of $SU(n)$ with $R$-charge $R$ can be written as follows:
\begin{align}
\PE\left[i_{\text{chir}}^{\bf n} \left(p, q,U\right)\right] &= \prod_{a=1}^{n} \Gpq{(pq)^{\frac{R}{2}}  z_a} \,, \quad 
\PE\left[i_{\text{chir}}^{\overline{\bf n}} \left(p, q,U\right)\right] = \prod_{a=1}^{n} \Gpq{(pq)^{\frac{R}{2}}  z^{-1}_a}  \,,\nn\\
\end{align}
where $z_a$ with $a=1,...,n$ and $\prod_{a=1}^{n} z_a=1$ are the fugacities parametrizing the Cartan subalgebra of $SU(n)$ and the elliptic gamma function is defined as
\be
\Gpq{x}\equiv \Gpq{x;p,q}  = \prod_{i,j=1}^\infty \frac{1-p^{i} q^{j} x^{-1}}{1- p^{i+1}  q^{j+1} x}\,.
\ee
We will also use the shorthand notation
\be
\Gpq{x^{\pm h}}=\Gpq{x^h}\Gpq{x^{-h}}\,.
\ee
On the other hand, the contribution of the vector multiplet in the adjoint representation of $SU(n)$ together with the $SU(n)$ Haar measure can be written as
\be
\frac{(p;p)_{\infty}^n(q;q)_{\infty}^n}{n!} \left.\oint_{\mathbb{T}^{n-1}} \prod_{a=1}^{n-1} \frac{\udl{z}_a}{2\pi i z_a} \prod^{n}_{a\neq b} \frac{1}{\Gpq{z_a z_b^{-1}}}\right|_{\prod_{a=1}^nz_a=1} \big(\cdots\big)\,,
\ee
where the $\big(\cdots\big)$ indicates the rest of the index which receives contribution from the chiral matter fields transforming in various representations of the gauge group. The integration contour is taken over a unitary circle in the complex plane for each element of the maximal torus of the gauge group. The prefactor $(p;p)_{\infty}^n(q;q)_{\infty}^n$ is the contribution of $n$ $U(1)$ free vector multiplets, one for each Cartan element of the gauge group $SU(n)$, where $(x;q) = \prod_{n=0}^\infty \left( 1-xq^n \right)$ is the $q$-Pochhammer symbol.  

For completeness, we report the parametrization we used for the characters of various representations of the groups $USp(2n)$ and $SO(n)$ that appeared in the main text. For $USp(2n)$ we mainly consider fundamental $\bf 2n$, adjoint $\bf n(2n+1)$ and (traceless) antisymmetric $\bf n(2n-1)-1$ representations
\begin{align}
\chi_{\bf 2n}^{USp(2n)}&=\sum_{a=1}^nz_a+z_a^{-1}\,,\nn\\
\chi_{\bf n(2n+1)}^{USp(2n)}&=n+\sum_{a=1}^n\left(z_a^2+z_a^{-2}\right)+\sum_{a<b}^n\left(z_az_b+z_az_b^{-1}+z_a^{-1}z_b+z_a^{-1}z_b^{-1}\right)\,,\nn\\
\chi_{\bf n(2n-1)-1}^{USp(2n)}&=n-1+\sum_{a<b}^n\left(z_az_b+z_az_b^{-1}+z_a^{-1}z_b+z_a^{-1}z_b^{-1}\right)\,.
\end{align}
For $SO(n)$ the characters of the representations are different depending on whether $n$ is even or odd, so it is useful to write $n=2k+\epsilon$ with $k$ the rank of the group and $\epsilon=0,1$. The main representations that we consider are the vector $\bf n$, the adjoint $\bf \frac{n(n-1)}{2}$ and the (traceless) symmetric representations $\bf \frac{n(n+1)-2}{2}$
\begin{align}
\chi_{\bf n}^{SO(n)}&=\epsilon+\sum_{a=1}^kz_a+z_a^{-1}\,,\nn\\
\chi_{\bf \frac{n(n-1)}{2}}^{SO(n)}&=k+\epsilon\sum_{a=1}^k\left(z_a+z_a^{-1}\right)+\sum_{a<b}^n\left(z_az_b+z_az_b^{-1}+z_a^{-1}z_b+z_a^{-1}z_b^{-1}\right)\,,\nn\\
\chi_{\bf \frac{n(n+1)-2}{2}}^{SO(n)}&=k+\epsilon-1+\epsilon\sum_{a=1}^k\left(z_a+z_a^{-1}\right)+\sum_{a=1}^k\left(z_a^2+z_a^{-2}\right)+\sum_{a<b}^k\left(z_az_b+z_az_b^{-1}+z_a^{-1}z_b+z_a^{-1}z_b^{-1}\right)\,.
\end{align}
Other useful characters are those for the symmetric $\bf \frac{n(n+1)}{2}$ and anti-symmetric $\bf \frac{n(n-1)}{2}$ rperesentations of $SU(n)$
\begin{align}
\chi_{\bf \frac{n(n+1)}{2}}^{SU(n)}&=\sum_{a\leq b}^nz_az_b\,,\nn\\
\chi_{\bf \frac{n(n-1)}{2}}^{SU(n)}&=\sum_{a< b}^nz_az_b\,,
\end{align}
where as usual for $SU(n)$ fugacities we have $\prod_{a=1}^nz_a=1$.

\section{Matching of 't Hooft anomalies}
\label{app:anomalies}
In this appendix we show the matching of the 't Hooft anomalies of the different $4d$ S-confining theories presented in Section \ref{4dSconfining}.

\subsection*{$\cU_1[n]$}
\label{app:anomU1}
The first thing we can compare on both sides is the central charges. Using the relation between the 't Hooft anomalies of the $U(1)_R$ symmetry and the $a$ and $c$ central charges \cite{Anselmi:1997am}
\begin{align}
a&=\frac{3}{32}\left(3\mathrm{Tr}\,R^3-\mathrm{Tr}\,R\right)=3 n^2+3 n+\frac{1}{8}\,,\nonumber\\
c&=\frac{1}{32}\left(9\mathrm{Tr}\,R^3-5\mathrm{Tr}\,R\right)=3 n^2+3 n+\frac{1}{4}\,,
\end{align}
we can determing the central charges of the gauge theory
\begin{align}
a_{\text{l.h.s.}} &= n (2 n + 1) a_0[2] + (n (2 n - 1) - 1) a_0\Bigl[\frac{4}{3}\Bigr] + (4 n + 2)(2 n) a_0\Bigl[\frac{1}{3}\Bigr] \\
&= \frac{1}{48} \left(8n^2 + 6n + 1\right) = a_0\Bigl[\frac{2}{3}\Bigr] \left(8n^2 + 6n + 1\right)\,, \label{aUSpLHS}
\end{align}
\begin{align}
c_{\text{l.h.s.}} &= n (2 n + 1) c_0[2] + (n (2 n - 1) - 1) c_0\Bigl[\frac{4}{3}\Bigr] + (4 n + 2)(2 n) c_0\Bigl[\frac{1}{3}\Bigr] \\
&= \frac{1}{24} \left(8n^2 + 6n + 1\right) = c_0\Bigl[\frac{2}{3}\Bigr] \left(8n^2 + 6n + 1\right)\,. \label{cUSpLHS}
\end{align}
The first term is the contribution of the gauginos, the second is the contribution of the antisymmetric traceless field $a$ and the last term comes from the fundamental $p$. The various contributions are written in terms of
\begin{align} 
a_0[x] &= \frac{3}{32} \left(3(x-1)^3 - (x-1)\right) \\
c_0[x] &= \frac{1}{32} \left(9(x-1)^3 - 5(x-1)\right)\,.
\end{align}

On the WZ side, the central charges are given by
\begin{align}
a_{\text{r.h.s.}} &= (2n+1) (4 n + 1) a_0\Bigl[\frac{2}{3}\Bigr]= a_0\Bigl[\frac{2}{3}\Bigr] \left(8n^2 + 6n + 1\right)\,, \label{aUSpRHS}
\end{align}
\begin{align}
c_{\text{r.h.s.}} &= (2n+1) (4 n + 1) c_0\Bigl[\frac{2}{3}\Bigr] = c_0\Bigl[\frac{2}{3}\Bigr] \left(8n^2 + 6n + 1\right)\,. \label{cUSpRHS}
\end{align}
In the first equality we have the contribution of the $USp(4n+2)$ antisymmetric $A$ with the trace part. We can indeed see the matching between \eqref{aUSpLHS}-\eqref{aUSpRHS} and \eqref{cUSpLHS}-\eqref{cUSpRHS}.

We can also consider the 't Hooft anomalies for the global symmetries
\begin{align}
\Tr\left(U(1)_R \, USp(4n+2)^2 \right)_{\text{l.h.s.}} &= \left(\frac{1}{3}-1\right) 2n\times \frac{1}{2} = -\frac{2n}{3} \,,\\
\Tr\left(U(1)_R \, USp(4n+2)^2 \right)_{\text{r.h.s.}} &= \left(\frac{2}{3}-1\right) 2n= -\frac{2n}{3}\,,
\end{align}
where we used that the Dynkin indices for the fundamental and for the antisymmetric representations of $USp(2N)$ are $\tfrac{1}{2}$ and $N-1$ respectively.

\subsection*{$\cO_1[n]$}
\label{app:anomS1}
Once again we can start by computing the central charges of the gauge theory
\begin{align}
a_{\text{l.h.s.}} &= \frac{1}{2} n (n - 1) a_0[2] + \frac{1}{2} n (n + 1) - 1) a_0\Bigl[\frac{4}{3}\Bigr] + n(2n-2) a_0\Bigl[\frac{1}{3}\Bigr] \\
&= \frac{1}{48} \left(2n^2 - 3n + 1\right) = a_0\Bigl[\frac{2}{3}\Bigr] \left(2n^2 - 3n + 1\right)\,, \label{aSOLHS}
\end{align}
\begin{align}
c_{\text{l.h.s.}} &= \frac{1}{2} n (n - 1) c_0[2] + \frac{1}{2} n (n + 1) - 1) c_0\Bigl[\frac{4}{3}\Bigr] + n(2n-2) c_0\Bigl[\frac{1}{3}\Bigr] \\
&= \frac{1}{24} \left(2n^2 - 3n + 1\right) = c_0\Bigl[\frac{2}{3}\Bigr] \left(2n^2 - 3n + 1\right)\,. \label{cSOLHS}
\end{align}
The first term is the contribution of the gauginos, the second is the contribution of the symmetric traceless field $s$ and the last term comes from the chiral $p$ in the vector representation.

On the WZ side, the central charges are given by
\begin{align}
a_{\text{r.h.s.}} &= (n-1) (2n - 1) a_0\Bigl[\frac{2}{3}\Bigr]= a_0\Bigl[\frac{2}{3}\Bigr] \left(2n^2 - 3n + 1\right)\,, \label{aSORHS}
\end{align}
\begin{align}
c_{\text{r.h.s.}} &= (n-1) (2n - 1) c_0\Bigl[\frac{2}{3}\Bigr] = c_0\Bigl[\frac{2}{3}\Bigr] \left(2n^2 - 3n + 1\right)\,. \label{cSORHS}
\end{align}
We can indeed see the matching between \eqref{aSOLHS}-\eqref{aSORHS} and \eqref{cSOLHS}-\eqref{cSORHS}.

We can also consider other 't Hooft anomalies
\begin{align}
\Tr\left(U(1)_R \, SO(2n-2)^2 \right)_{\text{l.h.s.}} &= \left(\frac{1}{3}-1\right) n \times 1 = -\frac{2n}{3} \\
\Tr\left(U(1)_R \, SO(2n-2)^2 \right)_{\text{r.h.s.}} &= \left(\frac{2}{3}-1\right) 2n = -\frac{2n}{3}\,,
\end{align}
where we used that the Dynkin indices for the vector and the symmetric representations of $SO(N)$ are 1 and $N+2$ respectively.

\subsection*{$\cU_2[n,h]$}
\label{app:anomU2}
The central charges of the gauge theory are
\begin{align}
a_{\text{l.h.s.}} &= n (2n + 1) a_0[2] + n (2n - 1)  a_0\Bigl[\frac{2}{3}\Bigr] + 2n(2n+8-4h) a_0\Bigl[\frac{2}{3}\Bigr] \\
&+ (2h)(2n) a_0\Bigl[\frac{h-n}{3h}\Bigr] + h(2h-1) a_0\Bigl[2-2\frac{h-n}{3h}\Bigr] + h(2h-1) a_0\Bigl[2-2\frac{h-n}{3h} -\frac{2}{3}\Bigr] \nn \\
&+ 2h (2n+8 -4h) a_0\Bigl[2-\frac{h-n}{3h} -\frac{2}{3}\Bigr] = 0\,, \label{aU2}
\end{align}
\begin{align}
c_{\text{l.h.s.}} &= n (2n + 1) c_0[2] + n (2n - 1) c_0\Bigl[\frac{2}{3}\Bigr] + 2n(2n+8-4h) c_0\Bigl[\frac{2}{3}\Bigr] \\
&+ (2h)(2n) c_0\Bigl[\frac{h-n}{3h}\Bigr] + h(2h-1) c_0\Bigl[2-2\frac{h-n}{3h}\Bigr] + h(2h-1) c_0\Bigl[2-2\frac{h-n}{3h} -\frac{2}{3}\Bigr] \nn \\
&+ 2h (2n+8 -4h) c_0\Bigl[2-\frac{h-n}{3h} -\frac{2}{3}\Bigr] = 0\,. \label{cU2}
\end{align}
We can see that they vanish, in accordance with our claim that the theory is trivial in the IR.

Also the other 't Hooft anomalies vanish
\begin{align}
\Tr\left(U(1)_R \, SU(2h)^2 \right) &= \left[2n \left(\frac{h-n}{3h}-1\right) + (2n + 8 - 4h) \left(2-\frac{2}{3}-\frac{h-n}{3h}-1\right) \right]\times\frac{1}{2} \nn \\
&+ \left[ \left(2-2\frac{h-n}{3h}-1\right) + \left(2-\frac{2}{3}-2\frac{h-n}{3h}-1\right) \right]  (h-1) = 0\,,
\end{align}
\begin{align}
\Tr\left(U(1)_R \, USp(2n+8-4h)^2 \right) &= \left[2n \left(\frac{2}{3}-1\right) \, + 2h \left(2-\frac{2}{3}-\frac{h-n}{3h}-1\right) \right] \times\frac{1}{2} = 0\,,
\end{align}
where we used that the Dynkin indices of the fundamental and of the antisymmetric representations of $SU(N)$ are $\frac{1}{2}$ and $\frac{N-2}{2}$, respectively.

\subsection*{$\cS_2[n,h]$}
\label{app:anomS2}
The central charges of the gauge theory are
\begin{align}
a_{\text{l.h.s.}} &= \frac{1}{2} n (n - 1) a_0[2] + \frac{1}{2} n(n+1) a_0\Bigl[\frac{2}{3}\Bigr] + n(n-8-2h) a_0\Bigl[\frac{2}{3}\Bigr] \\
&+ h\,n a_0\Bigl[\frac{h-n}{3h}\Bigr] + \frac{1}{2} h (h+1) a_0\Bigl[2-2\frac{h-n}{3h}\Bigr] + \frac{1}{2} h (h+1) a_0\Bigl[2-2\frac{h-n}{3h} -\frac{2}{3}\Bigr] \nn \\
&+ h (n-8 -2h) a_0\Bigl[2-\frac{h-n}{3h} -\frac{2}{3}\Bigr] = 0\,, \label{aS2}
\end{align}
\begin{align}
c_{\text{l.h.s.}} &= \frac{1}{2} n (n - 1) c_0[2] + \frac{1}{2} n(n+1) c_0\Bigl[\frac{2}{3}\Bigr] + n(n-8-2h) c_0\Bigl[\frac{2}{3}\Bigr] \\
&+ h\,n c_0\Bigl[\frac{h-n}{3h}\Bigr] + \frac{1}{2} h (h+1) c_0\Bigl[2-2\frac{h-n}{3h}\Bigr] + \frac{1}{2} h (h+1) c_0\Bigl[2-2\frac{h-n}{3h} -\frac{2}{3}\Bigr] \nn \\
&+ h (n-8 -2h) c_0\Bigl[2-\frac{h-n}{3h} -\frac{2}{3}\Bigr] = 0\,. \label{aS2}
\end{align}
The other 't Hooft anomalies are
\begin{align}
\Tr\left(U(1)_R \, SU(h)^2 \right) &= \left[n \left(\frac{h-n}{3h}-1\right) + (n - 8 - 2h) \left(2-\frac{2}{3}-\frac{h-n}{3h}-1\right) \right]\times\frac{1}{2} \nn \\
&+ \left[ \left(2-2\frac{h-n}{3h}-1\right) + \left(2-\frac{2}{3}-2\frac{h-n}{3h}-1\right) \right] \frac{h+2}{2}  = 0\,,
\end{align}
\begin{align}
\Tr\left(U(1)_R \, (SO(n+8-2h))^2 \right) &= \left[n \left(\frac{2}{3}-1\right) \, + h \left(2-\frac{2}{3}-\frac{h-n}{3h}-1\right) \right] \times 1 = 0\,,
\end{align}
where we used that the Dynkin index for the symmetric representation of $SU(N)$ is $\frac{N+2}{2}$.

\subsection*{$\cA_2[n,h]$}
\label{app:anomA2}
The central charges are
\begin{align}
a_{\text{l.h.s.}} &= (n^2 - 1) a_0[2] + n^2 a_0\Bigl[\frac{2}{3}\Bigr] + 2n(n-2h) a_0\Bigl[\frac{2}{3}\Bigr] \\
&+ 2h\,n a_0\Bigl[\frac{h-n}{3h}\Bigr] + h^2 a_0\Bigl[2-2\frac{h-n}{3h}\Bigr] + h^2 a_0\Bigl[2-2\frac{h-n}{3h} -\frac{2}{3}\Bigr] \nn \\
&+ 2h (n-2h) a_0\Bigl[2-\frac{h-n}{3h} -\frac{2}{3}\Bigr] \nn \\
&= a_0[0] = a_{\text{r.h.s.}}\,, \label{aA2}
\end{align}
\begin{align}
c_{\text{l.h.s.}} &= (n^2 - 1) c_0[2] + n^2 c_0\Bigl[\frac{2}{3}\Bigr] + 2n(n-2h) c_0\Bigl[\frac{2}{3}\Bigr] \\
&+ 2h\,n c_0\Bigl[\frac{h-n}{3h}\Bigr] + h^2 c_0\Bigl[2-2\frac{h-n}{3h}\Bigr] + h^2 c_0\Bigl[2-2\frac{h-n}{3h} -\frac{2}{3}\Bigr] \nn \\
&+ 2h (n-2h) c_0\Bigl[2-\frac{h-n}{3h} -\frac{2}{3}\Bigr] = c_0[0] = c_{\text{r.h.s.}}\,. \label{cA2}
\end{align}
Some other 't Hooft anomalies are
\begin{align}
&\Tr\left(U(1)_R \, SU(h)_p^2 \right) = \Tr\left(U(1)_R \, SU(h)_{\pt}^2 \right) = n \left(\frac{h-n}{3h}-1\right) \nn \\
&+h \left(2-2\frac{h-n}{3h}-1\right) +h \left(2-2\frac{h-n}{3h} \frac{2}{3}-1\right) + (n- 2h) \left(2-\frac{2}{3}-\frac{h-n}{3h}-1\right) = 0\,,
\end{align}
\begin{align}
\Tr\left(U(1)_R \, SU(n-2h)^2 \right) &= \left[2n \left(\frac{2}{3}-1\right) \, + 2h \left(2-\frac{2}{3}-\frac{h-n}{3h}-1\right) \right] \frac{1}{2} \nn \\
& = 0
\end{align}
Everything is compatible with the claim that in the IR the theory is given by 2 singlets of R-charge 0 and one of R-charge 2 which are uncharged under the non-abelian symmetries.

\section{Deconfinement derivation for $\cU_1[2]$} \label{ProofU1n2}
In this appendix we derive the S-confinement result $\cU_1[n=2]$. The first step is the same as in the general $n$ case of Subsection \ref{subsec:U1deriv}. We have to split the $10$ fundamental into $8+1+1$. We obtain the following quiver: 
\be \label{U1n2a} \scalebox{0.85}{\bpic[node distance=2cm,gSUnode/.style={circle,red,draw,minimum size=8mm},gUSpnode/.style={circle,blue,draw,minimum size=8mm},gSOnode/.style={circle,ForestGreen,draw,minimum size=8mm},fnode/.style={rectangle,draw,minimum size=8mm}]
\begin{scope}[shift={(-3,0)}]    
\node at (-2,1) {$\cU_1[2]:$};
\node[gUSpnode] (G1) at (0,0) {$4$};
\node[fnode, blue] (F1) at (2.5,0) {$10$};
\draw (G1) -- (F1) node[midway,above] {$p$};
\draw (0.3,0.3) to[out=90,in=0]  (0,0.7) to[out=180,in=90] (-0.3,0.3);
\node[right] at (0,-1.3) {$ \cW= a p p$};
\node at (0.6,0.9) {$a$, \textcolor{Green}{$\frac{4}{3}$}};
\node at (0.9,-0.4) {\textcolor{Green}{$\frac{1}{3}$}};
\node at (5,0) {$\equiv$};
\end{scope}
\begin{scope}[shift={(4.5,0)}]
\node[gUSpnode] (G2) at (0,0) {$4$};
\node[fnode, blue] (F2) at (2.5,0) {$8$};
\node[fnode, violet] (F3) at (1,-1.8) {$1$};
\node[fnode, red] (F4) at (-1,-1.8) {$1$};
\draw (G2) -- (F2);
\draw (G2) -- (F3);
\draw (G2) -- (F4);
\draw (0.3,0.3) to[out=90,in=0]  (0,0.7) to[out=180,in=90] (-0.3,0.3);
\node[right] at (-0.8,-3) {$ \cW= a q q + a p_1 p_2$};
\node at (0.6,0.9) {$a$, \textcolor{Green}{$\frac{4}{3}$}};
\node at (1.3,0.4) {$q$, \textcolor{Green}{$\frac{1}{3}$}};
\node at (1.1,-0.7) {$p_2$, \textcolor{Green}{$\frac{1}{3}$}};
\node at (-1.1,-0.7) {\textcolor{Green}{$\frac{1}{3}$}, $p_1$};
\end{scope}
\epic} \ee 
As in the general $n$ case, we deconfine the antisymmetric traceless field $a$ so to get
\be \label{U1n2b} \scalebox{0.85}{\bpic[node distance=2cm,gSUnode/.style={circle,red,draw,minimum size=8mm},gUSpnode/.style={circle,blue,draw,minimum size=8mm},gSOnode/.style={circle,ForestGreen,draw,minimum size=8mm},fnode/.style={rectangle,draw,minimum size=8mm}]
\begin{scope}[shift={(0,0)}]
\node[gUSpnode] (G1) at (-2,0) {$2$};
\node[gUSpnode] (G2) at (0,0) {$4$};
\node[fnode, blue] (F1) at (2.5,0) {$8$};
\node[fnode,red] (F2) at (-2,-1.8) {$1$};
\node[fnode] (F3) at (0,-1.8) {$1$};
\node[fnode,violet] (F4) at (1.6,-1.8) {$1$};
\draw (G1) -- (G2);
\draw (G2) -- (F1);
\draw (G1) -- (F2);
\draw (G1) -- (F3);
\draw (G2) -- (F3);
\draw (G2) -- (F4);
\node at (-1,0.4) {$b$, \textcolor{Green}{$\frac{2}{3}$}};
\node at (1.3,0.4) {$q$, \textcolor{Green}{$\frac{1}{3}$}};
\node at (1.4,-0.7) {$p_2$, \textcolor{Green}{$\frac{1}{3}$}};
\node at (-2.7,-0.8) {\textcolor{Green}{$-\frac{1}{3}$}, $v$};
\node at (-1.3,-0.8) {\textcolor{Green}{$-\frac{1}{3}$}, $d$};
\node at (0,-0.8) {\textcolor{Green}{$\frac{5}{3}$}, $e$};
\node at (0,-3) {$ \cW= bq qb + p_2 e + b d e + \Flip[v d; b b]$};
\end{scope}
\epic} \ee 

The mass term in the superpotential $p_2 e$ after the deconfinement is what makes the case $n=2$ specific. Indeed in the higher $n$ case \eqref{U1nb}, it is no longer a mass term for the fields $e$ and $p_2$. We can then integrate out the fields $p_2$ and $e$ to get:
\be \label{U1n2c} \scalebox{0.85}{\bpic[node distance=2cm,gSUnode/.style={circle,red,draw,minimum size=8mm},gUSpnode/.style={circle,blue,draw,minimum size=8mm},gSOnode/.style={circle,ForestGreen,draw,minimum size=8mm},fnode/.style={rectangle,draw,minimum size=8mm}]
\begin{scope}[shift={(-3,0)}]    
\node[gUSpnode] (G1) at (-2,0) {$2$};
\node[gUSpnode] (G2) at (0,0) {$4$};
\node[fnode, blue] (F1) at (2.5,0) {$8$};
\node[fnode, red] (F2) at (-2,-1.8) {$1$};
\node[fnode] (F3) at (0,-1.8) {$1$};
\draw (G1) -- (G2);
\draw (G2) -- (F1);
\draw (G1) -- (F2);
\draw (G1) -- (F3);
\node at (-1,0.4) {$b$, \textcolor{Green}{$\frac{2}{3}$}};
\node at (1.3,0.4) {$q$, \textcolor{Green}{$\frac{1}{3}$}};
\node at (-2.7,-0.8) {\textcolor{Green}{$-\frac{1}{3}$}, $v$};
\node at (-0.4,-0.8) {\textcolor{Green}{$-\frac{1}{3}$}, $d$};
\node at (0,-3) {$ \cW= bqqb + \Flip[v d; b b]$};
\node at (4,0) {$\equiv$};
\end{scope}
\begin{scope}[shift={(6,0)}]
\node[gUSpnode] (G3) at (-2,0) {$2$};
\node[gUSpnode] (G4) at (-0.3,0) {$4$};
\node[fnode, blue] (F4) at (1.4,0) {$8$};
\node[fnode] (F5) at (-3.7,0) {$2$};
\draw (G3) -- (G4);
\draw (G4) -- (F4);
\draw (G3) -- (F5);
\node at (-1.1,0.4) {$b$, \textcolor{Green}{$\frac{2}{3}$}};
\node at (0.6,0.4) {$q$, \textcolor{Green}{$\frac{1}{3}$}};
\node at (-2.8,0.4) {$f$, \textcolor{Green}{$-\frac{1}{3}$}};
\node at (-1.2,-1.1) {$ \cW= bq qb +  \Flip[ff; b b] $};
\end{scope}
\epic} \ee 

Then we dualize the $USp(4)$ gauge node using the IP duality \cite{Intriligator:1995ne}. Due to the term $bq qb$ in the superpotential, we don't produce a link between the $USp(2) \equiv SU(2)$ gauge node and $USp(8)$ flavor node, since it becomes a mass term for such field. We obtain the following quiver after integrating out the massive field:
\be \label{U1n2d} \scalebox{0.85}{\bpic[node distance=2cm,gSUnode/.style={circle,red,draw,minimum size=8mm},gUSpnode/.style={circle,blue,draw,minimum size=8mm},gSOnode/.style={circle,ForestGreen,draw,minimum size=8mm},fnode/.style={rectangle,draw,minimum size=8mm}]
\begin{scope}[shift={(0,0)}]
\node[gUSpnode] (G3) at (-2,0) {$2$};
\node[gUSpnode] (G4) at (-0.3,0) {$2$};
\node[fnode, blue] (F4) at (1.4,0) {$8$};
\node[fnode] (F5) at (-3.7,0) {$2$};
\draw (G3) -- (G4);
\draw (G4) -- pic[pos=0.7,sloped]{arrow} (F4);
\draw (G3) -- (F5);
\node at (-1.1,0.4) {$B$, \textcolor{Green}{$\frac{1}{3}$}};
\node at (0.6,0.4) {$Q$, \textcolor{Green}{$\frac{2}{3}$}};
\node at (-2.8,0.4) {$f$, \textcolor{Green}{$-\frac{1}{3}$}};
\node at (-1.2,-1.1) {$ \cW= BQQB + \Flip[QQ; ff] $};
\end{scope}
\epic} \ee

We now see that the $USp(2)$ on the left is connected to $4$ fundamental chirals. This situation is referred as the quantum deformed moduli space (QDMS) \cite{Seiberg:1994bz,Seiberg:1994pq}. It triggers an Higgsing that leads to a complete breaking of the the two $USp(2)$ gauge nodes. The low energy d.o.f.'s consist of a traceful antisymmetric field $A_1$ of the $USp(8)$ flavor symmetry, a bifundamental field $P$ between the $SU(2)$ and $USp(8)$ and a singlet $\eta$. The R-charges of all the fields are $\frac{2}{3}$. The WZ theory that we obtain is the following:
\be \label{U1n2e} \scalebox{0.85}{\bpic[node distance=2cm,gSUnode/.style={circle,red,draw,minimum size=8mm},gUSpnode/.style={circle,blue,draw,minimum size=8mm},gSOnode/.style={circle,ForestGreen,draw,minimum size=8mm},fnode/.style={rectangle,draw,minimum size=8mm}]
\begin{scope}[shift={(-3,0)}]    
\node[fnode] (F1) at (0,0) {$2$};
\node[fnode, blue] (F2) at (2,0) {$8$};
\draw (F1) -- (F2) node[midway] {$\times$};
\draw (2.3,0.4) to[out=90,in=0]  (2,0.8) to[out=180,in=90] (1.7,0.4);
\node at (0,-1.3) {$ \cW= A_1 P P + A_1^3 + \tr(A_1) A_1^2 + \eta P P$};
\node at (2.8,0.9) {$A_1$, \textcolor{Green}{$\frac{2}{3}$}};
\node at (1,-0.4) {$P$, \textcolor{Green}{$\frac{2}{3}$}};
\node at (5,0) {$\equiv$};
\end{scope}
\begin{scope}[shift={(5,0)}]
\node[fnode, blue] (F3) at (0,0) {$10$};
\draw (0.3,0.4) to[out=90,in=0]  (0,0.8) to[out=180,in=90] (-0.3,0.4);
\node[right] at (-1.8,-1.3) {$ \cW= A^3 + \tr(A) A^2$};
\node at (0.7,0.9) {$A$, \textcolor{Green}{$\frac{2}{3}$}};
\end{scope}
\epic} \ee 
The mapping between \eqref{U1n2d} and \eqref{U1n2e} is
\be \label{mapU1n2} 
\ba{c}
\Flipper[Q \, Q] \\
f \, B \, Q \\
B \, B
\ea
\quad \longleftrightarrow \quad
\ba{c}
A_1 \\
P \\
\eta 
\ea
\ee 

On the left of \eqref{U1n2e}, we have written the most general superpotential compatible with the R-charges. We also combined all the fields on the l.h.s.~so to form a traceful antisymmetric of $USp(10)$. The r.h.s.~of \eqref{U1n2e} is precisely the desired result.

\section{Kutasov--Schwimmer-like dualities}

In this appendix we recall the different dualities for gauge theories with classical gauge group, matter in rank-2 and fundamental representations and a non-trivial superpotential involving some power of the rank-2 matter. The first of these dualities is for an $SU(n)$ gauge theory with a field in the adjoint representation and is due to Kutasov and Schwimmer \cite{Kutasov:1995ve,Kutasov:1995np}. Later Intriligator proposed the variants for symplectic and special orthogonal gauge groups \cite{Intriligator:1995ff}. 

\subsection*{$SU(n)$ case with adjoint} \label{SUKutaosv}
We present the duality for the $SU(n)$ gauge theory with a field $X$ in the adjoint representation\footnote{The adjoint field can be both taken to be traceful or traceless. On the dual the adjoint would be correspondingly traceful or traceless. \label{commentTrace}}, $F$ flavors and a non-trivial superpotential $\cW = X^{k+1}$. This duality appeared in \cite{Kutasov:1995ve,Kutasov:1995np}. The quiver summarizing this duality is the following:
\be \label{KutasovSUadjoint} \scalebox{0.85}{\bpic[node distance=2cm,gSUnode/.style={circle,red,draw,minimum size=8mm},gUSpnode/.style={circle,blue,draw,minimum size=8mm},gSOnode/.style={circle,ForestGreen,draw,minimum size=8mm},fnode/.style={rectangle,draw,minimum size=8mm}]
\begin{scope}[shift={(-3,0)}]    
\node[gSUnode] (G1) at (0,0) {$n$};
\node[fnode] (F1) at (1,-1.8) {$F$};
\node[fnode] (F2) at (-1,-1.8) {$F$};
\draw (G1) -- pic[pos=0.7,sloped]{arrow} (F1);
\draw (G1) -- pic[pos=0.4,sloped]{arrow} (F2);
\draw (0.3,0.3) to[out=90,in=0]  (0,0.7) to[out=180,in=90] (-0.3,0.3);
\node at (0,-2.8) {$ \cW= X^{k+1}$};
\node at (0.9,0.9) {$X$, \textcolor{Green}{$\frac{2}{k+1}$}};
\node at (1.9,-0.8) {$\Qt$, \textcolor{Green}{$1-\frac{2}{k+1}\frac{n}{F}$}};
\node at (-0.9,-0.8) {$Q$};
\node at (5,0) {$\Llra$};
\end{scope}
\begin{scope}[shift={(4.5,0)}]
\node[gSUnode] (G2) at (0,0) {\scalebox{0.85}{$kF-n$}};
\node[fnode] (F3) at (1,-1.8) {$F$};
\node[fnode] (F4) at (-1,-1.8) {$F$};
\draw (G2) -- pic[pos=0.6,sloped, very thick]{arrow=latex reversed} (F3);
\draw (G2) -- pic[pos=0.4,sloped, very thick]{arrow=latex reversed} (F4);
\draw (0.3,0.6) to[out=90,in=0]  (0,1) to[out=180,in=90] (-0.3,0.6);
\node at (0,-2.8) {$ \cW= Y^{k+1} + \Flip[\sum_{j=1}^{k} \qt \, Y^{k-j} \, q]$};
\node at (0.8,1.2) {$Y$, \textcolor{Green}{$\frac{2}{k+1}$}};
\node at (-0.9,-0.8) {$\qt$};
\node at (2.1,-0.8) {$q$, \textcolor{Green}{$1-\frac{2}{k+1}\frac{kF-n}{F}$}};
\end{scope}
\epic} \ee 
The mapping of the chiral ring generators is the following:
\be \label{mapKutasovSUadjoint} 
\ba{c}
X^j \\
\Qt X^{l-1} Q
\ea
\quad \longleftrightarrow \quad
\ba{c}
Y^j \\
\Flipper[\qt Y^{k-l+1} q]
\ea
\quad
\ba{c}
j=2,\dots,k \\
l=1,\dots,k
\ea
\ee

\subsection*{$USp(2n)$ case with antisymmetric} \label{UspKutaosv}
The duality for the $USp(2n)$ case with an antisymmetric field $A$,\footnote{The same comment as in the footnote \ref{commentTrace} applies for the antisymmetric field.} $2F$ fundamentals and a non-trivial superpotential $\cW = A^{k+1}$ appeared in \cite{Intriligator:1995ff}. The duality is given by the following quiver:
\be \label{KutasovUSpAntisym} \scalebox{0.85}{\bpic[node distance=2cm,gSUnode/.style={circle,red,draw,minimum size=8mm},gUSpnode/.style={circle,blue,draw,minimum size=8mm},gSOnode/.style={circle,ForestGreen,draw,minimum size=8mm},fnode/.style={rectangle,draw,minimum size=8mm}]
\begin{scope}[shift={(-3,0)}]    
\node[gUSpnode] (G1) at (0,0) {$2n$};
\node[fnode] (F1) at (2,0) {$2F$};
\draw (G1) -- pic[pos=0.7,sloped]{arrow} (F1);
\draw (0.3,0.3) to[out=90,in=0]  (0,0.7) to[out=180,in=90] (-0.3,0.3);
\node at (0,-2.2) {$ \cW= A^{k+1}$};
\node at (0.8,0.9) {$A$, \textcolor{Green}{$\frac{2}{k+1}$}};
\node at (1,-0.4) {$Q$};
\node at (1,-1.1) {\textcolor{Green}{$1-\frac{2(n+k)}{(k+1)F}$}};
\node at (5,0) {$\Llra$};
\end{scope}
\begin{scope}[shift={(4.5,0)}]
\node[gUSpnode] (G2) at (0,0) {\scalebox{0.8}{$2k(F-2)-2n$}};
\node[fnode] (F3) at (2.8,0) {$2F$};
\draw (G2) -- pic[pos=0.7,sloped, very thick]{arrow=latex reversed} (F3);
\draw (0.5,1.1) to[out=90,in=0]  (0,1.7) to[out=180,in=90] (-0.5,1.1);
\node at (0,-2.2) {$ \cW= B^{k+1} + \Flip[\sum_{j=1}^{k} q \, B^{k-j} \, q] $};
\node at (1,1.8) {$B$, \textcolor{Green}{$\frac{2}{k+1}$}};
\node at (1.7,-0.4) {$q$};
\node at (2.2,-1.1) {\textcolor{Green}{$1-\frac{2(k(F-1)-n)}{(k+1)F}$}};
\end{scope}
\epic} \ee
The mapping of the chiral ring generators is the following:
\be \label{mapKutasovUSpAntisym} 
\ba{c}
A^j \\
Q A^{l-1} Q
\ea
\quad \longleftrightarrow \quad
\ba{c}
B^j \\
\Flipper[q B^{k-l+1} q]
\ea
\quad
\ba{c}
j=2,\dots,k \\
l=1,\dots,k
\ea
\ee

\subsection*{$SO(n)$ case with symmetric} \label{SOKutaosv}
The $SO(n)$ case with a field in the symmetric representation,\footnote{The same comment as in the footnote \ref{commentTrace} applies for the symmetric field.} $2F$ chirals in the vector representation and a non-trivial superpotential $ \cW= S^{k+1}$ appeared in \cite{Intriligator:1995ff}. The duality is given by the following quiver:
\be \label{KutasovSOSym} \scalebox{0.85}{\bpic[node distance=2cm,gSUnode/.style={circle,red,draw,minimum size=8mm},gUSpnode/.style={circle,blue,draw,minimum size=8mm},gSOnode/.style={circle,ForestGreen,draw,minimum size=8mm},fnode/.style={rectangle,draw,minimum size=8mm}]
\begin{scope}[shift={(-3,0)}]    
\node[gSOnode] (G1) at (0,0) {$2n$};
\node[fnode] (F1) at (2,0) {$2F$};
\draw (G1) -- pic[pos=0.7,sloped]{arrow} (F1);
\draw (0.3,0.3) to[out=90,in=0]  (0,0.7) to[out=180,in=90] (-0.3,0.3);
\node at (0,-2.2) {$ \cW= S^{k+1}$};
\node at (0.8,0.9) {$S$, \textcolor{Green}{$\frac{2}{k+1}$}};
\node at (1,-0.4) {$Q$};
\node at (1,-1.1) {\textcolor{Green}{$1-\frac{2(n-2k)}{(k+1)F}$}};
\node at (5,0) {$\Llra$};
\end{scope}
\begin{scope}[shift={(4.5,0)}]
\node[gSOnode] (G2) at (0,0) {\scalebox{0.8}{$2k(F-2)-2n$}};
\node[fnode] (F3) at (2.5,0) {$2F$};
\draw (G2) -- pic[pos=0.7,sloped, very thick]{arrow=latex reversed} (F3);
\draw (0.5,1.1) to[out=90,in=0]  (0,1.7) to[out=180,in=90] (-0.5,1.1);
\node at (0,-2.2) {$ \cW= T^{k+1} + \Flip[\sum_{j=1}^{k} q \, T^{k-j} \, q] $};
\node at (1,1.8) {$T$, \textcolor{Green}{$\frac{2}{k+1}$}};
\node at (1.7,-0.4) {$q$};
\node at (2.2,-1.1) {\textcolor{Green}{$1-\frac{2(k(F+2)-n)}{(k+1)F}$}};
\end{scope}
\epic} \ee
The mapping of the chiral ring generators is the following:
\be \label{mapKutasovSOSym} 
\ba{c}
S^j \\
Q S^{l-1} Q
\ea
\quad \longleftrightarrow \quad
\ba{c}
T^j \\
\Flipper[q T^{k-l+1} q]
\ea
\quad
\ba{c}
j=2,\dots,k \\
l=1,\dots,k
\ea
\ee

\section{$\cN=1$ conformal manifolds of $\cN=2$ necklace quivers}
\label{CMnecklace}
We consider the $\mathbb{Z}_k$ orbifold of $\cN=4$ $SU(N)$ SYM. The resulting theory is a $SU(N)^k$ quiver with a cubic superpotential consisting of $2k$ terms
\be 
\cW = \sum_{i=1}^k \lambda_i \mathrm{Tr}(\Phi_i Q_i \Qt_i - \Phi_i Q_{i+1} \Qt_{i+1} ) \,,
\ee
where $\Phi_i$ is the adjoint of each gauge group and $Q_i, \Qt_i$ are bifundamental hypermultiplets, with $\{Q_{k+1}, \Qt_{k+1}\} = \{Q_1, \Qt_1\}$. Each elementary chiral multiplet has the $\mathcal{N}=1$ R-charge $R=\frac{2}{3}$ on the whole conformal manifold, which includes the free theory.

The $\cN=2$ conformal manifold is $k$-complex dimensional and is parameterized by the $k$ gauge couplings. The precise $2k$ superpotential couplings are set by $\cN=2$ supersymmetry in terms of the gauge couplings $\tau_i$, which are related by supersymmetry to the couplings $\lambda_i$ in the $\mathcal{N}=1$ superpotential.

We now want to determine the $\cN=1$ conformal manifold, which contains the  $\cN=2$ conformal manifold as a submanifold. We use the method introduced in \cite{Kol:2002zt}, of quotienting the space of marginal deformations (that is the chiral ring operators with $R_0=2$) by the broken global symmetries.
If $k>3$, there are $k+1$ chiral ring operators with $R_0=2$ (if $k\leq 3$ there are additional operators which we discuss below)
\begin{align}
\sum_{i=1}^k \mathrm{Tr}\left(\Phi_i Q_i \Qt_i + \Phi_i Q_{i+1} \Qt_{i+1}\right)\,\qquad \mathrm{Tr}(\Phi_i^3)\,,\quad i=1,\ldots,k\,.
\end{align}
 
The global symmetry, in $\cN=1$ language, on the $\cN=2$ conformal submanifold is 
\be 
U(1)_F \times U(1)^k_{baryonic} \times U(1)_{R_0}\,, 
\ee
where $U(1)_F$ acts with charges $+2,-1,-1$ on $\Phi_i, Q_i, \Qt_i$, respectively, the $i^{th}$ baryonic $U(1)$ acts with charge $\pm 1$ on $Q_i, \Qt_i$ leaving the remaining fields uncharged, and $U(1)_{R_0}$ assigns the canonical R-charge $R_0=\frac{2}{3}$ to all the chirals. We conclude that the barionic $U(1)$'s are not broken by the marginal operators, while the $U(1)_F$ is broken. 
 
Hence, there are $k+1-1=k$ additional $\cN=1$ directions on the conformal manifold. The full $\cN=1$ conformal manifold is $2k$-dimensional for $k>3$.
One deformation is the $\beta$-deformation, which exists for any SCFT on D3-branes at toric Calabi--Yau singularities \cite{Benvenuti:2005wi}, while the other operators are linear combinations of the $\mathrm{Tr}(\Phi_i^3)$ operators.

\subsubsection*{Special cases with $k\leq 3$}

Let us briefly comment on the cases with $k\leq 3$, where additional direction in the $\mathcal{N}=1$ conformal manifold arise.
\begin{itemize}
\item If $k=1$ the theory is $\cN=4$ SYM and the $\cN=1$ conformal manifold is $3$-dimensional, as we briefly reviewed in the previous subsection.

\item If $k=2$ there are $2$ additional marginal operators in the chiral ring
\be
\mathrm{Tr}(\Phi_1 Q_1 \Qt_2) \sim \mathrm{Tr}(\Phi_2 Q_2 \Qt_1)\,,\qquad \mathrm{Tr}(\Phi_1 Q_2 \Qt_1) \sim \mathrm{Tr}(\Phi_2 Q_1 \Qt_2)\,,
\ee
where the equivalences are due to the F-term relations. These operators break one of the $2$ baryonic symmetries, hence they provide one additional direction in the conformal manifold, which is $5$-dimensional.

\item If $k=3$ the operators 
\be
\mathrm{Tr}(Q_1 Q_2 Q_3)\,,\qquad \mathrm{Tr}(\Qt_1 \Qt_2 \Qt_3)\,,
\ee
are chiral ring operators with R-charge $R_0=2$. These operators break one of the $3$ baryonic symmetries, hence they provide one additional direction in the conformal manifold, which is $7$-dimensional.
\end{itemize}

\bibliographystyle{ytphys}
\bibliography{refs}

\providecommand{\href}[2]{#2}\begingroup\raggedright\begin{thebibliography}{10}

\bibitem{Gadde:2015xta}
A.~Gadde, S.~S. Razamat, and B.~Willett, ``{''Lagrangian'' for a Non-Lagrangian
  Field Theory with $\mathcal N=2$ Supersymmetry},''
  \href{http://dx.doi.org/10.1103/PhysRevLett.115.171604}{{\em Phys. Rev.
  Lett.} {\bfseries 115} no.~17, (2015) 171604},
  \href{http://arxiv.org/abs/1505.05834}{{\ttfamily arXiv:1505.05834
  [hep-th]}}.

\bibitem{Maruyoshi:2016tqk}
K.~Maruyoshi and J.~Song, ``{Enhancement of Supersymmetry via Renormalization
  Group Flow and the Superconformal Index},''
  \href{http://dx.doi.org/10.1103/PhysRevLett.118.151602}{{\em Phys. Rev.
  Lett.} {\bfseries 118} no.~15, (2017) 151602},
  \href{http://arxiv.org/abs/1606.05632}{{\ttfamily arXiv:1606.05632
  [hep-th]}}.

\bibitem{Maruyoshi:2016aim}
K.~Maruyoshi and J.~Song, ``{$ \mathcal{N}=1 $ deformations and RG flows of $
  \mathcal{N}=2 $ SCFTs},''
  \href{http://dx.doi.org/10.1007/JHEP02(2017)075}{{\em JHEP} {\bfseries 02}
  (2017) 075}, \href{http://arxiv.org/abs/1607.04281}{{\ttfamily
  arXiv:1607.04281 [hep-th]}}.

\bibitem{Agarwal:2016pjo}
P.~Agarwal, K.~Maruyoshi, and J.~Song, ``{$ \mathcal{N} $ =1 Deformations and
  RG flows of $ \mathcal{N} $ =2 SCFTs, part II: non-principal deformations},''
  \href{http://dx.doi.org/10.1007/JHEP12(2016)103}{{\em JHEP} {\bfseries 12}
  (2016) 103}, \href{http://arxiv.org/abs/1610.05311}{{\ttfamily
  arXiv:1610.05311 [hep-th]}}. [Addendum: JHEP 04, 113 (2017)].

\bibitem{Benvenuti:2017lle}
S.~Benvenuti and S.~Giacomelli, ``{Supersymmetric gauge theories with decoupled
  operators and chiral ring stability},''
  \href{http://dx.doi.org/10.1103/PhysRevLett.119.251601}{{\em Phys. Rev.
  Lett.} {\bfseries 119} no.~25, (2017) 251601},
  \href{http://arxiv.org/abs/1706.02225}{{\ttfamily arXiv:1706.02225
  [hep-th]}}.

\bibitem{Benvenuti:2017kud}
S.~Benvenuti and S.~Giacomelli, ``{Abelianization and sequential confinement in
  $2+1$ dimensions},'' \href{http://dx.doi.org/10.1007/JHEP10(2017)173}{{\em
  JHEP} {\bfseries 10} (2017) 173},
  \href{http://arxiv.org/abs/1706.04949}{{\ttfamily arXiv:1706.04949
  [hep-th]}}.

\bibitem{Agarwal:2017roi}
P.~Agarwal, A.~Sciarappa, and J.~Song, ``{$ \mathcal{N} $ =1 Lagrangians for
  generalized Argyres-Douglas theories},''
  \href{http://dx.doi.org/10.1007/JHEP10(2017)211}{{\em JHEP} {\bfseries 10}
  (2017) 211}, \href{http://arxiv.org/abs/1707.04751}{{\ttfamily
  arXiv:1707.04751 [hep-th]}}.

\bibitem{Benvenuti:2017bpg}
S.~Benvenuti and S.~Giacomelli, ``{Lagrangians for generalized Argyres-Douglas
  theories},'' \href{http://dx.doi.org/10.1007/JHEP10(2017)106}{{\em JHEP}
  {\bfseries 10} (2017) 106}, \href{http://arxiv.org/abs/1707.05113}{{\ttfamily
  arXiv:1707.05113 [hep-th]}}.

\bibitem{Giacomelli:2017ckh}
S.~Giacomelli, ``{RG flows with supersymmetry enhancement and geometric
  engineering},'' \href{http://dx.doi.org/10.1007/JHEP06(2018)156}{{\em JHEP}
  {\bfseries 06} (2018) 156}, \href{http://arxiv.org/abs/1710.06469}{{\ttfamily
  arXiv:1710.06469 [hep-th]}}.

\bibitem{Agarwal:2018ejn}
P.~Agarwal, K.~Maruyoshi, and J.~Song, ``{A
  \textquotedblleft{}Lagrangian\textquotedblright{} for the E$_{7}$
  superconformal theory},''
  \href{http://dx.doi.org/10.1007/JHEP05(2018)193}{{\em JHEP} {\bfseries 05}
  (2018) 193}, \href{http://arxiv.org/abs/1802.05268}{{\ttfamily
  arXiv:1802.05268 [hep-th]}}.

\bibitem{Agarwal:2018oxb}
P.~Agarwal, ``{On dimensional reduction of 4d N=1 Lagrangians for
  Argyres-Douglas theories},''
  \href{http://dx.doi.org/10.1007/JHEP03(2019)011}{{\em JHEP} {\bfseries 03}
  (2019) 011}, \href{http://arxiv.org/abs/1809.10534}{{\ttfamily
  arXiv:1809.10534 [hep-th]}}.

\bibitem{Zafrir:2019hps}
G.~Zafrir, ``{An $ \mathcal{N} $ = 1 Lagrangian for the rank 1 E$_{6}$
  superconformal theory},''
  \href{http://dx.doi.org/10.1007/JHEP12(2020)098}{{\em JHEP} {\bfseries 12}
  (2020) 098}, \href{http://arxiv.org/abs/1912.09348}{{\ttfamily
  arXiv:1912.09348 [hep-th]}}.

\bibitem{Garozzo:2020pmz}
I.~Garozzo, N.~Mekareeya, M.~Sacchi, and G.~Zafrir, ``{Symmetry enhancement and
  duality walls in 5d gauge theories},''
  \href{http://dx.doi.org/10.1007/JHEP06(2020)159}{{\em JHEP} {\bfseries 06}
  (2020) 159}, \href{http://arxiv.org/abs/2003.07373}{{\ttfamily
  arXiv:2003.07373 [hep-th]}}.

\bibitem{Zafrir:2020epd}
G.~Zafrir, ``{An $ \mathcal{N} $ = 1 Lagrangian for an $ \mathcal{N} $ = 3
  SCFT},'' \href{http://dx.doi.org/10.1007/JHEP01(2021)062}{{\em JHEP}
  {\bfseries 01} (2021) 062}, \href{http://arxiv.org/abs/2007.14955}{{\ttfamily
  arXiv:2007.14955 [hep-th]}}.

\bibitem{Etxebarria:2021lmq}
I.~n.~G. Etxebarria, B.~Heidenreich, M.~Lotito, and A.~K. Sorout,
  ``{Deconfining $ \mathcal{N} $ = 2 SCFTs or the art of brane bending},''
  \href{http://dx.doi.org/10.1007/JHEP03(2022)140}{{\em JHEP} {\bfseries 03}
  (2022) 140}, \href{http://arxiv.org/abs/2111.08022}{{\ttfamily
  arXiv:2111.08022 [hep-th]}}.

\bibitem{Kang:2023pot}
M.~J. Kang, C.~Lawrie, K.-H. Lee, and J.~Song, ``{Emergent N=4 supersymmetry
  from N=1},'' \href{http://arxiv.org/abs/2302.06622}{{\ttfamily
  arXiv:2302.06622 [hep-th]}}.

\bibitem{Berkooz:1995km}
M.~Berkooz, ``{The Dual of supersymmetric SU(2k) with an antisymmetric tensor
  and composite dualities},''
  \href{http://dx.doi.org/10.1016/0550-3213(95)00400-M}{{\em Nucl. Phys. B}
  {\bfseries 452} (1995) 513--525},
  \href{http://arxiv.org/abs/hep-th/9505067}{{\ttfamily arXiv:hep-th/9505067}}.

\bibitem{Pouliot:1995me}
P.~Pouliot, ``{Duality in SUSY SU(N) with an antisymmetric tensor},''
  \href{http://dx.doi.org/10.1016/0370-2693(95)01427-6}{{\em Phys. Lett. B}
  {\bfseries 367} (1996) 151--156},
  \href{http://arxiv.org/abs/hep-th/9510148}{{\ttfamily arXiv:hep-th/9510148}}.

\bibitem{Luty:1996cg}
M.~A. Luty, M.~Schmaltz, and J.~Terning, ``{A Sequence of duals for Sp(2N)
  supersymmetric gauge theories with adjoint matter},''
  \href{http://dx.doi.org/10.1103/PhysRevD.54.7815}{{\em Phys. Rev. D}
  {\bfseries 54} (1996) 7815--7824},
  \href{http://arxiv.org/abs/hep-th/9603034}{{\ttfamily arXiv:hep-th/9603034}}.

\bibitem{Giacomelli:2017vgk}
S.~Giacomelli and N.~Mekareeya, ``{Mirror theories of 3d $ \mathcal{N} $ = 2
  SQCD},'' \href{http://dx.doi.org/10.1007/JHEP03(2018)126}{{\em JHEP}
  {\bfseries 03} (2018) 126}, \href{http://arxiv.org/abs/1711.11525}{{\ttfamily
  arXiv:1711.11525 [hep-th]}}.

\bibitem{Pasquetti:2019uop}
S.~Pasquetti and M.~Sacchi, ``{From 3$d$ dualities to 2$d$ free field
  correlators and back},''
  \href{http://dx.doi.org/10.1007/JHEP11(2019)081}{{\em JHEP} {\bfseries 11}
  (2019) 081}, \href{http://arxiv.org/abs/1903.10817}{{\ttfamily
  arXiv:1903.10817 [hep-th]}}.

\bibitem{Pasquetti:2019tix}
S.~Pasquetti and M.~Sacchi, ``{3d dualities from 2d free field correlators:
  recombination and rank stabilization},''
  \href{http://dx.doi.org/10.1007/JHEP01(2020)061}{{\em JHEP} {\bfseries 01}
  (2020) 061}, \href{http://arxiv.org/abs/1905.05807}{{\ttfamily
  arXiv:1905.05807 [hep-th]}}.

\bibitem{Sacchi:2020pet}
M.~Sacchi, ``{New 2d $ \mathcal{N} $ = (0, 2) dualities from four
  dimensions},'' \href{http://dx.doi.org/10.1007/JHEP12(2020)009}{{\em JHEP}
  {\bfseries 12} (2020) 009}, \href{http://arxiv.org/abs/2004.13672}{{\ttfamily
  arXiv:2004.13672 [hep-th]}}.

\bibitem{Benvenuti:2020gvy}
S.~Benvenuti, I.~Garozzo, and G.~Lo~Monaco, ``{Sequential deconfinement in 3d $
  \mathcal{N} $ = 2 gauge theories},''
  \href{http://dx.doi.org/10.1007/JHEP07(2021)191}{{\em JHEP} {\bfseries 07}
  (2021) 191}, \href{http://arxiv.org/abs/2012.09773}{{\ttfamily
  arXiv:2012.09773 [hep-th]}}.

\bibitem{Bajeot:2022kwt}
S.~Bajeot and S.~Benvenuti, ``{S-confinements from deconfinements},''
  \href{http://dx.doi.org/10.1007/JHEP11(2022)071}{{\em JHEP} {\bfseries 11}
  (2022) 071}, \href{http://arxiv.org/abs/2201.11049}{{\ttfamily
  arXiv:2201.11049 [hep-th]}}.

\bibitem{Bottini:2022vpy}
L.~E. Bottini, C.~Hwang, S.~Pasquetti, and M.~Sacchi, ``{Dualities from
  dualities: the sequential deconfinement technique},''
  \href{http://dx.doi.org/10.1007/JHEP05(2022)069}{{\em JHEP} {\bfseries 05}
  (2022) 069}, \href{http://arxiv.org/abs/2201.11090}{{\ttfamily
  arXiv:2201.11090 [hep-th]}}.

\bibitem{Bajeot:2022lah}
S.~Bajeot and S.~Benvenuti, ``{Sequential deconfinement and self-dualities in
  4d$ \mathcal{N} $ = 1 gauge theories},''
  \href{http://dx.doi.org/10.1007/JHEP10(2022)007}{{\em JHEP} {\bfseries 10}
  (2022) 007}, \href{http://arxiv.org/abs/2206.11364}{{\ttfamily
  arXiv:2206.11364 [hep-th]}}.

\bibitem{Benvenuti:2021nwt}
S.~Benvenuti and G.~Lo~Monaco, ``{A toolkit for ortho-symplectic dualities},''
  \href{http://arxiv.org/abs/2112.12154}{{\ttfamily arXiv:2112.12154
  [hep-th]}}.

\bibitem{Bajeot:2022wmu}
S.~Bajeot and S.~Benvenuti, ``{4d N=1 dualities from 5d dualities},''
  \href{http://arxiv.org/abs/2212.11217}{{\ttfamily arXiv:2212.11217
  [hep-th]}}.

\bibitem{Hwang:2020wpd}
C.~Hwang, S.~Pasquetti, and M.~Sacchi, ``{4d mirror-like dualities},''
  \href{http://dx.doi.org/10.1007/JHEP09(2020)047}{{\em JHEP} {\bfseries 09}
  (2020) 047}, \href{http://arxiv.org/abs/2002.12897}{{\ttfamily
  arXiv:2002.12897 [hep-th]}}.

\bibitem{Bottini:2021vms}
L.~E. Bottini, C.~Hwang, S.~Pasquetti, and M.~Sacchi, ``{4d S-duality wall and
  SL(2, \ensuremath{\mathbb{Z}}) relations},''
  \href{http://dx.doi.org/10.1007/JHEP03(2022)035}{{\em JHEP} {\bfseries 03}
  (2022) 035}, \href{http://arxiv.org/abs/2110.08001}{{\ttfamily
  arXiv:2110.08001 [hep-th]}}.

\bibitem{Hwang:2021ulb}
C.~Hwang, S.~Pasquetti, and M.~Sacchi, ``{Rethinking mirror symmetry as a local
  duality on fields},''
  \href{http://dx.doi.org/10.1103/PhysRevD.106.105014}{{\em Phys. Rev. D}
  {\bfseries 106} no.~10, (2022) 105014},
  \href{http://arxiv.org/abs/2110.11362}{{\ttfamily arXiv:2110.11362
  [hep-th]}}.

\bibitem{Comi:2022aqo}
R.~Comi, C.~Hwang, F.~Marino, S.~Pasquetti, and M.~Sacchi, ``{The
  $SL(2,\mathbb{Z})$ dualization algorithm at work},''
  \href{http://arxiv.org/abs/2212.10571}{{\ttfamily arXiv:2212.10571
  [hep-th]}}.

\bibitem{Intriligator:1995ne}
K.~A. Intriligator and P.~Pouliot, ``{Exact superpotentials, quantum vacua and
  duality in supersymmetric SP(N(c)) gauge theories},''
  \href{http://dx.doi.org/10.1016/0370-2693(95)00618-U}{{\em Phys. Lett. B}
  {\bfseries 353} (1995) 471--476},
  \href{http://arxiv.org/abs/hep-th/9505006}{{\ttfamily arXiv:hep-th/9505006}}.

\bibitem{Aharony:1997gp}
O.~Aharony, ``{IR duality in d = 3 N=2 supersymmetric USp(2N(c)) and U(N(c))
  gauge theories},''
  \href{http://dx.doi.org/10.1016/S0370-2693(97)00530-3}{{\em Phys. Lett. B}
  {\bfseries 404} (1997) 71--76},
  \href{http://arxiv.org/abs/hep-th/9703215}{{\ttfamily arXiv:hep-th/9703215}}.

\bibitem{2003math......9252R}
E.~M. {Rains}, ``{Transformations of elliptic hypergometric integrals},''
  \href{http://dx.doi.org/10.48550/arXiv.math/0309252}{{\em arXiv Mathematics
  e-prints} (Sept., 2003) math/0309252},
  \href{http://arxiv.org/abs/math/0309252}{{\ttfamily arXiv:math/0309252
  [math.QA]}}.

\bibitem{spiridonov2004theta}
V.~Spiridonov, ``Theta hypergeometric integrals,'' {\em St. Petersburg
  Mathematical Journal} {\bfseries 15} no.~6, (2004) 929--967.

\bibitem{Spiridonov:2009za}
V.~P. Spiridonov and G.~S. Vartanov, ``{Elliptic Hypergeometry of
  Supersymmetric Dualities},''
  \href{http://dx.doi.org/10.1007/s00220-011-1218-9}{{\em Commun. Math. Phys.}
  {\bfseries 304} (2011) 797--874},
  \href{http://arxiv.org/abs/0910.5944}{{\ttfamily arXiv:0910.5944 [hep-th]}}.

\bibitem{Spiridonov:2011hf}
V.~P. Spiridonov and G.~S. Vartanov, ``{Elliptic hypergeometry of
  supersymmetric dualities II. Orthogonal groups, knots, and vortices},''
  \href{http://dx.doi.org/10.1007/s00220-013-1861-4}{{\em Commun. Math. Phys.}
  {\bfseries 325} (2014) 421--486},
  \href{http://arxiv.org/abs/1107.5788}{{\ttfamily arXiv:1107.5788 [hep-th]}}.

\bibitem{Csaki:1996zb}
C.~Csaki, M.~Schmaltz, and W.~Skiba, ``{Confinement in N=1 SUSY gauge theories
  and model building tools},''
  \href{http://dx.doi.org/10.1103/PhysRevD.55.7840}{{\em Phys. Rev. D}
  {\bfseries 55} (1997) 7840--7858},
  \href{http://arxiv.org/abs/hep-th/9612207}{{\ttfamily arXiv:hep-th/9612207}}.

\bibitem{Seiberg:1994bz}
N.~Seiberg, ``{Exact results on the space of vacua of four-dimensional SUSY
  gauge theories},'' \href{http://dx.doi.org/10.1103/PhysRevD.49.6857}{{\em
  Phys. Rev. D} {\bfseries 49} (1994) 6857--6863},
  \href{http://arxiv.org/abs/hep-th/9402044}{{\ttfamily arXiv:hep-th/9402044}}.

\bibitem{Kutasov:1995ve}
D.~Kutasov, ``{A Comment on duality in N=1 supersymmetric nonAbelian gauge
  theories},'' \href{http://dx.doi.org/10.1016/0370-2693(95)00392-X}{{\em Phys.
  Lett. B} {\bfseries 351} (1995) 230--234},
  \href{http://arxiv.org/abs/hep-th/9503086}{{\ttfamily arXiv:hep-th/9503086}}.

\bibitem{Kutasov:1995np}
D.~Kutasov and A.~Schwimmer, ``{On duality in supersymmetric Yang-Mills
  theory},'' \href{http://dx.doi.org/10.1016/0370-2693(95)00676-C}{{\em Phys.
  Lett. B} {\bfseries 354} (1995) 315--321},
  \href{http://arxiv.org/abs/hep-th/9505004}{{\ttfamily arXiv:hep-th/9505004}}.

\bibitem{Intriligator:1995ff}
K.~A. Intriligator, ``{New RG fixed points and duality in supersymmetric
  SP(N(c)) and SO(N(c)) gauge theories},''
  \href{http://dx.doi.org/10.1016/0550-3213(95)00296-5}{{\em Nucl. Phys. B}
  {\bfseries 448} (1995) 187--198},
  \href{http://arxiv.org/abs/hep-th/9505051}{{\ttfamily arXiv:hep-th/9505051}}.

\bibitem{Gaiotto:2012xa}
D.~Gaiotto, L.~Rastelli, and S.~S. Razamat, ``{Bootstrapping the superconformal
  index with surface defects},''
  \href{http://dx.doi.org/10.1007/JHEP01(2013)022}{{\em JHEP} {\bfseries 01}
  (2013) 022}, \href{http://arxiv.org/abs/1207.3577}{{\ttfamily arXiv:1207.3577
  [hep-th]}}.

\bibitem{Benini:2017dud}
F.~Benini, S.~Benvenuti, and S.~Pasquetti, ``{SUSY monopole potentials in 2+1
  dimensions},'' \href{http://dx.doi.org/10.1007/JHEP08(2017)086}{{\em JHEP}
  {\bfseries 08} (2017) 086}, \href{http://arxiv.org/abs/1703.08460}{{\ttfamily
  arXiv:1703.08460 [hep-th]}}.

\bibitem{Xie:2012hs}
D.~Xie, ``{General Argyres-Douglas Theory},''
  \href{http://dx.doi.org/10.1007/JHEP01(2013)100}{{\em JHEP} {\bfseries 01}
  (2013) 100}, \href{http://arxiv.org/abs/1204.2270}{{\ttfamily arXiv:1204.2270
  [hep-th]}}.

\bibitem{Cecotti:2012jx}
S.~Cecotti and M.~Del~Zotto, ``{Infinitely many N=2 SCFT with ADE flavor
  symmetry},'' \href{http://dx.doi.org/10.1007/JHEP01(2013)191}{{\em JHEP}
  {\bfseries 01} (2013) 191}, \href{http://arxiv.org/abs/1210.2886}{{\ttfamily
  arXiv:1210.2886 [hep-th]}}.

\bibitem{Cecotti:2013lda}
S.~Cecotti, M.~Del~Zotto, and S.~Giacomelli, ``{More on the N=2 superconformal
  systems of type $D_p(G)$},''
  \href{http://dx.doi.org/10.1007/JHEP04(2013)153}{{\em JHEP} {\bfseries 04}
  (2013) 153}, \href{http://arxiv.org/abs/1303.3149}{{\ttfamily arXiv:1303.3149
  [hep-th]}}.

\bibitem{Kang:2021lic}
M.~J. Kang, C.~Lawrie, and J.~Song, ``{Infinitely many 4D N=2 SCFTs with a=c
  and beyond},'' \href{http://dx.doi.org/10.1103/PhysRevD.104.105005}{{\em
  Phys. Rev. D} {\bfseries 104} no.~10, (2021) 105005},
  \href{http://arxiv.org/abs/2106.12579}{{\ttfamily arXiv:2106.12579
  [hep-th]}}.

\bibitem{Kang:2021ccs}
M.~J. Kang, C.~Lawrie, K.-H. Lee, and J.~Song, ``{Infinitely many 4D N=1 SCFTs
  with a=c},'' \href{http://dx.doi.org/10.1103/PhysRevD.105.126006}{{\em Phys.
  Rev. D} {\bfseries 105} no.~12, (2022) 126006},
  \href{http://arxiv.org/abs/2111.12092}{{\ttfamily arXiv:2111.12092
  [hep-th]}}.

\bibitem{Kang:2022zsl}
M.~J. Kang, C.~Lawrie, K.-H. Lee, M.~Sacchi, and J.~Song, ``{Higgs branch,
  Coulomb branch, and Hall-Littlewood index},''
  \href{http://dx.doi.org/10.1103/PhysRevD.106.106021}{{\em Phys. Rev. D}
  {\bfseries 106} no.~10, (2022) 106021},
  \href{http://arxiv.org/abs/2207.05764}{{\ttfamily arXiv:2207.05764
  [hep-th]}}.

\bibitem{Kang:2022vab}
M.~J. Kang, C.~Lawrie, K.-H. Lee, and J.~Song, ``{Operator spectroscopy for 4D
  SCFTs with a=c},'' \href{http://dx.doi.org/10.1103/PhysRevD.107.066018}{{\em
  Phys. Rev. D} {\bfseries 107} no.~6, (2023) 066018},
  \href{http://arxiv.org/abs/2210.06497}{{\ttfamily arXiv:2210.06497
  [hep-th]}}.

\bibitem{Buican:2015hsa}
M.~Buican and T.~Nishinaka, ``{Argyres\textendash{}Douglas theories, S$^1$
  reductions, and topological symmetries},''
  \href{http://dx.doi.org/10.1088/1751-8113/49/4/045401}{{\em J. Phys. A}
  {\bfseries 49} no.~4, (2016) 045401},
  \href{http://arxiv.org/abs/1505.06205}{{\ttfamily arXiv:1505.06205
  [hep-th]}}.

\bibitem{Giacomelli:2020ryy}
S.~Giacomelli, N.~Mekareeya, and M.~Sacchi, ``{New aspects of Argyres--Douglas
  theories and their dimensional reduction},''
  \href{http://dx.doi.org/10.1007/JHEP03(2021)242}{{\em JHEP} {\bfseries 03}
  (2021) 242}, \href{http://arxiv.org/abs/2012.12852}{{\ttfamily
  arXiv:2012.12852 [hep-th]}}.

\bibitem{Closset:2020afy}
C.~Closset, S.~Giacomelli, S.~Schafer-Nameki, and Y.-N. Wang, ``{5d and 4d
  SCFTs: Canonical Singularities, Trinions and S-Dualities},''
  \href{http://dx.doi.org/10.1007/JHEP05(2021)274}{{\em JHEP} {\bfseries 05}
  (2021) 274}, \href{http://arxiv.org/abs/2012.12827}{{\ttfamily
  arXiv:2012.12827 [hep-th]}}.

\bibitem{Hwang:2021xyw}
C.~Hwang, S.~S. Razamat, E.~Sabag, and M.~Sacchi, ``{Rank $Q$ E-string on
  spheres with flux},''
  \href{http://dx.doi.org/10.21468/SciPostPhys.11.2.044}{{\em SciPost Phys.}
  {\bfseries 11} no.~2, (2021) 044},
  \href{http://arxiv.org/abs/2103.09149}{{\ttfamily arXiv:2103.09149
  [hep-th]}}.

\bibitem{Pasquetti:2019hxf}
S.~Pasquetti, S.~S. Razamat, M.~Sacchi, and G.~Zafrir, ``{Rank $Q$ E-string on
  a torus with flux},''
  \href{http://dx.doi.org/10.21468/SciPostPhys.8.1.014}{{\em SciPost Phys.}
  {\bfseries 8} no.~1, (2020) 014},
  \href{http://arxiv.org/abs/1908.03278}{{\ttfamily arXiv:1908.03278
  [hep-th]}}.

\bibitem{Aharony:2013dha}
O.~Aharony, S.~S. Razamat, N.~Seiberg, and B.~Willett, ``{3d dualities from 4d
  dualities},'' \href{http://dx.doi.org/10.1007/JHEP07(2013)149}{{\em JHEP}
  {\bfseries 07} (2013) 149}, \href{http://arxiv.org/abs/1305.3924}{{\ttfamily
  arXiv:1305.3924 [hep-th]}}.

\bibitem{Aharony:2013kma}
O.~Aharony, S.~S. Razamat, N.~Seiberg, and B.~Willett, ``{3$d$ dualities from
  4$d$ dualities for orthogonal groups},''
  \href{http://dx.doi.org/10.1007/JHEP08(2013)099}{{\em JHEP} {\bfseries 08}
  (2013) 099}, \href{http://arxiv.org/abs/1307.0511}{{\ttfamily arXiv:1307.0511
  [hep-th]}}.

\bibitem{Sacchi:2021afk}
M.~Sacchi, O.~Sela, and G.~Zafrir, ``{Compactifying 5d superconformal field
  theories to 3d},'' \href{http://dx.doi.org/10.1007/JHEP09(2021)149}{{\em
  JHEP} {\bfseries 09} (2021) 149},
  \href{http://arxiv.org/abs/2105.01497}{{\ttfamily arXiv:2105.01497
  [hep-th]}}.

\bibitem{Sacchi:2021wvg}
M.~Sacchi, O.~Sela, and G.~Zafrir, ``{On the 3d compactifications of 5d SCFTs
  associated with SU(N + 1) gauge theories},''
  \href{http://dx.doi.org/10.1007/JHEP05(2022)053}{{\em JHEP} {\bfseries 05}
  (2022) 053}, \href{http://arxiv.org/abs/2111.12745}{{\ttfamily
  arXiv:2111.12745 [hep-th]}}.

\bibitem{Sacchi:2023rtp}
M.~Sacchi, O.~Sela, and G.~Zafrir, ``{Trinions for the $3d$ compactification of
  the $5d$ rank 1 $E_{N_f+1}$ SCFTs},''
  \href{http://arxiv.org/abs/2301.06561}{{\ttfamily arXiv:2301.06561
  [hep-th]}}.

\bibitem{Gaiotto:2014kfa}
D.~Gaiotto, A.~Kapustin, N.~Seiberg, and B.~Willett, ``{Generalized Global
  Symmetries},'' \href{http://dx.doi.org/10.1007/JHEP02(2015)172}{{\em JHEP}
  {\bfseries 02} (2015) 172}, \href{http://arxiv.org/abs/1412.5148}{{\ttfamily
  arXiv:1412.5148 [hep-th]}}.

\bibitem{Jafferis:2011ns}
D.~Jafferis and X.~Yin, ``{A Duality Appetizer},''
  \href{http://arxiv.org/abs/1103.5700}{{\ttfamily arXiv:1103.5700 [hep-th]}}.

\bibitem{Spiridonov:2014cxa}
V.~P. Spiridonov and G.~S. Vartanov, ``{Vanishing superconformal indices and
  the chiral symmetry breaking},''
  \href{http://dx.doi.org/10.1007/JHEP06(2014)062}{{\em JHEP} {\bfseries 06}
  (2014) 062}, \href{http://arxiv.org/abs/1402.2312}{{\ttfamily arXiv:1402.2312
  [hep-th]}}.

\bibitem{BADpaper}
S.~Giacomelli, C.~Hwang, F.~Marino, S.~Pasquetti, and M.~Sacchi, ``{to
  appear},''.

\bibitem{Leigh:1995ep}
R.~G. Leigh and M.~J. Strassler, ``{Exactly marginal operators and duality in
  four-dimensional N=1 supersymmetric gauge theory},''
  \href{http://dx.doi.org/10.1016/0550-3213(95)00261-P}{{\em Nucl. Phys. B}
  {\bfseries 447} (1995) 95--136},
  \href{http://arxiv.org/abs/hep-th/9503121}{{\ttfamily arXiv:hep-th/9503121}}.

\bibitem{Kol:2002zt}
B.~Kol, ``{On conformal deformations},''
  \href{http://dx.doi.org/10.1088/1126-6708/2002/09/046}{{\em JHEP} {\bfseries
  09} (2002) 046}, \href{http://arxiv.org/abs/hep-th/0205141}{{\ttfamily
  arXiv:hep-th/0205141}}.

\bibitem{Benvenuti:2005wi}
S.~Benvenuti and A.~Hanany, ``{Conformal manifolds for the conifold and other
  toric field theories},''
  \href{http://dx.doi.org/10.1088/1126-6708/2005/08/024}{{\em JHEP} {\bfseries
  08} (2005) 024}, \href{http://arxiv.org/abs/hep-th/0502043}{{\ttfamily
  arXiv:hep-th/0502043}}.

\bibitem{Green:2010da}
D.~Green, Z.~Komargodski, N.~Seiberg, Y.~Tachikawa, and B.~Wecht, ``{Exactly
  Marginal Deformations and Global Symmetries},''
  \href{http://dx.doi.org/10.1007/JHEP06(2010)106}{{\em JHEP} {\bfseries 06}
  (2010) 106}, \href{http://arxiv.org/abs/1005.3546}{{\ttfamily arXiv:1005.3546
  [hep-th]}}.

\bibitem{Kol:2010ub}
B.~Kol, ``{On Conformal Deformations II},''
  \href{http://arxiv.org/abs/1005.4408}{{\ttfamily arXiv:1005.4408 [hep-th]}}.

\bibitem{Intriligator:2003jj}
K.~A. Intriligator and B.~Wecht, ``{The Exact superconformal R symmetry
  maximizes a},'' \href{http://dx.doi.org/10.1016/S0550-3213(03)00459-0}{{\em
  Nucl. Phys. B} {\bfseries 667} (2003) 183--200},
  \href{http://arxiv.org/abs/hep-th/0304128}{{\ttfamily arXiv:hep-th/0304128}}.

\bibitem{Jafferis:2010un}
D.~L. Jafferis, ``{The Exact Superconformal R-Symmetry Extremizes Z},''
  \href{http://dx.doi.org/10.1007/JHEP05(2012)159}{{\em JHEP} {\bfseries 05}
  (2012) 159}, \href{http://arxiv.org/abs/1012.3210}{{\ttfamily arXiv:1012.3210
  [hep-th]}}.

\bibitem{Xie:2016evu}
D.~Xie, W.~Yan, and S.-T. Yau, ``{Chiral algebra of the Argyres-Douglas theory
  from M5 branes},'' \href{http://dx.doi.org/10.1103/PhysRevD.103.065003}{{\em
  Phys. Rev. D} {\bfseries 103} no.~6, (2021) 065003},
  \href{http://arxiv.org/abs/1604.02155}{{\ttfamily arXiv:1604.02155
  [hep-th]}}.

\bibitem{Xie:2015rpa}
D.~Xie and S.-T. Yau, ``{4d N=2 SCFT and singularity theory Part I:
  Classification},'' \href{http://arxiv.org/abs/1510.01324}{{\ttfamily
  arXiv:1510.01324 [hep-th]}}.

\bibitem{Anselmi:1997am}
D.~Anselmi, D.~Z. Freedman, M.~T. Grisaru, and A.~A. Johansen,
  ``{Nonperturbative formulas for central functions of supersymmetric gauge
  theories},'' \href{http://dx.doi.org/10.1016/S0550-3213(98)00278-8}{{\em
  Nucl. Phys. B} {\bfseries 526} (1998) 543--571},
  \href{http://arxiv.org/abs/hep-th/9708042}{{\ttfamily arXiv:hep-th/9708042}}.

\bibitem{Gadde:2011uv}
A.~Gadde, L.~Rastelli, S.~S. Razamat, and W.~Yan, ``{Gauge Theories and
  Macdonald Polynomials},''
  \href{http://dx.doi.org/10.1007/s00220-012-1607-8}{{\em Commun. Math. Phys.}
  {\bfseries 319} (2013) 147--193},
  \href{http://arxiv.org/abs/1110.3740}{{\ttfamily arXiv:1110.3740 [hep-th]}}.

\bibitem{Song:2017oew}
J.~Song, D.~Xie, and W.~Yan, ``{Vertex operator algebras of Argyres-Douglas
  theories from M5-branes},''
  \href{http://dx.doi.org/10.1007/JHEP12(2017)123}{{\em JHEP} {\bfseries 12}
  (2017) 123}, \href{http://arxiv.org/abs/1706.01607}{{\ttfamily
  arXiv:1706.01607 [hep-th]}}.

\bibitem{Kinney:2005ej}
J.~Kinney, J.~M. Maldacena, S.~Minwalla, and S.~Raju, ``{An Index for 4
  dimensional super conformal theories},''
  \href{http://dx.doi.org/10.1007/s00220-007-0258-7}{{\em Commun. Math. Phys.}
  {\bfseries 275} (2007) 209--254},
  \href{http://arxiv.org/abs/hep-th/0510251}{{\ttfamily arXiv:hep-th/0510251}}.

\bibitem{Romelsberger:2005eg}
C.~Romelsberger, ``{Counting chiral primaries in N = 1, d=4 superconformal
  field theories},''
  \href{http://dx.doi.org/10.1016/j.nuclphysb.2006.03.037}{{\em Nucl. Phys. B}
  {\bfseries 747} (2006) 329--353},
  \href{http://arxiv.org/abs/hep-th/0510060}{{\ttfamily arXiv:hep-th/0510060}}.

\bibitem{Dolan:2008qi}
F.~A. Dolan and H.~Osborn, ``{Applications of the Superconformal Index for
  Protected Operators and q-Hypergeometric Identities to N=1 Dual Theories},''
  \href{http://dx.doi.org/10.1016/j.nuclphysb.2009.01.028}{{\em Nucl. Phys. B}
  {\bfseries 818} (2009) 137--178},
  \href{http://arxiv.org/abs/0801.4947}{{\ttfamily arXiv:0801.4947 [hep-th]}}.

\bibitem{Rastelli:2016tbz}
L.~Rastelli and S.~S. Razamat, ``{The supersymmetric index in four
  dimensions},'' \href{http://dx.doi.org/10.1088/1751-8121/aa76a6}{{\em J.
  Phys. A} {\bfseries 50} no.~44, (2017) 443013},
  \href{http://arxiv.org/abs/1608.02965}{{\ttfamily arXiv:1608.02965
  [hep-th]}}.

\bibitem{Seiberg:1994pq}
N.~Seiberg, ``{Electric - magnetic duality in supersymmetric nonAbelian gauge
  theories},'' \href{http://dx.doi.org/10.1016/0550-3213(94)00023-8}{{\em Nucl.
  Phys. B} {\bfseries 435} (1995) 129--146},
  \href{http://arxiv.org/abs/hep-th/9411149}{{\ttfamily arXiv:hep-th/9411149}}.

\end{thebibliography}\endgroup
\end{document}